\documentclass[12pt]{article}
\pdfoutput=1
\usepackage{amsmath,amssymb,amsthm,mathtools,bm,graphicx,float,array,multirow,multicol,rotfloat,caption,subcaption,enumerate,geometry,mathdots,adjustbox,booktabs,parskip,mathtools,tikz,tikz-cd,csquotes,rotating,empheq,mathrsfs,xfrac}
\usepackage{pdflscape}
\usepackage{dsfont}
\usepackage{pifont}
\usepackage[all]{xy}
\usepackage[shortlabels]{enumitem}
\usepackage[normalem]{ulem}
\usepackage[vcentermath]{youngtab}
\usepackage[boxsize=.8em]{ytableau}
\usepackage[numbers,sort&compress]{natbib}
\usepackage[hidelinks,linktocpage]{hyperref}
\usepackage[capitalise,sort]{cleveref}
\usepackage{longtable}
\usepackage{hhline}
\numberwithin{equation}{section}
\graphicspath{{figs/}}
\makeatletter
\def\myrotate{\ifodd\c@page\else+\fi 90}
\g@addto@macro{\landscape}{\PLS@Rotate{\myrotate}}
\usetikzlibrary{arrows,shapes.misc,positioning,decorations.pathmorphing,decorations.markings,decorations.pathreplacing,matrix,patterns,backgrounds,calc}
\tikzset{
scale cd/.style={every label/.append style={scale=#1}, cells={nodes={scale=#1}}},
gauge/.style={rounded rectangle, draw=black!100, thick, minimum size=2mm}, 
gaugeD/.style={rounded rectangle, draw=black!100,double,thick,minimum size=2mm},  
empty/.style={rounded rectangle, draw=white!100, thick, minimum size=2mm}, 
flavor/.style={rectangle, draw=black!100, thick, minimum size=2mm},
flavorD/.style={rectangle, draw=black!100, double,thick, minimum size=2mm},
node/.style={circle, thick, draw=black!100,fill=white!100,  minimum size=2mm, inner sep=0pt},
sqnode/.style={rectangle
, thick, draw=black!100,fill=white!100,  minimum size=2mm, inner sep=0pt
},
sonode/.style={circle, thick, draw=black!100,fill=red!100,  minimum size=2mm, inner sep=0pt},
spnode/.style={circle, thick, draw=black!100,fill=blue!100,  minimum size=2mm, inner sep=0pt},
fnode/.style={rectangle, thick, draw=black!100,fill=white!100,  minimum size=2mm, inner sep=0pt},
tnode/.style={rounded rectangle, outer sep=0pt, thick, minimum size=2mm},
brace/.style={decoration={brace, mirror},decorate}
}
\hypersetup{colorlinks=true}
\hypersetup{linkcolor=black}
\hypersetup{citecolor=black}
\hypersetup{urlcolor=black}
\theoremstyle{plain}
\newtheorem*{thm*}{Theorem}

\makeatother

\DeclareSymbolFont{largesymbolsA}{U}{jkpexa}{m}{n}
\SetSymbolFont{largesymbolsA}{bold}{U}{jkpexa}{bx}{n}
\DeclareMathSymbol{\varprod}{\mathop}{largesymbolsA}{16}

\DeclareMathOperator{\U}{U}
\DeclareMathOperator{\SU}{SU}
\DeclareMathOperator{\SO}{SO}

\DeclareMathOperator{\USp}{USp}

\DeclareMathOperator{\diag}{diag}
\newcommand{\CC}{\mathbb{C}}

\newcommand{\ZZ}{\mathbb{Z}}

\newcommand{\ID}{\mathds{1}}

\newcommand{\coma}{\, , \quad}
\newcommand{\fstop}{\, .}

\newcommand{\PE}{\text{PE}}

\newcommand{\HS}{\text{HS}}
\newcommand{\sfstop}[2]{\begin{tikzpicture}[baseline=#2]
    \node at (0,0) {#1};
\end{tikzpicture}}

\def\su{\mathfrak{su}}

\def\so{\mathfrak{so}}

\def\usp{\mathfrak{usp}}

\newcommand{\Lpagenumber}{\ifdim\textwidth=\linewidth\else\bgroup
	\dimendef\margin=0 
	\ifodd\value{page}\margin=\oddsidemargin
	\else\margin=\evensidemargin
	\fi
	\raisebox{\dimexpr -3\topmargin-\headheight-\headsep-0.5\linewidth}[0pt][0pt]{%
		\rlap{\hspace{\dimexpr \margin+\textheight+3\footskip}%
			\llap{\rotatebox{90}{\hspace{-4.5cm}\thepage\hfill}}}}%
	\egroup\fi}
\AddToHook{shipout/background}{\Lpagenumber}%

\def\fnote#1#2{\begingroup\def\thefootnote{#1}\footnote{#2}
     \addtocounter{footnote}{-1}\endgroup}

\tikzset{
    ncbar angle/.initial=90,
    ncbar/.style={
        to path=(\tikztostart)
        -- ($(\tikztostart)!#1!\pgfkeysvalueof{/tikz/ncbar angle}:(\tikztotarget)$)
        -- ($(\tikztotarget)!($(\tikztostart)!#1!\pgfkeysvalueof{/tikz/ncbar angle}:(\tikztotarget)$)!\pgfkeysvalueof{/tikz/ncbar angle}:(\tikztostart)$)
        -- (\tikztotarget)
    },
    ncbar/.default=0.5cm,
}

\tikzset{square left brace/.style={ncbar=0.2cm}}
\tikzset{square right brace/.style={ncbar=-0.2cm}}

\hypersetup{
	pdftitle={Discrete Gauging of 6d SCFTs and Wreathed 3d N=4 Quivers},    
	pdfauthor={\textcopyright\ Craig Lawrie, Thekla Lepper, Alessandro Mininno},     
	pdfsubject={hep-th},   
	pdfcreator={pdfLaTex},   
	pdfproducer={LaTex}, 
	pdfkeywords={},
	colorlinks=true
  ,urlcolor=blue
  ,anchorcolor=blue
  ,citecolor=blue
  ,filecolor=blue
  ,linkcolor=blue
  ,menucolor=blue
  ,linktocpage=true
}

\crefname{figure}{Figure}{Figures}
\crefname{table}{Table}{Tables}
\crefname{definition}{Definition}{Definitions}
\crefname{proposition}{Proposition}{Propositions}
\crefname{claim}{Claim}{Claims}
\crefname{conjecture}{Conjecture}{Conjectures}

\allowdisplaybreaks[1]

\begin{document}

\begin{titlepage}
\vspace*{-3cm} 
\begin{flushright}
{\tt DESY-25-025 
}\\ 
\end{flushright}
\begin{center}
\vspace{1.7cm}
{\LARGE\bfseries Discrete Gauging of 6d SCFTs \\[0.5em] and Wreathed 3d \boldmath{$\mathcal{N}=4$} Quivers}
\vspace{1cm}

{\large
Craig Lawrie,$^1$ Thekla Lepper,$^2$ and Alessandro Mininno$^{3,a}$}\\
\vspace{.8cm}
{$^1$ Deutsches Elektronen-Synchrotron DESY,\\ Notkestr.~85, 22607 Hamburg, Germany}\par
\vspace{.2cm}
{$^2$ Dipartimento di Fisica, Università di Torino, and INFN Sezione di Torino \\
Via P. Giuria 1, I-10125 Torino, Italy}\par
\vspace{.2cm}
{$^3$ Department of Physics, University of Wisconsin--Madison,\\1150 University Avenue, Madison, WI 53706, USA}\par
\vspace{.3cm}

\scalebox{.8}{\tt craig.lawrie1729@gmail.com, 
thekla.lepper@unito.it,
mininno@physics.wisc.edu}\par
\vspace{1.2cm}

\fnote{}{${}^a$ Part of this project was conducted when the author was affiliated to II. Institut f\"ur Theoretische Physik, Universit\"at Hamburg, Luruper Chaussee 149, 22607 Hamburg, Germany.}

\end{center}

\vspace{-0.5cm}

\begin{abstract}
    We study the Higgs branch moduli space of certain 6d $(1,0)$ SCFTs after gauging their $\mathbb{Z}_2$ Green--Schwarz automorphism. We explain how to read the flavor symmetry of such SCFTs directly from the 6d construction, and we confirm the expectation by computing the Coulomb branch Hilbert series of their $\mathbb{Z}_2$-wreathed 3d $\mathcal{N}=4$ magnetic quiver. To perform the latter computation, we explicitly introduce a methodology to determine such Hilbert series for $\mathbb{Z}_2$-wreathed orthosymplectic quivers.  
\end{abstract}

\vspace{1cm}
\vfill 
\end{titlepage}

\tableofcontents
\newpage

\section{Introduction}

In recent years, there has been a proliferation of constructions and (attempted) classifications of higher-dimensional superconformal field theories (SCFTs), typically driven via geometric approaches.\footnote{See \cite{Akhond:2021xio,Argyres:2022mnu} for recent reviews of some of the various constructions of SCFTs in various dimensions.} Often, discretely-gauged SCFTs play a crucial role in testing the robustness of such constructions and the techniques for extracting physical properties.
Consider $\mathcal{T}$ an SCFT with a discrete global symmetry $\Gamma$, obtained from such a geometric or top-down approach, and then consider the new theory $\widetilde{\mathcal{T}}$ obtained by gauging $\Gamma$. Then, the theories $\widetilde{\mathcal{T}}$ can source novel properties that are absent in their non-discretely-gauged cousins. For example, in the world of 4d $\mathcal{N}=2$ SCFTs, such discrete gaugings are known \cite{Argyres:2017tmj,Bourton:2018jwb,Bourget:2018ond,Argyres:2018wxu} to violate the conjecture \cite{Beem:2014zpa,Tachikawa:2013kta} that the Coulomb branch chiral ring is freely generated, which leads to a failure of the Shapere--Tachikawa \cite{Shapere:2008zf} formula for the central charges in terms of the Coulomb branch operators. In this paper, we explore a similar discrete-gauging operation on the 6d $(1,0)$ SCFTs obtained via the geometric construction \cite{Heckman:2013pva,Heckman:2015bfa} in F-theory. We focus on the $\frac{1}{2}$-BPS operator sector known as the Higgs branch chiral ring of particular 6d SCFTs, which realize a discrete global symmetry from a $\mathbb{Z}_2$ Green--Schwarz (GS) automorphism \cite{Apruzzi:2017iqe}, and analyze the consequences of the gauging of such a symmetry.

The 6d $(1,0)$ SCFTs are generically strongly-coupled, and thus it can be challenging to extract their properties. More generally, let $\mathcal{T}$ denote an arbitrary eight-supercharge theory in dimensions $\geq 3$, for which we would like to analyze the Higgs branch. For example, one might want to study what the interacting fixed points along subloci inside the Higgs branch are, or know what $\frac{1}{2}$-BPS operators belong to the subsector of the theory known as the Higgs branch chiral ring. Moreover, it is interesting to know if the Higgs branch chiral ring is finitely generated and what its generators and the relations among them are. One way to make progress on such questions is to find a 3d $\mathcal{N}=4$ Lagrangian theory, $\mathcal{T}_M$, such that the Coulomb branch of $\mathcal{T}_M$ is isomorphic to the Higgs branch of $\mathcal{T}$:\footnote{More generally, the Higgs branch may be isomorphic to the union of Coulomb branches of a collection of magnetic quivers, but this situation will not arise in this paper.}
\begin{equation}
    \operatorname{HB} \left[ \mathcal{T} \right] \,\,\cong\,\, \operatorname{CB} \left[ \mathcal{T}_M \right] \,.
\end{equation}
Any such $\mathcal{T}_M$ is known as a magnetic quiver for the Higgs branch of $\mathcal{T}$ \cite{Hanany:1996ie, Hanany:1997gh, Ferlito:2017xdq}. For many of the 6d $(1,0)$ SCFTs, there are known $\mathcal{T}_M$, see, for example, \cite{Mekareeya:2017jgc,Hanany:2018uhm,Hanany:2018vph,Cabrera:2019izd,Cabrera:2019dob,Fazzi:2022hal,Lawrie:2023uiu,Lawrie:2024zon,Lawrie:2024wan}.

Once a magnetic quiver is known, we can take advantage of the Lagrangian description to study the Coulomb branch, for which there are a variety of well-developed tools. For example, if we are interested in the spectrum of operators belonging to the Coulomb branch chiral ring, as well as the generators and relations, then we can use the monopole formula \cite{Cremonesi:2013lqa} to determine the Coulomb branch Hilbert series. Alternatively, if we are interested in the interacting fixed points on the Coulomb branch, and the vacuum expectation values (VEVs) that must be given to trigger the corresponding renormalization group flows, we can use a quiver subtraction algorithm \cite{Cabrera:2018ann} or the decay and fission algorithm \cite{Bourget:2023dkj,Bourget:2024mgn,Lawrie:2024wan}. In this way, we learn detailed information about the Higgs branch of (non-Lagrangian) higher-dimensional SCFTs.

We now suppose that $\mathcal{T}$ denotes a 6d $(1,0)$ SCFT with discrete global symmetry $\Gamma$, and with magnetic quiver for the Higgs branch $\mathcal{T}_M$. Furthermore, we suppose that we can gauge $\Gamma$,\footnote{In particular, we are assuming that there are no discrete anomalies for this $\Gamma$. Evaluating the 6d 't Hooft anomalies for discrete symmetries, perhaps along the lines of \cite{Hsieh:2018ifc} in 4d, is an interesting problem, but beyond the scope of this work. Note that mixed anomalies between $\Gamma$ and other global symmetries are not a problem; gauging $\Gamma$ then leads to a theory with non-invertible symmetries as discussed in \cite{Lawrie:2023tdz,Apruzzi:2024cty}.} leading to a new 6d $(1,0)$ SCFT which we denote as $\widetilde{\mathcal{T}}$. To understand the structure of the Higgs branch of this discretely-gauged theory, it would be useful to find a magnetic quiver for the Higgs branch: $\widetilde{\mathcal{T}}_M$. In this paper, building upon previous work in lower-dimensional field theories \cite{Arias-Tamargo:2019jyh,Hanany:2018cgo,Bourget:2020bxh,Arias-Tamargo:2022nlf,Hanany:2023uzn,Giacomelli:2024sex,Grimminger:2024mks}, we propose that the magnetic quiver for the Higgs branch of the discretely-gauged theory is the $\Gamma$ wreathing of the magnetic quiver of the SCFT before discrete gauging:
\begin{equation}\label{eqn:bigboy}
    \widetilde{\mathcal{T}}_M = \mathcal{T}_M \wr \Gamma \,.
\end{equation}

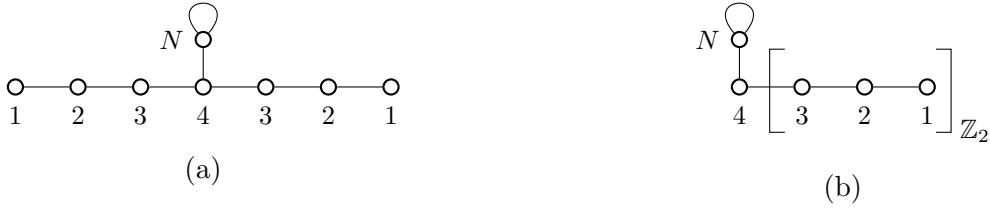
\begin{figure}[t]
    \centering
    \begin{subfigure}[t]{0.49\textwidth}
    \centering
    \begin{tikzpicture}[baseline=0,font=\footnotesize]
        \node[node, label=below:{$1$}] (A1) {};
      \node[node, label=below:{$2$}] (A2) [right=6mm of A1] {};
      \node[node, label=below:{$3$}] (A3) [right=6mm of A2] {};
      \node[node, label=below:{$4$}] (N4) [right=6mm of A3] {};
      \node[node, label=left:{$N$}] (Nu) [above=4mm of N4] {};
      \node[node, label=below:{$3$}] (A4) [right=6mm of N4] {};
      \node[node, label=below:{$2$}] (A5) [right=6mm of A4] {};
      \node[node, label=below:{$1$}] (A6) [right=6mm of A5]{};
    
        \draw (A1.east) -- (A2.west);
        \draw (A2.east) -- (A3.west);
        \draw (A3.east) -- (N4.west);
        \draw (N4.east) -- (A4.west);
        \draw (A4.east) -- (A5.west);
        \draw (A5.east) -- (A6.west);  
        \draw (N4.north) -- (Nu.south);
      \draw (Nu) to[out=130, in=410, looseness=12] (Nu);
    \end{tikzpicture}
    \caption{}
    \label{fig:exquiv1}
    \end{subfigure}\hfill
    \begin{subfigure}[t]{0.49\textwidth}
    \centering
 \begin{tikzpicture}[baseline=0,font=\footnotesize]
      \node[node, label=below:{$4$}] (N4)  {};
      \node[node, label=left:{$N$}] (Nu) [above=4mm of N4] {};
      \node[node, label=below:{$3$}] (A4) [right=6mm of N4] {};
      \node[node, label=below:{$2$}] (A5) [right=6mm of A4] {};
      \node[node, label=below:{$1$}] (A6) [right=6mm of A5]{};
        \draw (N4.east) -- (A4.west);
        \draw (A4.east) -- (A5.west);
        \draw (A5.east) -- (A6.west);  
        \draw (N4.north) -- (Nu.south);
      \draw (Nu) to[out=130, in=410, looseness=12] (Nu);
      \draw 
 (0.6,0.5) to [square right brace] (0.6,-0.6);
\draw
(2.6,0.5) to [square left brace] (2.6,-0.6);
       \node (Z2) [below right=3mm of A6] {$\ZZ_2$};
    \end{tikzpicture}
    \caption{}
     \label{fig:exquiv2}
    \end{subfigure}
    \caption{In Figure \ref{fig:exquiv1}, we depict an example of a unitary quiver that has two identical tails attached to the central node, and thus has a $\mathbb{Z}_2$ quiver automorphism which can be wreathed. In Figure \ref{fig:exquiv2}, we depict the wreathed quiver obtained by $\mathbb{Z}_2$ wreathing of the quiver in Figure \ref{fig:exquiv1}.}
    \label{fig:wreathingexamples}
\end{figure}

To define what it means for a quiver to admit a wreathing, we must first define the group-theoretic notion of the wreath product. In general, consider a reductive Lie group $G$, and then the wreath product of $G$ by $\Gamma \subseteq S_k$ is defined as
\begin{equation}
    G\wr \Gamma \equiv \left(\prod_{i=1}^kG_i\right)\rtimes \Gamma\coma
\end{equation}
where $\times$ is the Cartesian product of sets, and $G_i$ with $i=1,\ldots,k$ are $k$ copies of the original group $G$. An element of $G\wr \Gamma$ is denoted $(g,\sigma)$ given by $k$ elements $g_i\in G_i$ and a permutation $\sigma \in \Gamma$. In this way, $G\wr \Gamma$ can be seen as the direct product of $k$ copies of $G$, which can be permuted by $\Gamma$.  Since in this paper we are interested in $\ZZ_2\simeq S_2$ wreathing, we explain briefly how it works for an $S_2$ wreathing of a group $G$. One first considers the direct product of two $G$ groups, i.e., 
\begin{equation}
    G^2 = G \times G \,,
\end{equation} and we consider the action 
\begin{equation}
    f\,:\,S_2 \rightarrow \mathrm{Aut}(G^2) \,,
\end{equation}
such that given an element $g_i \in G_i$, then 
$f$ acts as $f(\sigma)(g_1,g_2) = (g_{\sigma(1)},g_{\sigma(2)})$, where $\sigma \in S_2$, for all elements $g_i \in G$. We note that, if $G$ is of order $|G|$, then $G\wr S_2$ is of order $2|G|^2$. 
Now, consider a quiver gauge theory with gauge group which contains two copies of a particular subgroup $G_w$, $G \supseteq G_w \times G_w$, and such that there exists a $\mathbb{Z}_2$ automorphism of the quiver that exchanges the two copies of $G_w$. Then, we say that the quiver admits a $\mathbb{Z}_2$ wreathing, and the wreathing replaces
\begin{equation}
    G_w \times G_w \quad \rightarrow \quad G_w \wr \mathbb{Z}_2 \,.
\end{equation}
Since $G_w \wr \mathbb{Z}_2$ is a Lie group, it is a priori perfectly sensible to consider the gauge theory associated with it. The hypermultiplet spectrum remains unchanged under this operation. To give a particularly relevant example: a star-shaped quiver with $k$ identical tails admits a wreathing by any subgroup of $S_k$. In Figure \ref{fig:wreathingexamples}, we demonstrate an example of a 3d $\mathcal{N}=4$ quiver that admits a $\mathbb{Z}_2$ wreathing and denote the resulting wreathed quiver.

\subsubsection*{Summary of the Results}

In this work, we focus specifically on 6d $(1,0)$ SCFTs, $\mathcal{T}$, with eight supercharges that arise on the worldvolume of M5-branes probing an orbifold singularity, known as conformal matter \cite{DelZotto:2014hpa}, which possess a $\mathbb{Z}_2$ discrete global symmetry. When the $\mathbb{Z}_2$ is gauged leading to the theory $\widetilde{\mathcal{T}}$, we can extract/conjecture properties of $\widetilde{\mathcal{T}}$, such as the continuous flavor symmetry, by studying the action of the $\mathbb{Z}_2$ on the known Higgs branch chiral ring generators of $\mathcal{T}$. There are some subtleties in this analysis which we explain, for example, under certain circumstances an $\mathfrak{so}(2m)$ global symmetry of $\mathcal{T}$ can be transformed into a $\mathfrak{u}(m)$ global symmetry of $\widetilde{\mathcal{T}}$.

The main focus of this work is on, so-called, $(D, D)$ conformal matter, where the magnetic quivers for the Higgs branches are known to involve orthosymplectic gauge nodes. We develop a procedure to compute the Coulomb branch Hilbert series of general $\mathbb{Z}_2$-wreathed orthosymplectic quivers, and we apply this technique to the $\mathbb{Z}_2$-wreathing of the magnetic quivers associated with the $\mathcal{T}$. We see that the flavor symmetry of $\widetilde{\mathcal{T}}$ predicted from the action of the $\mathbb{Z}_2$ on the 6d Higgs branch operators is consistent with the Coulomb symmetry of the wreathed magnetic quiver as determined from the Hilbert series, and thus we find strong evidence for equation \eqref{eqn:bigboy} in this class of theories.

\subsubsection*{Structure of the Paper}

The structure of this paper is as follows. First, in Section \ref{sec:6d}, we briefly review the geometric construction of 6d $(1,0)$ SCFTs, in particular for the conformal matter theories, and explain how their Higgs branches are related to the Coulomb branches of (wreathed) 3d $\mathcal{N}=4$ quiver gauge theories. 
In Section \ref{sec:allSectionDGofCM}, we discuss the discrete gauging of conformal matter, which will serve as a reference for all the examples discussed in this work. This is one of the novel aspects of this work, and in Section \ref{sec:discretegaugingCM} we show how to determine the flavor symmetry of the theory after discrete gauging directly from the tensor branch analysis in 6d. In Section \ref{sec:wreathedCB}, we explain how the discrete gauging is reflected in the wreathing of the 3d magnetic quiver for the Higgs branch, and Section \ref{sec:wreathedprefactors} contains an explanation of how to compute the Hilbert series for wreathed theories with orthosymplectic groups. We test our proposal in Section \ref{sec:A3D3}, focusing on examples that admit both unitary and orthosymplectic quivers. More examples are discussed in Section \ref{sec:moreexamples}, in which we confirm the expected flavor symmetry after discrete gauging, which we introduced in Section \ref{sec:discretegaugingCM} by explicitly computing the Hilbert series of the $\ZZ_2$-wreathed magnetic quivers. We discuss some implications and directions for future study in Section \ref{sec:disc}. In Appendix \ref{sec:prefactors}, we review how to compute the classical contribution for the Coulomb branch Hilbert series.

\subsubsection*{Conventions and Notation}

\begin{enumerate}
    \item In the following, we consider both unitary and orthosymplectic quivers, and we follow the usual language in which \tikz{\node[gauge] {};} represents a gauge node and its label is the rank of the unitary gauge group, while \tikz{\node[flavor] {};} denotes a flavor node and its label denotes the number of fundamentals. When the gauge or flavor node is $\mathfrak{so}$, we will color the node in {\color{red}{red}}, while we will color the $\mathfrak{usp}$ nodes in {\color{blue}{blue}}. The line connecting two nodes represents a hypermultiplet in the bifundamental representation of the two groups. If the line edge corresponds to multiple hypermultiplets, we will specify this. A loop attached to a unitary gauge node indicates an adjoint-valued hypermultiplet, whereas a loop attached to a symplectic gauge node refers to an antisymmetric hypermultiplet.
    \item We adopt the notions of excess number, balance, underbalance, and overbalance as in \cite{Gaiotto:2008ak}.  For a $\U(N)$ gauge group with $N_f$ hypermultiplets transforming in the fundamental representation, the excess number for such a gauge group is defined as
\begin{equation}
\texttt{e}_{\U(N)} = N_f - 2N~.
\end{equation}
If, in a given quiver theory, all the excess numbers are $\texttt{e}_{\U(N)} \geq 0$, the theory is said to be \textit{good}, because all the monopole operators are above the unitarity bound. If a gauge group has $\texttt{e}_{\U(N)}=0$, we call that node \textit{balanced}; while if $\texttt{e}_{\U(N)} >0$, the gauge group is said to be overbalanced. If $\texttt{e}_{\U(N)} <0$, the gauge group is said to be underbalanced. In particular, if any of the gauge groups has $\texttt{e}_{\U(N)}=-1$ the theory is called \textit{ugly} because there is a monopole operator saturating the unitarity bound; otherwise, the theory is \textit{bad}, because it admits monopole operators below the unitarity bound.\footnote{In some cases, it may happen that monopoles valued in the magnetic lattice of multiple gauge groups make an apparent good theory bad. This is the case for some affine Dynkin-shaped unitary quiver theories \cite{Cremonesi:2014xha}. A similar discussion can be applied to some orthosymplectic magnetic quivers, and recently an adjustment to the balance notion for $\SO$ groups has been proposed in \cite{Lawrie:2024wan}.}  The excess number for $\SO$ and $\USp$ gauge groups with $N_f$ (full) hypermultiplets transforming in the fundamental representation is defined, respectively as
\begin{equation}
    \texttt{e}_{\SO(N)} = N_f-N+1\coma \texttt{e}_{\USp(2N)} = N_f - 2N-1\fstop
\end{equation}
For orthosymplectic nodes, having any of the gauge group with $\texttt{e}_{\SO(N)/\USp(2N)}=-1$ already signals that the theory is \textit{bad}. 
\end{enumerate}

\section{Review of 6d Conformal Matter}
\label{sec:6d}

In this section, we review the construction of 6d $(1,0)$ SCFTs, known as conformal matter \cite{DelZotto:2014hpa}, which are theories living on the worldvolume of M5-branes probing an orbifold singularity. We discuss their realization from the M-theory and F-theory perspectives, and how to determine their flavor symmetries from their tensor branch effective field theories. We also explain how to construct the corresponding 3d $\mathcal{N}=4$ magnetic quivers from their Higgs branches. 

\subsection{The Higgs Branch of Conformal Matter}

Conformal matter refers to the class of 6d $(1,0)$ SCFTs that live on the worldvolume of M5-branes probing an orbifold singularity \cite{DelZotto:2014hpa}. More specifically, rank $N$ $(\mathfrak{g}, \mathfrak{g})$ conformal matter, where $\mathfrak{g}$ is an ADE Lie algebra, lives on the worldvolume of a stack of $N$ M5-branes probing a $\mathbb{C}^2/\Gamma$ singularity, where $\Gamma$ is the finite subgroup of $\SU(2)$ associated with $\mathfrak{g}$ via the McKay correspondence \cite{MR604577}. We denote these theories as
\begin{equation}
    A_{N-1}^{\mathfrak{g}} \,.
\end{equation}
Such theories have a flavor symmetry which is, at least, 
\begin{equation}
    \mathfrak{g} \oplus \mathfrak{g} \,,
\end{equation}
which we refer to as the left and right flavor factors. In fact, the conformal matter theories are the parents of a whole family of SCFTs obtained from Higgs branch renormalization group flows. These flows are triggered via giving nilpotent vacuum expectation values to the moment map operators of the $\mathfrak{g} \oplus \mathfrak{g}$ flavor algebra. Let $O_L$ and $O_R$ denote nilpotent orbits of $\mathfrak{g}$, then we can refer to the child theories of this family as
\begin{equation}\label{eqn:higgsedCM}
    A_{N-1}^{\mathfrak{g}}(O_L, O_R) \,.
\end{equation}

Now that we have defined a class of 6d $(1,0)$ SCFTs, we would like to understand some of their physical properties. In this paper, we are particularly interested in the structure of the Higgs branch.\footnote{
Various properties of the Higgs branches of conformal matter theories have been studied from diverse perspectives. For nilpotent Higgsing of the moment map operators, see, for example, \cite{DelZotto:2015rca,Ohmori:2015pua,Ohmori:2015pia,Hanany:2018vph,Baume:2023onr,Distler:2022yse,DKL,Bourget:2022tmw,Hanany:2022itc,Baume:2021qho,Lawrie:2024zon}. There are also Higgs branch renormalization group flows which change $\mathfrak{g}$ (such as those triggered by end-to-end operators \cite{Baume:2020ure,Heckman:2020otd,Razamat:2019mdt,Bergman:2020bvi,Baume:2022cot,DKL}) and $N$.}
A magnetic quiver for the Higgs branch of $A_{N-1}^{\su(K)}(O_L, O_R)$, that is, where $\Gamma = \mathbb{Z}_K$ in the M-theory description, has been proposed in \cite{Hanany:2018vph}, utilizing the magnetic phase of the brane description of the 6d $(1,0)$ SCFTs in Type IIA string theory. It is proposed that
\begin{equation}\label{eqn:AtypeHB}
    \operatorname{HB}\left[ A_{N-1}^{\su(K)}(O_L, O_R) \right] \quad \cong \quad \operatorname{CB}\left[ \begin{gathered}
    \begin{tikzpicture}[baseline=0,font=\footnotesize]
    \node[tnode] (T1) {$T_{O_L}[\su(K)]$};
    \node[node, label=below:{$K$}] (N3) [right=6mm of T1] {};
    \node[tnode] (T2) [right=6mm of N3] {$T_{O_R}[\su(K)]$};
    \node[node, label=right:{$N$}] (Nu) [above=5mm of N3] {};
    
    \draw (T1.east) -- (N3.west);
    \draw (N3.east) -- (T2.west);    
    \draw (N3.north) -- (Nu.south);
    \draw (Nu) to[out=130, in=410, looseness=12] (Nu);
\end{tikzpicture}
    \end{gathered} \right] \,,
\end{equation}
where $T_O[\su(K)]$ are the 3d $\mathcal{N}=4$ theories introduced in \cite{Gaiotto:2008ak}; 
these are, in fact, linear Lagrangian theories, as we explain presently. Nilpotent orbits of $\su(K)$ are in one-to-one correspondence with integer partitions of $K$;\footnote{See \cite{Collingwood_1993} for the canonical reference on nilpotent orbits of semi-simple Lie algebras.} let 
\begin{equation}
    P = [p_1, p_2, \cdots, p_K] \qquad \text{ with } \qquad \sum_{i=1}^K p_i = K \,,
\end{equation}
written in weakly-decreasing order and zero-extended to be of length $K$, be the integer partition associated to the nilpotent orbit $O$. Then, the Lagrangian theory $T_{O}[\su(K)]$ is simply the linear quiver
\begin{equation}
    \begin{gathered}
    \begin{tikzpicture}[baseline=0,font=\footnotesize]
        \node[node, label=below:{$n_1$}] (A3) {};
      \node[node, label=below:{$n_2$}] (A4) [right=8mm of A3] {};
      \node[tnode] (N3) [right=6mm of A4] {$\cdots$};
        \node[node, label=below:{$n_{K-2}$}] (A5) [right=6mm of N3] {};
        \node[node, label=below:{$n_{K-1}$}] (A6) [right=8mm of A5] {};
        \node[flavor, label=below:{$K$}] (A7) [right=8mm of A6] {};
    
        \draw (A3.east) -- (A4.west);
      \draw (A4.east) -- (N3.west);
      \draw (N3.east) -- (A5.west);    
      \draw (A5.east) -- (A6.west);
      \draw (A6.east) -- (A7.west);
    \end{tikzpicture} 
  \end{gathered} \,,
\end{equation}
where the $n_i$ are specified by the choice of partition as
\begin{equation}
    n_i = \sum_{j=K+1-i}^K p_j \,.
\end{equation}

A similar proposal for the magnetic quiver of the Higgs branch of the $(D, D)$ conformal matter SCFTs, $A_{N-1}^{\so(2K)}(O_L, O_R)$, was also put forward in \cite{Hanany:2018vph,Hanany:2022itc}. In the M-theory construction, this is where $\Gamma$ is the binary dihedral group. They proposed that 
\begin{equation}\label{eqn:DtypeHB}
    \operatorname{HB}\left[ A_{N-1}^{\so(2K)}(O_L, O_R) \right] \quad \cong \quad \operatorname{CB}\left[ \begin{gathered}
    \begin{tikzpicture}[baseline=0,font=\footnotesize]
    \node[tnode] (T1) {$T_{O_L}[\so(2K)]$};
    \node[node, label=below:{$2K$},fill=red] (N3) [right=6mm of T1] {};
    \node[tnode] (T2) [right=6mm of N3] {$T_{O_R}[\so(2K)]$};
    \node[node, label=right:{$2N$},fill=blue] (Nu) [above=5mm of N3] {};
    
    \draw (T1.east) -- (N3.west);
    \draw (N3.east) -- (T2.west);    
    \draw (N3.north) -- (Nu.south);
    \draw (Nu) to[out=130, in=410, looseness=12] (Nu);
\end{tikzpicture}
    \end{gathered} \right] \,,
\end{equation}
via the engineering of the SCFTs in Type IIA string theory.\footnote{Note: there is an important choice of global structure that must be made here; we return to this point in Section \ref{sec:higherformsym}.} The brane analysis in \cite{Hanany:2018vph,Hanany:2022itc} requires that the partitions associated with $O_L$ and $O_R$ be \emph{special partitions}; this is in contrast to the geometric engineering approach (see Section \ref{sec:GS}), where SCFTs are associated with all pairs of D-partitions, as in equation \eqref{eqn:higgsedCM} \cite{Heckman:2016ssk,Mekareeya:2016yal}. While we do not have a derivation of equation \eqref{eqn:DtypeHB} for non-special partitions, all the obvious cross-checks are passed, and thus we assume that equation \eqref{eqn:DtypeHB} is valid in general. The $T_O[\so(2K)]$ theories again have a simple Lagrangian description. Each nilpotent orbit of $\so(2K)$ can be associated with a D-partition of $2K$;\footnote{A D-partition of $2K$ is an integer partition of $2K$ such that every even element appears with even multiplicity. A D-partition which is \emph{very even} has only even elements.} however, this is not a unique association: each very even D-partition is, in fact, associated with two distinct nilpotent orbits. This particular subtlety is not well-understood for the $T_O[\so(2K)]$ theories, and thus in this paper we will perforce ignore this distinction.\footnote{Recent work in four and six dimensions has emphasized the importance of and determined the physical distinction between the nilpotent orbits associated to the same very even D-partition \cite{Distler:2022nsn,Distler:2022yse}.} Let
\begin{equation}
    P = [p_1, p_2, \cdots, p_{2K}] \qquad \text{ with } \qquad \sum_{i=1}^{2K} p_i = 2K \,,
\end{equation}
be a D-partition of $2K$, written in weakly-decreasing order and zero-extended to be of length $2K$. Let $O$ be the nilpotent orbit associated with the D-partition $P$, then $T_O[\so(2K)]$ has the following Lagrangian description as an alternating sequence of orthosymplectic nodes:
\begin{equation}
    \begin{gathered}
    \begin{tikzpicture}[baseline=0,font=\footnotesize]
        \node[node, label=below:{$n_1$},fill=red] (A3) {};
      \node[node, label=below:{$n_2$},fill=blue] (A4) [right=9mm of A3] {};
      \node[tnode] (N3) [right=6mm of A4] {$\cdots$};
        \node[node, label=below:{$n_{2K-2}$},fill=red] (A5) [right=6mm of N3] {};
        \node[node, label=below:{$n_{2K-1}$},fill=blue] (A6) [right=9mm of A5] {};
        \node[flavor, label=below:{$2K$},fill=red] (A7) [right=9mm of A6] {};
    
        \draw (A3.east) -- (A4.west);
      \draw (A4.east) -- (N3.west);
      \draw (N3.east) -- (A5.west);    
      \draw (A5.east) -- (A6.west);
      \draw (A6.east) -- (A7.west);
    \end{tikzpicture} 
  \end{gathered} \,.
\end{equation}
Here, the $n_i$ are specified by
\begin{equation}
    n_i = \left\{
    \renewcommand*{\arraystretch}{2.5}
    \begin{array}{lr}
        \displaystyle 2 \left\lfloor \, \sum_{j=2K+1-i}^{2K} \frac{p_j}{2} \, \right\rfloor  &\text{ if $i$ odd,} \\
        \displaystyle 2 \left\lceil \, \sum_{j=2K+1-i}^{2K} \frac{p_j}{2} \, \right\rceil &\text{ if $i$ even.}
    \end{array} \right.
\end{equation}

Examining the magnetic quivers for Higgsed $(A, A)$ and $(D, D)$ conformal matter in equations \eqref{eqn:AtypeHB} and \eqref{eqn:DtypeHB}, respectively, we notice that when $O_L = O_R$ the magnetic quiver exhibits a $\ZZ_2$ diagram automorphism. The physical manifestation of this $\ZZ_2$ in the 6d $(1,0)$ SCFTs is the subject of Section \ref{sec:GS}. 

\subsection{Conformal Matter from F-theory}

We have discussed the realization of the conformal matter theories as living on the worldvolume of M5-branes, that is, from an M-theory perspective. We now turn to the F-theory perspective. We briefly review the construction of the conformal matter theories via the atomic/geometric  perspective in F-theory \cite{Heckman:2013pva,Heckman:2015bfa}. See \cite{Heckman:2018jxk} for a detailed review of the atomic construction.

We begin with the $(A, A)$ conformal matter theories. Let $\widetilde{Y}$ be a non-compact elliptically-fibered Calabi--Yau threefold $\pi : \widetilde{Y} \rightarrow \widetilde{B}$ such that $\widetilde{B}$ contains a linear chain of $N-1$ pairwise intersecting smooth complex rational curves of self-intersection $(-2)$ and each supporting a split singular fiber of Kodaira--Neron type $I_K$ at the generic point, and such that the fibers are minimal over the intersection points of the curves. We can denote such a Calabi--Yau threefold via the shorthand notation:
\begin{equation}\label{eqn:cceg}
    \underbrace{\overset{\su(K)}{2}\cdots\overset{\su(K)}{2}}_{N-1} \,,
\end{equation}
where each $2$ denotes a $\mathbb{P}^1 \subset  \widetilde{B}$ with self-intersection number $(-2)$ and a neighboring $2$ indicates that the corresponding complex curves intersect with intersection number $+1$; the $\su(K)$ over the $2$ indicates that over the generic point of that $\mathbb{P}^1$ there is supported a split singular fiber of Kodaira type $I_K$. We often refer to geometries denoted as in equation \eqref{eqn:cceg} as a ``curve configuration'' or, for reasons we will see anon, as a ``tensor branch geometry''. There exists a contraction map
\begin{equation}
    \rho \, : \, \widetilde{Y} \rightarrow Y \,,
\end{equation}
which simultaneously shrinks the volume of all compact curves in $\widetilde{B}$ to zero \cite{artin1962}. We see that the base of the elliptic fibration $Y$ is simply $\mathbb{C}^2/\ZZ_N$. Compactification of F-theory on $Y$ yields the $(A, A)$ conformal matter theory $A_{N-1}^{\su(K)}$. 

We can similarly ask about the result of the compactification of F-theory on $\widetilde{Y}$. Instead of an SCFT, this yields an SQFT, which is known as the effective field theory on the tensor branch of the SCFT associated with $Y$. This consists of vector multiplets (associated with each gauge algebra $\mathfrak{g}$ supported over a compact curve), hypermultiplets (associated with the intersections of compact and (non-)compact curves), and tensor multiplets (one associated with each compact curve) where the scalar in the tensor multiplet is given a vacuum expectation value proportional to the volume of the associated curve. Each tensor multiplet contains an anti-self-dual 2-form, which couples to a tensionful string of the 6d theory: tuning the VEVs of the scalars to be zero causes these strings to become tensionless, and their non-trivial dynamics generates the SCFT \cite{Witten:1995zh,Seiberg:1996qx}. 
As an SQFT, it is possible to compute a variety of the physical properties of the effective field theory at the generic point of the tensor branch; a part of the power of this geometric construction of 6d $(1,0)$ SCFTs is the ability to compute quantities at the generic point of the tensor branch which can be tracked to the superconformal field theory at the origin of the tensor branch. Such quantities are typically related to global symmetries, as discussed in \cref{sec:GS,sec:flavor}, or anomalies.

Next, we turn to understanding how the Higgsed theories are constructed in F-theory. Write the partitions of $K$ associated with nilpotent orbits of $\su(K)$, $O_L$ and $O_R$, respectively, in multiplicative form as
\begin{equation}
    \left[ K^{m_K}, \cdots, 2^{m_2}, 1^{m_1}\right] \qquad \text{and} \qquad \left[ K^{m_K'}, \cdots, 2^{m_2'}, 1^{m_1'}\right] \,.
\end{equation}
Then, the tensor branch geometry $\widetilde{Y}$ associated with the SCFT $A_{N-1}^{\su(K)}(O_L, O_R)$ is
\begin{equation}\label{eqn:sugeneric}
    \underset{[m_1]}{\overset{\su(k_1)}{2}}\,\,
    \underset{[m_2]}{\overset{\su(k_2)}{2}}\,\,
    \cdots
    \underset{[m_K]}{\overset{\su(k_K)}{2}}\,\,
    \underbrace{\underset{\phantom{[m_K]}}{\overset{\su(K)}{2}} \,\,\cdots \,\,\overset{\su(K)}{2}}_{N-2K-1} \,\,
    \underset{[m_K']}{\overset{\su(k_K')}{2}}\,\,
    \cdots
    \underset{[m_2']}{\overset{\su(k_2')}{2}}\,\,
    \underset{[m_1']}{\overset{\su(k_1')}{2}}\,.
\end{equation}
The $k_i$ and $k_i'$ are fixed by gauge-anomaly cancellation, which necessitates that
\begin{equation}
    2k_i - k_{i-1} - k_{i+1} - m_i = 0 \,,
\end{equation}
where we have defined $k_0 = 0$ and $k_{K+1} = K$ for convenience, and analogously for $k_i'$ in terms of $m_i'$. In this paper, we are interested in theories $A_{N-1}^{\su(K)}(O, O)$, where we have set $O_L = O_R = O$, and satisfying 
\begin{equation}\label{eqn:notbad}
    N - 2\ell \geq 0 \,,
\end{equation}
where $\ell$ is the largest integer such that $m_\ell$ in the partition corresponding to $O$ is non-zero.\footnote{When equation \eqref{eqn:notbad} is not satisfied, the Higgsing is ``bad''; see \cite{Hassler:2019eso,Distler:2022kjb,LMS} for discussions of such cases.}

A similar method to construct the $\widetilde{Y}$ such that, after applying the contraction map $\rho: \widetilde{Y} \rightarrow Y$, F-theory compactified on $Y$ engineers the $(D, D)$ conformal matter SCFTs $A_{N-1}^{\so(2K)}(O_L, O_R)$ is known. The $\widetilde{Y}$ generically consists of an alternating chain of pairwise intersecting $(-1)$- and $(-4)$-curves with non-split $I_{2K-8}$ fibers supported over former and split $I_{K-4}^*$ over the latter. Let the nilpotent orbits $O_L$, $O_R$ of $\so(2K)$ be associated with the following D-partitions of $2K$, written in multiplicative form:
\begin{equation}
    [ (2K)^{m_{2K}}, \cdots, 2^{m_2}, 1^{m_1}] \qquad \text{and} \qquad [ (2K)^{m_{2K}'}, \cdots, 2^{m_2'}, 1^{m_1'}] \,,
\end{equation}
respectively.\footnote{As we have already emphasized, certain D-partitions are in fact associated with two distinct nilpotent orbits. The difference in the tensor branch geometries between two such nilpotent orbits is related to $\theta$-angles, which we suppress in the following. See \cite{Distler:2022yse} for a careful analysis of the curve configuration when Higgsing by nilpotent orbits associated with very even D-partitions.} Then, the curve configuration is
\begin{equation}\label{eqn:socmHiggsed}
   \scalebox{0.836}{ $\displaystyle\underset{[m_1]}{\overset{\usp(2k_1)}{1}}\,\, 
    \underset{[m_2]}{\overset{\so(k_2)}{4}}\,\,
    \underset{[m_3]}{\overset{\usp(2k_3)}{1}}\,\, 
    \underset{[m_4]}{\overset{\so(k_4)}{4}}\,\,
    \cdots
    \underset{[m_{2K-1}]}{\overset{\usp(2k_{2K-1})}{1}}\,\, 
    \underbrace{\underset{\phantom{[m_2]}}{\overset{\so(2K)}{4}}\,\, 
    \overset{\usp(2K-8)}{1}\,\, 
    \cdots 
    \overset{\so(2K)}{4}}_{(N-2K + 1)\,\,  (-4)\text{-curves}}
    \underset{[m_{2K-1}']}{\overset{\usp(2k_{2K-1}')}{1}}\,\, 
    \cdots
    \underset{[m_4']}{\overset{\so(k_4')}{4}}\,\,
    \underset{[m_3']}{\overset{\usp(2k_3')}{1}}\,\, 
    \underset{[m_2']}{\overset{\so(k_2')}{4}}\,\,
    \underset{[m_1']}{\overset{\usp(2k_1')}{1}} 
    \,.$}
\end{equation}
The $k_i$, $k_i'$ are again fixed by anomaly cancellation; this imposes that 
\begin{equation}\label{eqn:soanomaly}
    \begin{aligned}
        4k_i + 16 - k_{i+1} - k_{i-1} - m_i  &= 0 \quad \text{ if } \quad i \text{ odd } \,, \\
        k_i - 8 - k_{i-1} - k_{i+1} - \frac{m_i}{2} &= 0 \quad \text{ if } \quad i \text{ even } \,,
    \end{aligned}
\end{equation}
and similarly for the $k_i'$ in terms of the $m_i'$. In writing these conditions, we made the following convenient definitions: $k_0 = 0$ and $k_{2K} = 2K$. Sometimes, solving the anomaly cancellation conditions in equation \eqref{eqn:soanomaly}, yields some $k_i$ which are negative; when this happens, we need to apply the following replacement rules to determine the correct tensor branch geometry:
{\allowdisplaybreaks[4]
    \begin{alignat}{2}
        \overset{\usp(-6)}{1}\,\,\overset{\so(3)}{4}\,\,\overset{\usp(-4)}{1}\,\,\overset{\so(5)}{4}\,\,\overset{\usp(-2)}{1}\,\,\overset{\so(7)}{4} \cdots &\longrightarrow && \,\,2\,\,\overset{\su(2)}{2}\,\,\overset{\mathfrak{g}_2}{3} \cdots \,,\nonumber \\
        \overset{\usp(-6)}{1}\,\,\overset{\so(4)}{4}\,\,\overset{\usp(-2)}{1}\,\,\overset{\so(7)}{4} \cdots &\longrightarrow &&\,\, \overset{\su(2)}{2}\,\,\overset{\mathfrak{g}_2}{3} \cdots \,, \nonumber\\
        \overset{\usp(-6)}{1}\,\,\overset{\so(4)}{4}\,\,\overset{\usp(-2)}{1}\,\,\overset{\so(8)}{4} \cdots &\longrightarrow &&\,\, \overset{\su(2)}{2}\,\,\overset{\so(7)}{3} \cdots \,, \nonumber\\
        \overset{\usp(-4)}{1}\,\,\overset{\so(5)}{4}\,\,\overset{\usp(-2)}{1}\,\,\overset{\so(7)}{4} \cdots &\longrightarrow &&\,\, \overset{\su(2)}{2}\,\,\overset{\so(7)}{3} \cdots \,, \label{eqn:replacementrules}\\
        \overset{\usp(-4)}{1}\,\,\overset{\so(6)}{4} \cdots &\longrightarrow &&\,\, \overset{\su(3)}{3} \cdots \,, \nonumber\\
        \overset{\usp(-4)}{1}\,\,\overset{\so(7)}{4} \cdots &\longrightarrow &&\,\, \overset{\mathfrak{g}_2}{3} \cdots \,, \nonumber\\
        \overset{\usp(-4)}{1}\,\,\overset{\so(8)}{4} \cdots &\longrightarrow &&\,\, \overset{\so(7)}{3} \cdots \,, \nonumber\\
        \overset{\usp(-2)}{1}\,\,\overset{\mathfrak{g}}{4} \cdots &\longrightarrow &&\,\, \overset{\mathfrak{g}}{3} \cdots \,,\nonumber
    \end{alignat}
}
where the final row is a catch-all replacement for any such special orthogonal $\mathfrak{g}$ that appears when solving equation \eqref{eqn:soanomaly}.
Similarly to the $(A, A)$ conformal matter case, in this paper we are only interested in $(D, D)$ theories which are Higgsed by the same nilpotent orbit and are not bad, which means that equation \eqref{eqn:notbad} must also be satisfied for the $A_{N-1}^{\so(2K)}(O, O)$ that we consider.

Another object of interest is known as the ``partial tensor branch theory''. This is the SQFT obtained by taking the curve configuration at the generic point of the tensor branch for some conformal matter, and taking the volumes of all $(-1)$-curves to zero, until there are no $(-1)$-curves remaining.\footnote{For $(A, A)$ conformal matter, there are no $(-1)$-curves, and thus the partial tensor branch theory is the same as the theory at the generic point of the tensor branch.} This theory consists of a collection of vector multiplets coupled to strongly-coupled SCFT sectors associated with the collapsed curves.  The torus-compactification of the partial tensor branch SQFT is known to have an alternative realization \cite{Ohmori:2015pia,Ohmori:2015pua,Baume:2021qho} via a punctured sphere in class $\mathcal{S}$ \cite{Gaiotto:2009we,Gaiotto:2009hg}. Therefore, the analysis of \cite{Benini:2010uu} provides the magnetic quiver for the Higgs branch of this partial tensor branch theory in 6d. We have
\begin{equation}
    \operatorname{HB}\left[ \operatorname{PTB} \left[ A_{N-1}^{\su(K)}(O_L, O_R) \right] \right] \quad \cong \quad 
    \operatorname{CB}\left[
    \begin{gathered}
    \begin{tikzpicture}[baseline=0,font=\footnotesize]
    \node[tnode] (T1) {$T_{O_L}[\su(K)]$};
    \node[node, label=below:{$K$}] (N3) [right=6mm of T1] {};
    \node[tnode] (T2) [right=6mm of N3] {$T_{O_R}[\su(K)]$};
    \node[node, label=left:{$1$}] (Nul) [above left=6mm and 5mm of N3] {};
    \node[node, label=right:{$1$}] (Nur) [above right=6mm and 5mm of N3] {};
    \node[tnode] (T3) [above=5mm of N3] {$\cdots$};
    \draw[thick] [decorate,decoration={brace,amplitude=5pt},xshift=0, yshift=0cm]
([yshift=0.3cm]Nul.west) -- ([yshift=0.3cm]Nur.east) node [black,midway,above,yshift=2mm] 
{$N$};
    \draw (T1.east) -- (N3.west);
    \draw (N3.east) -- (T2.west);    
    \draw (N3.45) -- (Nur.225);
    \draw (N3.135) -- (Nul.315);
\end{tikzpicture}
    \end{gathered}\right] \,,
\end{equation}
and
\begin{equation}\label{eqn:ripto}
    \operatorname{HB}\left[ \operatorname{PTB} \left[ A_{N-1}^{\so(2K)}(O_L, O_R) \right] \right] \quad \cong \quad 
    \operatorname{CB}\left[
    \begin{gathered}
    \begin{tikzpicture}[baseline=0,font=\footnotesize]
    \node[tnode] (T1) {$T_{O_L}[\so(2K)]$};
    \node[node, label=below:{$2K$}, fill=red] (N3) [right=6mm of T1] {};
    \node[tnode] (T2) [right=6mm of N3] {$T_{O_R}[\so(2K)]$};
    \node[node, label=left:{$2$}] (Nul) [above left=6mm and 5mm of N3, fill=blue] {};
    \node[node, label=right:{$2$}] (Nur) [above right=6mm and 5mm of N3, fill=blue] {};
    \node[tnode] (T3) [above=5mm of N3] {$\cdots$};
    \draw (T1.east) -- (N3.west);
    \draw (N3.east) -- (T2.west);    
    \draw (N3.45) -- (Nur.225);
    \draw (N3.135) -- (Nul.315);
    \draw[thick] [decorate,decoration={brace,amplitude=5pt},xshift=0, yshift=0cm]
([yshift=0.3cm]Nul.west) -- ([yshift=0.3cm]Nur.east) node [black,midway,above,yshift=2mm] 
{$N$};
\end{tikzpicture}
    \end{gathered}\right] \,,
\end{equation}
where we have written $\operatorname{PTB}[\mathcal{T}]$ to denote the partial tensor branch SQFT associated with the conformal matter SCFT $\mathcal{T}$. We can see that the magnetic quiver for the Higgs branch of the partial tensor branch theory has a $S_N$ diagram automorphism, which permutes the $N$ either $\mathfrak{u}(1)$ or $\usp(2)$ nodes, and when $O_L = O_R$ there also exists a $\ZZ_2$ diagram automorphism that swaps the two horizontal tails. Note that the magnetic quiver for the Higgs branch of the SCFTs, as given in equations \eqref{eqn:AtypeHB} and \eqref{eqn:DtypeHB}, is obtained by wreathing the $S_N$ automorphism of the PTB Higgs branches, following the proposal in \cite{Hanany:2018vph}.

\subsection{Flavor Symmetry of Conformal Matter}
\label{sec:flavor}

As we have explained, a non-compact elliptically-fibered Calabi--Yau threefold satisfying the appropriate conditions, for which the relevant details are captured by the tensor branch curve configuration, gives rise to a 6d $(1,0)$ SCFT via F-theory geometric engineering. For the purposes of this paper, we are interested in both the discrete global symmetries and the continuous global symmetries; in this section, we review how the latter are determined from the tensor branch curve configuration. The algorithm is explained for non-Abelian symmetries in \cite{Baume:2021qho} and for Abelian symmetries in \cite{Apruzzi:2020eqi,Lee:2018ihr}; we review only the details relevant for conformal matter theories here.

We begin with a discussion of the non-Abelian flavor symmetries. Cancellation of gauge anomalies implies that when the Lie algebra $\mathfrak{g}$ is supported over a compact curve of self-intersection $(-n)$, then the number and representations under $\mathfrak{g}$ of the hypermultiplets in the effective field theory are completely fixed by the pair $(\mathfrak{g}, n)$.\footnote{In some instances, the pair $(\mathfrak{g}, n)$ does not uniquely fix the matter spectrum, however this does not occur in the conformal matter theories that we discuss in this paper.} Consider a curve configuration involving a singular fiber associated with the Lie algebra $\mathfrak{g}$ over a $(-n)$-curve. Assume that gauge-anomaly cancellation implies that there must be $m$ hypermultiplets in the irreducible representation $\bm{R}$ of $\mathfrak{g}$,\footnote{Anomaly cancellation may require matter in multiple distinct irreducible representations of $\mathfrak{g}$, in which case there may be a contribution of a product of simple Lie algebras to the global symmetry.} and furthermore assume that $m' \leq m$ of these hypermultiplets are not trapped at intersections of the $(-n)$-curve with other compact curves. Then, there is the following non-Abelian flavor symmetry factor:\footnote{We note that these flavor symmetry factors arise when $\bm{R}$ is an irreducible representation of the gauge group, not just the gauge algebra. As an example of where this is not the case: let $\mathfrak{g} = \su(3)$ and $G = \SU(3) \ltimes \ZZ_2$, i.e., a semi-direct product of the simply-connected $\SU(3)$ and its $\ZZ_2$ outer-automorphism; the $\bm{3}$ and $\overline{\bm{3}}$ are not irreducible representations of $G$, but the $\bm{3} \oplus \overline{\bm{3}}$ is.}
\begin{equation}\label{eqn:nonABgen}
    \mathfrak{f} = \begin{cases}
        \su(m') &\quad \text{if $\bm{R}$ is complex,} \\
        \usp(2m') &\quad \text{if $\bm{R}$ is real,} \\
        \so(2m') &\quad \text{if $\bm{R}$ is pseudo-real.}
    \end{cases}
\end{equation}
Recall that in the presence of an odd number of half-hypermultiplets, which exist only when $\bm{R}$ is pseudo-real, then $m'$ can be half-odd-integer.
An exception to this general rule in equation \eqref{eqn:nonABgen} occurs when considering a configuration involving $\su(2)$ supported on a $(-2)$-curve, see \cite{Baume:2021qho,Heckman:2015bfa,Morrison:2016djb,Ohmori:2015pia}. For example, for such configurations, we may need to know the following special rules:
\begin{equation}
  \begin{aligned}
    2m' = 8 \quad &\Rightarrow \quad \mathfrak{f} = \so(7) \,, \\
    2m' = 7 \quad &\Rightarrow \quad \mathfrak{f} = \mathfrak{g}_2 \,, \\
    2m' = 6 \quad &\Rightarrow \quad \mathfrak{f} = \su(3) \,.
  \end{aligned}
\end{equation}
These non-Abelian flavor factors we have just described can be considered as ``classical'' flavor symmetries that rotate a number of hypermultiplets in the same representation. The second interesting class of non-Abelian flavor symmetries was referred to as ``E-string flavor'' in \cite{Baume:2021qho}. This arises as follows; consider a tensor branch curve configuration of the form
\begin{equation}
    \cdots \overset{\mathfrak{g}_L}{n_L} \,\, 1 \,\, \overset{\mathfrak{g}_R}{n_R} \cdots \,.
\end{equation}
The rules for constructing Calabi--Yau threefolds associated with SCFTs impose that 
\begin{equation}
    \mathfrak{g}_L \oplus \mathfrak{g}_R \subseteq \mathfrak{e}_8 \,,
\end{equation}
and that a $(-1)$-curve can intersect at most two compact curves. Let $\mathfrak{f}$ be the non-Abelian part of the commutant of the embedding of $\mathfrak{g}_L \oplus \mathfrak{g}_R$ inside $\mathfrak{e}_8$:
\begin{equation}\label{eqn:commutant}
    \mathfrak{f} \,\, = \,\, \operatorname{Commutant}(\mathfrak{e}_8, \mathfrak{g}_L \oplus \mathfrak{g}_R) \,.
\end{equation}
Then, $\mathfrak{f}$ is a factor in the non-Abelian global symmetry of the associated SCFT. Most of the relevant commutants for determining the E-string flavor were listed in \cite{Baume:2021qho}.

For Abelian global symmetries, the situation is more complicated. A priori, whenever there is a hypermultiplet in a complex representation, including bifundamental hypermultiplets, there is a classical $\mathfrak{u}(1)$ symmetry that rotates that hypermultiplet. However, these symmetries often suffer from ABJ anomalies \cite{Apruzzi:2020eqi,Lee:2018ihr}. A systematic analysis of which particular linear combinations of classical $\mathfrak{u}(1)$ symmetries survive these ABJ anomalies, in terms of the tensor branch curve configuration, has appeared in \cite{Apruzzi:2020eqi}. However, we are interested in the total rank of the surviving Abelian factors, and this requires knowledge of only a few simple rules from \cite{Apruzzi:2020eqi}. Consider a curve configuration of the form
\begin{equation}
    \underset{[m_1]}{\overset{\su(k_1)}{2}} \,\, 
    \underset{[m_2]}{\overset{\su(k_2)}{2}} \,\, \cdots \,\, 
    \underset{[m_{N-2}]}{\overset{\su(k_{N-2})}{2}} \,\, 
    \underset{[m_{N-1}]}{\overset{\su(k_{N-1})}{2}} \, \,,
\end{equation}
where $m_i$ denotes the number of dangling hypermultiplets associated with each gauge algebra. We assume that all $k_i \geq 2$ and that at least one $k_i \geq 3$. Let $\ell$ denote the number of $m_i$ such that $m_i \geq 1$; then, after taking into account the ABJ anomalies, the Abelian part of the global symmetry of the configuration is
\begin{equation}
    \mathfrak{f} \, = \, \mathfrak{u}(1)^{\ell - 1} \,.
\end{equation}
There are two other ways that $\mathfrak{u}(1)$ symmetries may arise that we have to consider in this paper. We may have a $\usp(2K)$ algebra supported over a $(-1)$-curve, such that there are $m' = 1$ dangling hypermultiplets in the fundamental representation of the $\usp(2K)$; then there is a classical $\so(2) \cong \mathfrak{u}(1)$ global symmetry, which does not suffer from an ABJ anomaly. Similarly, we may have an undecorated $(-1)$-curve such that the commutant in equation \eqref{eqn:commutant} contains Abelian factors. As there are only some configurations with E-string flavor that we consider in this paper, we list them here:
\begin{equation}\label{eqn:commutants}
    \begin{aligned}
        \cdots \overset{\su(3)}{3}\,\,1\,\,\overset{\su(3)}{3}\cdots &\quad \Rightarrow \quad \mathfrak{f} = \su(3) \oplus \su(3)  \,, \\
        \cdots \overset{\mathfrak{g}_2}{3}\,\,1\,\,\overset{\mathfrak{g}_2}{3}\cdots &\quad \Rightarrow \quad \mathfrak{f} = \su(2) \,, \\
        \cdots \overset{\so(7)}{3}\,\,1\,\,\overset{\so(7)}{3}\cdots &\quad \Rightarrow \quad \mathfrak{f} = \mathfrak{u}(1) \,, \\
        \cdots \overset{\su(3)}{3}\,\,1\,\,\overset{\so(9)}{4}\cdots &\quad \Rightarrow \quad \mathfrak{f} = \mathfrak{u}(1) \,, \\
        \cdots \overset{\mathfrak{g}_2}{3}\,\,1\,\,\overset{\so(9)}{4}\cdots &\quad \Rightarrow \quad \mathfrak{f} = \varnothing \,, \\
        \cdots \overset{\so(7)}{3}\,\,1\,\,\overset{\so(9)}{4}\cdots &\quad \Rightarrow \quad \mathfrak{f} = \varnothing \,.
    \end{aligned}
\end{equation}
Each such $\mathfrak{u}(1)$ persists to a global symmetry of the quantum theory. 

\subsection{Comment on Higher-form Symmetries}
\label{sec:higherformsym}

We would like to briefly comment on the arising of 1-form symmetries when considering the 3d magnetic quivers for the Higgs branches of 6d conformal matter. We have explained that such magnetic quivers are given as in equations \eqref{eqn:DtypeHB} and \eqref{eqn:ripto}. In fact, there is a slight ambiguity in these descriptions that we clarify here. More generally, in \cite{Benini:2010uu,Cremonesi:2014vla}, the 3d mirror for class $\mathcal{S}$ \cite{Gaiotto:2009we,Gaiotto:2009gz} theories of types $A$ and $D$ were given as a diagonal gauging of a collection of the 3d $T_{O}[\mathfrak{g}]$ theories where a similar subtlety arises when $\mathfrak{g}$ is a D-type algebra.

An important observation that was made in \cite{Cremonesi:2014vla} is that if one wants to match the Higgs branch Hilbert series/Hall--Littlewood index of the 4d theory,\footnote{Here, we will assume that all punctures in the class $\mathcal{S}$ description are untwisted. When twisted punctures are incorporated the Hall--Littlewood index and the Higgs branch Hilbert series are generically not the same \cite{Kang:2022zsl}, and magnetic quivers for the Higgs branches have been proposed in \cite{Beratto:2020wmn,Kang:2022zsl}.} for the class $\mathcal{S}$ theories of type $D$, with the Coulomb branch Hilbert series of the proposed magnetic quiver, then it is necessary that the global form of the central node of the magnetic quiver is $G/\ZZ_2$. The gauging of a $\ZZ_2$ $1$-form symmetry in 3d creates a dual $0$-form symmetry, which affects the local operator spectrum, and thus the Hilbert series.

In the computation of the Hilbert series, we will explicitly refine by the $\ZZ_2$ fugacity $\omega$ for the $\ZZ_2^{[0]}$ $0$-form symmetry that the 3d mirror theories have. This is a possible refinement, and allows us to visualize both the choices of global form for the central node of the quivers. Either choice of $\omega$ is sensible when purely considering a 3d orthosymplectic quiver and its wreathing, however, the Hilbert series associated to the Higgs branch of the 4d class $\mathcal{S}$ theory corresponds to unrefining the Hilbert series by setting $\omega = 1$. We expect that the same choice of global form for the central node is the relevant choice for the magnetic quiver of the Higgs branch of 6d conformal matter.

\section{Discrete Gauging of Conformal Matter}
\label{sec:allSectionDGofCM}

Now that we have introduced the 6d $(1,0)$ SCFTs of interest in this work, we explain when they have discrete global symmetries, associated with so-called Green--Schwarz automorphisms. Then, using the tensor branch description, we conjecture the continuous flavor symmetry of the new 6d $(1,0)$ SCFTs obtained by gauging these discrete global symmetries. The predictions of the flavor symmetry obtained in Section \ref{sec:discretegaugingCM} will then be confirmed from the computation of the Hilbert series of the $\ZZ_2$-wreathed magnetic quivers in \cref{sec:A3D3,sec:moreexamples}.

\subsection{Discrete Symmetries and Green--Schwarz Automorphisms}
\label{sec:GS}

We have discussed how the continuous global symmetries of the 6d $(1,0)$ SCFTs can be determined from the tensor branch curve configuration associated to the Calabi--Yau geometry engineering the SCFT in F-theory. Discrete global symmetries of the SCFTs can also be obtained from the tensor branch description, as explained in \cite{Apruzzi:2017iqe}.

The tensor branch of 6d $(1,0)$ SCFTs is given by a collection of $\mathcal{N}=1$ tensor multiplets. The bosonic sector of these multiplets is formed by a scalar and an anti-self-dual 2-form. The conformal fixed point corresponds to taking the expectation values of all the scalars to zero simultaneously, which in turn corresponds to making the strings charged under the anti-self-dual 2-form tensionless. The metric on the tensor branch moduli space parametrized by the scalars in the tensor multiplets is obtained by considering the Dirac pairing $\mathcal{A}$ of all the string charges, defined as 
\begin{equation}
    \mathcal{A}\,:\, \Lambda \rightarrow \Lambda\coma
\end{equation}
where $\Lambda$ is the lattice of string charges. In order to uniquely determine the geometry of the tensor branch moduli space, one needs to find what is the group of transformations that leaves the Dirac pairing invariant. This group is also known as the automorphism group $\operatorname{Aut}(\Lambda)$ which receives two source contributions \cite{Apruzzi:2017iqe}
\begin{equation}
    \operatorname{Aut}(\Lambda) = \operatorname{Aut}(\Lambda_\text{end})  \times \operatorname{Aut}(\usp(2Q)) \,,
\end{equation}
where 
\begin{equation}
  \operatorname{Aut}(\Lambda_\text{end})  = \mathcal{O}_\text{end} \ltimes \mathcal{W}_\text{end} \coma
\end{equation}
with $\Lambda_\text{end}$ being the charge lattice of the endpoint configuration of curves (also known as the partial tensor branch description in Section \ref{sec:6d}), while $Q$ is the number of blowdowns of $(-1)$-curves which must be performed to go from the generic point of the tensor branch to the endpoint curve configuration. $\mathcal{O}_\text{end}$ represents the candidate global discrete symmetry, while $\mathcal{W}_\text{end}$ is the maximal normal subgroup of $\operatorname{Aut}(\Lambda_\text{end})$. For each family of 6d $(1,0)$ SCFTs that exist in the geometric construction via F-theory, $\mathcal{O}_\text{end}$ was determined from the $\mathbb{C}^2/\Gamma$ singularity of the base in \cite{Apruzzi:2017iqe}, in particular, if $\Gamma$ is a finite ADE group then 
\begin{equation}
    \mathcal{O}_\text{end} = \operatorname{Out}(\mathfrak{g}_\Gamma) \,,
\end{equation}
that is, the group of outer-automorphisms of the ADE Lie algebra $\mathfrak{g}_\Gamma$ associated to $\Gamma$ via the McKay correspondence.

As explained in \cite{Apruzzi:2017iqe}, the singular F-theory base $\mathbb{C}^2/\Gamma$, where $\Gamma$ is one of the appropriate finite subgroups of $\U(2)$, enjoys $\mathcal{O}_\text{end}$ as an isometry. The fact that $\mathcal{O}_\text{end}$ is an isometry of the base of the Calabi--Yau (both the base at the generic point of the tensor branch, and at the origin of the tensor branch) means that $\mathcal{O}_\text{end}$ is a candidate global symmetry of the associated SCFT. Of course, an isometry of the base of the compactification space does not necessarily uplift to an isometry of the full compactification space; the latter depends on the compatibility of the elliptic fibration.

To explore the uplift of the isometry to the Calabi--Yau threefold, we explore a pertinent example for this work. Consider the curve configuration 
\begin{equation}\label{eqn:sheila}
    [\mathfrak{su}(K)] \underbrace{\,\overset{\mathfrak{su}(K)}{2}\cdots\overset{\mathfrak{su}(K)}{2}\,}_{N-1} [\mathfrak{su}(K)] \,,
\end{equation}
associated to rank $N$ $(\mathfrak{su}(K), \mathfrak{su}(K))$ conformal matter, where we have written the classical flavor symmetries arising on the left and the right, and where we assume that $N > 2$. As the base geometry is simply $\mathbb{C}^2/\mathbb{Z}_N$, we know from \cite{Apruzzi:2017iqe} that 
\begin{equation}
    \mathcal{O}_\text{end} = \operatorname{Out}(\mathfrak{su}(N)) = \mathbb{Z}_2 \,.
\end{equation}
The action of the $\mathbb{Z}_2$ acts on the $(-2)$-curves in the configuration exactly as the outer-automorphism acts on the Dynkin diagram of the $\mathfrak{su}(N)$ algebra: that is, by simultaneously swapping the $k$-th curve from the left with the $k$-th curve from the right. It is clear that this is compatible with the structure of the elliptic fibration, since the singular fibers supported over the exchanged $(-2)$-curves are identical. Similarly, the singular fibers supported over the non-compact curves, which contribute the $\mathfrak{su}(K) \oplus \mathfrak{su}(K)$ continuous global symmetry, are permuted. Since the action of $\mathcal{O}_\text{end}$ uplifts on an isometry of the full Calabi--Yau threefold, we have identified a discrete global symmetry of the corresponding SCFT.

Further, we note that there can be isometries of the Calabi--Yau threefold that act as a trivial isometry on the base, but act in a non-trivial way only on the fiber. A good example of this phenomenon is the curve configuration
\begin{equation}
    \overset{\mathfrak{su}(K)}{2} \,,
\end{equation}
where $\mathcal{O}_\text{end} = 1$ as it is the group of outer-automorphisms of the $\mathfrak{su}(2)$ Dynkin diagram. Nevertheless, if we group the fundamental hypermultiplets required by anomaly cancellation together in the following way
\begin{equation}
    [\mathfrak{su}(K)] \, \overset{\mathfrak{su}(K)}{2} \, [\mathfrak{su}(K)] \,,
\end{equation}
which is the natural restriction to $N = 2$ of the configuration in equation \eqref{eqn:sheila}, then we can observe a $\mathbb{Z}_2$ that acts only in the fiber by exchanging the left and right flavor algebras. In fact, the flavor symmetry is enhanced
\begin{equation}
    \mathfrak{su}(K) \oplus \mathfrak{su}(K) \quad \rightarrow \quad \mathfrak{su}(2K) \,,
\end{equation}
and this $\mathbb{Z}_2$ discrete symmetry simply becomes a part of the enhanced continuous symmetry. 

\subsection{Flavor Symmetry of Discretely-gauged Conformal Matter}
\label{sec:discretegaugingCM}

We have now discussed when a 6d $(1,0)$ SCFT is expected to evince a discrete global symmetry by studying the effective description at the generic point of the tensor branch. If we assume that this discrete symmetry can be gauged, then we would like to determine some properties of these purported discretely-gauged 6d $(1,0)$ SCFTs. Here, we will focus on the flavor symmetry after discrete gauging; in particular, we will conjecture the flavor symmetry after discrete gauging by studying the tensor branch configuration, similarly to what we did in Section \ref{sec:flavor} for the SCFTs before discrete gauging.

Before returning to the 6d SCFTs themselves, let us first briefly discuss the representation theory of some of the disconnected gauge groups that we are interested in. Let us suppose that we have an algebra
\begin{equation}\label{eqn:babyyoda}
    \mathfrak{g} = \mathfrak{su}(K) \oplus \mathfrak{su}(K) \,.
\end{equation}
We wish to consider the global form of the gauge group to be
\begin{equation}\label{eqn:swiper}
    \SU(K) \times \SU(K) \ltimes \ZZ_2 \,,
\end{equation}
where the $\ZZ_2$ acts as the automorphism that swaps the two $\SU(N)$ factors. We are interested in the representation theory of this global form; in particular the (anti-)fundamental representation of one of the factors does not uplift to a representation of the group as the $\ZZ_2$ acts as
\begin{equation}
    \ZZ_2 \, : \quad (\bm{K}, \bm{1}) \rightarrow (\bm{1}, \bm{\overline{K}}) \,.
\end{equation}
Instead, we can see that 
\begin{equation}
    (\bm{K}, \bm{1}) \oplus (\bm{1}, \bm{\overline{K}}) \,,
\end{equation}
form (complex) irreducible representation of the group in equation \eqref{eqn:swiper}. To see how this analysis can reveal the flavor symmetry, we consider a 4d $\mathcal{N}=2$ conformal gauge theory with gauge algebra as in equation \eqref{eqn:babyyoda} and hypermultiplets in the following representations
\begin{equation}
    \begin{gathered}
        K \times (\bm{\overline{K}, \bm{1}}) \coma 1 \times (\bm{K}, \bm{\overline{K}}) \coma K \times (\bm{1}, \bm{K}) \,.
    \end{gathered}
\end{equation}
If the global form of the gauge group is taken as in equation \eqref{eqn:swiper}, then we expect that there exists an $\mathfrak{su}(K)$ global symmetry rotating the $(\bm{\overline{K}}, \bm{1}) \oplus (\bm{1}, \bm{K})$ irreducible representation of the group. This is in contradistinction to the $\mathfrak{su}(K) \times \mathfrak{su}(K)$ non-Abelian flavor symmetry that arises where the global form of the gauge group is simply $\SU(K) \times \SU(K)$.

Let us first explore the tensor branches associated with (Higgsed) $(A, A)$ conformal matter. We consider first the case where the number of M5-branes engineering the Higgsed conformal matter theory is odd, in which case the tensor branch curve configuration takes the following form:
\begin{equation}
    \underset{[m_1]}{\overset{\su(k_1)}{2}} \,\, 
    \underset{[m_2]}{\overset{\su(k_2)}{2}} \,\, \cdots \,\, 
    \underset{[m_q]}{\overset{\su(k_q)}{2}} \,\,
    \underset{[m_q]}{\overset{\su(k_q)}{2}} \,\, \cdots \,\,
    \underset{[m_{2}]}{\overset{\su({k_{2}})}{2}} \,\, 
    \underset{[m_{1}]}{\overset{\su({k_{
    1}})}{2}} \, \,.
\end{equation}
We can see that this configuration admits a $\ZZ_2$ GS automorphism. After the discrete gauging, the hypermultiplet spectrum indicates that the non-Abelian part of the flavor symmetry is
\begin{equation}
    \bigoplus_{i=1}^q \mathfrak{su}(m_i) \,.
\end{equation}
The Abelian global symmetries again require more care about the existence of ABJ anomalies. As in Section \ref{sec:flavor}, we assume that all $k_i \geq 2$, and that at least one $k_i \geq 3$. If an ABJ-anomaly-free $\mathfrak{u}(1)$ is localized entirely on one side of the tensor branch configuration, then we expect that it is identified with its mirror on the other side of the configuration due to the $\ZZ_2$ automorphism collecting together the hypermultiplet representations on the left and right. There also exists an anomaly-free $\mathfrak{u}(1)$ in the non-discretely-gauged theory which involves the generator of the $\mathfrak{u}(1)$ which rotates the $(\bm{k_q}, \bm{\overline{k_q}})$ bifundamental; as such, this anomaly-free $\mathfrak{u}(1)$ is not identified with any other $\mathfrak{u}(1)$ after the discrete-gauging. Therefore, if we let $\ell$ denote the number of $m_i$ with $m_i \geq 1$, then the total Abelian flavor symmetry is expected to be
\begin{equation}
    \mathfrak{u}(1)^{\ell} \,.
\end{equation}

Next, we consider the tensor branches that can be associated with Higgsed $(A, A)$ conformal matter, where the total number of M5-branes is even. We can write the tensor branch curve configuration as
\begin{equation}\label{eqn:illy}
    \underset{[m_1]}{\overset{\su(k_1)}{2}} \,\, 
    \underset{[m_2]}{\overset{\su(k_2)}{2}} \,\, \cdots \,\, 
    \underset{[m_q]}{\overset{\su(k_q)}{2}} \,\,
    \underset{[m_{q+1}]}{\overset{\su(k_{q+1})}{2}} \,\,
    \underset{[m_q]}{\overset{\su(k_q)}{2}} \,\, \cdots \,\,
    \underset{[m_{2}]}{\overset{\su({k_{2}})}{2}} \,\, 
    \underset{[m_{1}]}{\overset{\su({k_{
    1}})}{2}} \, \,,
\end{equation}
The only difference from the $N$ odd case is that the central $(-2)$-curve acts as a pivot for the Green--Schwarz automorphism. To determine the classical flavor symmetry attached to that central curve, after discrete gauging of the $\mathbb{Z}_2$, we look at the hypermultiplet spectrum. First, we note that $m_{q+1}$ must be even:
\begin{equation}
    m_{q+1} = 2k_{q+1} - 2k_q = 2p \,.
\end{equation}
We can consider this as $p$ dangling hypermultiplets in the fundamental representation of $\mathfrak{su}(k_{q+1})$ and $p$ in the anti-fundamental representation. Since the fundamental and anti-fundamental transform non-trivially under the $\mathbb{Z}_2$, we observe that we have $p$ hypermultiplets in the irreducible representation 
\begin{equation}
    \bm{k_{q+1}} \oplus \overline{\bm{k_{q+1}}} \,,
\end{equation}
of the $SU(k_{q+1}) \ltimes \mathbb{Z}_2$ gauge group. Since this is a real representation, we expect the hypermultiplets to be rotated by a classical symplectic symmetry.

Putting everything together, the hypermultiplet spectrum indicates that the non-Abelian part of the flavor symmetry after discrete gauging of the theory in equation \eqref{eqn:illy} is
\begin{equation}
    \mathfrak{usp}(m_{q+1}) \oplus  \bigoplus_{i=1}^q \mathfrak{su}(m_i) \,.
\end{equation}
A similar analysis of the Abelian symmetries applies as in the case where the number of M5-branes is odd. Again, let $\ell$ be the number of $m_i$ such that $m_i \geq 1$ for $i \leq q$. Due to the hypermultiplet spectrum, we expect that the total number of ABJ-anomaly-free Abelian symmetries in the discretely-gauged theory is
\begin{equation}
    \mathfrak{u}(1)^{\ell} \,.
\end{equation}

In this analysis, we have determined the presence of a $\mathfrak{usp}(m_{q+1})$ global symmetry arising from the central curve after discrete gauging by studying the hypermultiplet spectrum. In more complicated cases, such as with E-string flavor, we may not have access to a classical spectrum, and thus we would like to be able to understand the discretely-gauged flavor symmetry directly by thinking about the Higgs branch chiral ring operators of the non-discretely-gauged theory. In particular, note that the $\mathfrak{su}(m_{q+1})$ symmetry on the central node of a Higgsed conformal matter theory actually arises via the enhancement
\begin{equation}\label{eqn:gnorc}
    \mathfrak{su}\left(\frac{m_{q+1}}{2}\right) \oplus \mathfrak{su}\left(\frac{m_{q+1}}{2}\right) \quad \rightarrow \quad 
    \mathfrak{su}\left(m_{q+1}\right) \,.
\end{equation}
The $\ZZ_2$ GS automorphism acts as follows. Consider the abstract theory:\footnote{Which we assume is good, in the sense of \cite{LMS}.}
\begin{equation}
    A_{N}^{\mathfrak{su}(K)} \left( [K^{n_K}, \cdots, 2^{n_2}, 1^{n_1}], [K^{n_K}, \cdots, 2^{n_2}, 1^{n_1}] \right) \,.
\end{equation}
The GS automorphism identifies the moment maps of the $\mathfrak{su}(n_i)$ global symmetry arising from the left partition, with the same global symmetry arising from the right partition. When there is an enhancement, such as in equation \eqref{eqn:gnorc}, we also need to take care of the effect of the discrete-gauging on the extra moment maps. In this case, we have
\begin{equation}
    \begin{aligned}
        \mathfrak{su}(m_{q+1}) \,\, &\rightarrow \,\, \mathfrak{su}\left(\frac{m_{q+1}}{2}\right) \oplus \mathfrak{su}\left(\frac{m_{q+1}}{2}\right) \\
        \textbf{adj} \,\, &\rightarrow \,\, (\textbf{adj}, \bm{1}) \oplus (\bm{1}, \textbf{adj}) \oplus (\bm{1}, \bm{1}) \oplus \left(\bm{\frac{m_{q+1}}{2}}, \overline{\bm{\frac{m_{q+1}}{2}}}\right) \oplus \left(\overline{\bm{\frac{m_{q+1}}{2}}}, \bm{\frac{m_{q+1}}{2}}\right) \,.
    \end{aligned}
\end{equation}
The identification of the moment maps leaves behind a single $\mathfrak{su}\left(\frac{m_{q+1}}{2}\right)$ flavor symmetry, and we find that the matter content is obtained via the taking of the symmetric projection under the $\ZZ_2$ that swaps the flavor factors.\footnote{Here we have chosen a very specific $\mathbb{Z}_2$ action. In particular, different choices of $\mathbb{Z}_2$ action leads to a different projection, for example, conjugating the $\ZZ_2$ action that lead to equation \eqref{eq:symproject} by an element of $\mathfrak{su}(m_{q+1})$ can break the flavor symmetry to different real subgroups, like $\mathfrak{so}(m_{q+1})$. In this work, we considered the projection to $\mathfrak{usp}(m_{q+1})$ because it is the one that more naturally matches with the wreathing of those magnetic quivers that admit both a unitary and orthosymplectic realizations. The way in which we realized the wreathing in unitary magnetic quivers gives the branching rule in equation \eqref{eq:usptosu}, and we found the equivalent realization on orthosymplectic quivers. The aim of this work is to propose a procedure to wreath orthosymplectic magnetic quivers and give an explanation at the level of discretely-gauged conformal matter, but we are not claiming that we have exhausted all allowed discrete gaugings.} That is, the additional moment maps become:
\begin{equation}\label{eq:symproject}
    \text{Sym}^2 \left(\bm{\frac{m_{q+1}}{2}}, \overline{\bm{\frac{m_{q+1}}{2}}}\right)_{\ZZ_2} \oplus \overline{\text{Sym}^2 \left(\bm{\frac{m_{q+1}}{2}}, \overline{\bm{\frac{m_{q+1}}{2}}}\right)_{\ZZ_2}} \oplus \bm{1} \,,
\end{equation}
where the final $\bm{1}$ is the flavor singlet. That is, in the discretely-gauged theory we have moment maps charged under the following representations of $\mathfrak{su}\left(\frac{m_{q+1}}{2}\right)$:
\begin{equation}
    \textbf{adj} \oplus \textbf{Sym}^2 \oplus \overline{\textbf{Sym}^2} \oplus \bm{1} \,.
\end{equation}
Since this is simply the branching rule of the adjoint representation under
\begin{equation}\label{eq:usptosu}
    \mathfrak{usp}(m_{q+1}) \quad \rightarrow \quad \mathfrak{su}\left(\frac{m_{q+1}}{2}\right) \,,
\end{equation}
we observe an enhancement to a $\mathfrak{usp}(m_{q+1})$ flavor symmetry. We note that this derivation did not involve the hypermultiplet spectrum on the tensor branch, only the data of how the $\ZZ_2$ GS automorphism acts on the moment maps of the (non-Abelian) flavor symmetries.

We emphasize that, thus far, we have discussed the classical global symmetry arising from the analysis of the gauge theories living at the generic point of the tensor branch of the SCFTs under study. We propose that if the classical flavor symmetry at the generic point of the tensor branch ascends to a flavor symmetry of the SCFT in the non-discretely-gauged case, then it also ascends to a flavor symmetry of the SCFT in the discretely-gauged case. 

Now that we have discussed the case of Higgsed $(A, A)$ conformal matter, we turn to the Higgsed $(D, D)$ conformal matter. We consider a generic tensor branch curve configuration, for a conformal matter theory of odd rank, which takes the following abstract form:
\begin{equation}\label{eqn:DDhyp}
   \displaystyle\underset{[m_1]}{\overset{\usp(2k_1)}{1}}\,\, 
    \underset{[m_2]}{\overset{\so(k_2)}{4}}\,\,
    \underset{[m_3]}{\overset{\usp(2k_3)}{1}}\,\, 
    \cdots
    \underset{[m_{q-1}]}{\overset{\so(k_{q-1})}{4}}\,\,
    \underset{[m_{q}]}{\overset{\usp(2k_{q})}{1}}\,\, 
    \underset{[m_{q-1}]}{\overset{\so(k_{q-1}')}{4}}\,\,
    \cdots
    \underset{[m_3]}{\overset{\usp(2k_3)}{1}}\,\, 
    \underset{[m_2]}{\overset{\so(k_2)}{4}}\,\,
    \underset{[m_1]}{\overset{\usp(2k_1)}{1}} 
    \,,
\end{equation}
where $m_i$ and $k_i$ are related such that the anomaly cancellation conditions are all satisfied. We can see from this tensor branch that there exists a $\ZZ_2$ GS automorphism that extends to a discrete symmetry of the 6d SCFT at the origin of the tensor branch. 

When considering the Higgsed $(D, D)$ conformal matter, each simple flavor symmetry factor as described in Section \ref{sec:flavor} can be considered as localized on a single curve in the tensor branch configuration. This is in contrast to $(A, A)$ conformal matter, where the ABJ-anomaly-free $\mathfrak{u}(1)$ symmetries can be a linear combination of $\mathfrak{u}(1)$s distributed over a chain of curves. Since the $\ZZ_2$ discrete symmetry arising from the GS automorphism swaps the tensor multiplets and gauge algebras associated to the curves on the left and the right, it is clear that the gauging modifies the flavor symmetries as follows:
\begin{equation}
    \bigoplus_{i=1}^{q-1} \mathfrak{f}_i(m_i)^{\oplus 2} \quad \rightarrow \quad \bigoplus_{i=1}^{q-1} \mathfrak{f}_i(m_i) \,,
\end{equation}
where 
\begin{equation}
    \mathfrak{f}_i(m_i) = \begin{cases}
        \mathfrak{so}(m_i) \quad &\text{ if $i$ odd,} \\
        \mathfrak{usp}(m_i) \quad &\text{ if $i$ even.}
    \end{cases}
\end{equation}
We note that in equation \eqref{eqn:DDhyp}, we can have a configuration where we need to apply the replacement rules in equation \eqref{eqn:replacementrules} to observe the true tensor branch configuration. Regardless, our conclusion holds that the $\ZZ_2$ discrete gauging identifies the flavor symmetries that are localized on the curves to the left and to the right of the central $(-1)$-curve.

For the central $(-1)$-curve, we use the same derivation involving the moment maps that we used for the central $(-2)$-curve in the $(A, A)$ conformal matter case. We first assume that $k_q \geq 0$, so that we are considering a classical flavor symmetry as opposed to a flavor symmetry arising from the E-string. The flavor symmetry attached to the central $(-1)$-curve is enhanced as follows:
\begin{equation}
    \mathfrak{so}\left(\frac{m_q}{2} \right) \oplus \mathfrak{so}\left(\frac{m_q}{2} \right) \quad\rightarrow\quad \mathfrak{so}(m_q) \,,
\end{equation}
where the $\ZZ_2$ GS automorphism acts to identify the moment maps of the two factors on the left. Under the associated decomposition, the branching rule of the adjoint representation is as follows:
\begin{equation}
\begin{aligned}
     \so(m_{q+1}) \quad &\rightarrow \quad \mathfrak{so}\left(\frac{m_q}{2}\right)\oplus \mathfrak{so}\left(\frac{m_q}{2}\right)\\
     \textbf{adj} \quad &\rightarrow \quad (\textbf{adj}, \bm{1}) \oplus (\bm{1}, \textbf{adj}) \oplus \left(\bm{\frac{m_q}{2}}, \bm{\frac{m_q}{2}}\right) \,.
\end{aligned}
\end{equation}
The $\ZZ_2$ then projects onto the symmetric subspace of the additional matter in the bifundamental representation, leading to additional moment maps in the representations
\begin{equation}
    \text{Sym}^2 \left(\bm{\frac{m_q}{2}}, \bm{\frac{m_q}{2}}\right)_{\ZZ_2} \oplus \bm{1} \fstop
\end{equation}
In short, after discrete gauging, we find an $\mathfrak{so}\left(\frac{m_q}{2}\right)$ global symmetry with moment maps in the representation
\begin{equation}
    \textbf{adj} \oplus \textbf{Sym}^2 \oplus \bm{1} \,.
\end{equation}
This is nothing other than the branching rule of the adjoint representation under
\begin{equation}
    \mathfrak{u}\left(\frac{m_q}{2}\right)\quad \rightarrow \quad \mathfrak{so}\left(\frac{m_q}{2}\right)
\end{equation} 
and thus we would expect such an enhancement of the global symmetry.  

Even if we will not consider Higgsed $(D, D)$ conformal matter theories where the rank
is even, in which case the central curve is a $(-4)$-curve rather than a
$(-1)$-curve, we can nevertheless comment on their flavor symmetries. Again, it seems apparent that the flavor symmetries on
the left and right of the $(-4)$-curve are identified under the $\ZZ_2$
discrete gauging, so, it seems straightforward to extend the analysis identifying the moment maps for the central $(-4)$-curve as we did both for the central $(-2)$- and $(-1)$-curves. Since we will not study examples of magnetic quivers for the Higgs branch of such theories, we will refrain from commenting further.

Thus, we have determined the putative flavor symmetry for the $\ZZ_2$ discrete gauging of almost all examples of Higgsed $(A, A)$ and $(D, D)$ conformal matter that admit such a $\ZZ_2$ symmetry. There are only a small, finite handful of SCFTs where the flavor symmetry after discrete-gauging is unclear; these occur when performing
$\ZZ_2$ discrete gauging on certain theories where the central $(-1)$-curve is
undecorated and, thus, there exists E-string flavor attached to it.  To discuss these special cases, we begin by reminding the reader that the
generators of the Higgs branch chiral ring of the rank one $(D, D)$ conformal
matter theory, with tensor branch
\begin{equation}
  \overset{\mathfrak{usp}(2p)}{1} \,,
\end{equation}
are \cite{Hanany:2018uhm,Ferlito:2017xdq,Distler:2022yse}, under the $\mathfrak{so}(4p + 16) \oplus \mathfrak{su}(2)_R$ global symmetry,
\begin{equation}
  \mu \,:\,\, (\textbf{adj}, \bm{3}) \,, \qquad \mu^+ \,:\,\, (\bm{S^+},\, \bm{p + 3}) \,.
\end{equation}
The latter is an extra generator in the spinor representation of the classical
flavor symmetry.\footnote{A priori, there is the freedom in whether to choose the positive or negative chirality spinor representation as the additional generator. In this case, the two choices are related via an outer-automorphism of the $\mathfrak{so}(4p+16)$ global symmetry, and thus give rise to equivalent theories. However, when considering tensor branch configuration for more general Higgsed $(D, D)$ conformal matter, it is important to keep track of these choices, as they may not all be equivalent. See \cite{Distler:2022yse} for a detailed discussion.} Typically, when gauging on the left and the right, there will be a gauge-invariant remnant of the moment map $\mu$, which will contribute a moment map of the gauged theory. However, we can formally consider an ``analytic continuation'' to $p \leq 0$. For $p = 0$, we see that the $\mu^+$ is itself a moment map operator, and provides the enhancement
\begin{equation}
    \mathfrak{so}(16) \quad \rightarrow \quad \mathfrak{e}_8 \,.
\end{equation}
After gauging, the $\mu^+$ will not typically leave behind a gauge-invariant moment map operator, and attempting to construct a gauge-invariant operator out of $\mu^+$ generally leads to operators with large R-charge. However, when $p = -2$, which is formally allowed by the replacement rules in equation \eqref{eqn:replacementrules}, the $\mu^+$ operator has trivial R-charge, and thus can be combined with the $\mu^+$ of the central curve with $p=0$ to form gauge-invariant operators transforming in the $\bm{3}$ of the $\mathfrak{su}(2)_R$. When this situation occurs, we need to understand how the $\ZZ_2$ GS automorphism acts on this sector of the $\frac{1}{2}$-BPS operator spectrum. The only three 6d $(1,0)$ SCFTs that come from $(D, D)$ conformal matter and admit a $\ZZ_2$ GS automorphism that have this feature are associated to the tensor branch configurations
\begin{equation}
    \overset{\mathfrak{su}(2)}{2}\,\,\overset{\mathfrak{g}_2}{3}\,\,1\,\,\overset{\mathfrak{g}_2}{3}\,\,\overset{\mathfrak{su}(2)}{2} \,, \qquad \overset{\mathfrak{g}_2}{3}\,\,1\,\,\overset{\mathfrak{g}_2}{3} \,,\qquad \overset{\mathfrak{su}(3)}{3}\,\,1\,\,\overset{\mathfrak{su}(3)}{3} \,.
\end{equation}
Since these are rather exceptional cases we only briefly discuss one of them here.

As mentioned, these configurations have an enhanced flavor symmetry from that we might
expect when we think of the theory naively as a Higgsing of a $(D, D)$
conformal matter theory. To see this explicitly, we now delve into the latter theory more deeply. We consider the following conformal
matter theory with its associated tensor branch curve configuration:
\begin{equation}
  \begin{gathered}
    A_2^{\mathfrak{so}(8)}([3^2, 1^2], [3^2, 1^2])
  \end{gathered} \,\,: \qquad
  \begin{gathered}
    \overset{\mathfrak{su}(3)}{3} \,\,1\,\, \overset{\mathfrak{su}(3)}{3} \,.
  \end{gathered}
\end{equation}
As described in Section \ref{sec:flavor}, this SCFT has an $\mathfrak{su}(3)
\oplus \mathfrak{su}(3)$ flavor symmetry which can be determined by looking at
the commutant of the $\mathfrak{su}(3) \oplus \mathfrak{su}(3)$ gauge group
inside the $\mathfrak{e}_8$ flavor symmetry associated with the single
$(-1)$-curve. In fact, we can reproduce this flavor symmetry from the conformal
matter perspective. Naively, before applying the replacement rules in 
equation \eqref{eqn:replacementrules}, the tensor branch configuration corresponding to this SCFT is
\begin{equation}\label{eqn:faun}
	\underset{[2]}{\overset{\mathfrak{usp}(-4)}{1}} \,\, \overset{\mathfrak{so}(6)}{4} \,\, \underset{[4]}{1} \,\, \overset{\mathfrak{so}(6)}{4} \,\, \underset{[2]}{\overset{\mathfrak{usp}(-4)}{1}} \,.
\end{equation}
We have chosen to write the number of additional half-hypermultiplets directly,
rather than the classical flavor symmetries rotating them. This theory has a sequence of two enhancements of the flavor symmetries:
\begin{equation}
    \mathfrak{so}(2) \oplus \mathfrak{so}(2) \oplus \mathfrak{so}(2) \oplus \mathfrak{so}(2)  \quad \rightarrow \quad  \mathfrak{so}(2) \oplus \mathfrak{so}(4) \oplus \mathfrak{so}(2) \quad \rightarrow \quad  \mathfrak{su}(3) \oplus \mathfrak{su}(3) \,,
\end{equation}
where we have first written the manifest global symmetries from the nilpotent orbits, the enhanced symmetry due to the shortness of the tensor branch configuration, and finally the enhancement arising from the non-perturbative nature of the E-string. We now attempt to understand these enhancements, in particular the last one, in the non-discretely-gauged theory. 

Let us now try to write down 
gauge-invariant operators of the theory with tensor branch as in equation \eqref{eqn:faun} transforming in the $\bm{3}$ of the R-symmetry and built out 
of these operators. We obtain the following moment map operators:
\begin{equation}
  \begin{array}{c|cccc}
    \text{Operator} & \mathfrak{so}(2) & \mathfrak{su}(2) & \mathfrak{su}(2) & \mathfrak{so}(2) \\\hhline{=|====}
    \mu_L & \bm{1}_0 & \bm{1} & \bm{1} & \bm{1}_0 \\
    \mu_R & \bm{1}_0 & \bm{1} & \bm{1} & \bm{1}_0 \\
    \mu_C & \bm{1}_0 & \bm{3} & \bm{1} & \bm{1}_0 \\
    \mu_C & \bm{1}_0 & \bm{1} & \bm{3} & \bm{1}_0 \\
    \mu_L^+ \otimes \mu_C^+ \otimes \mu_R^+ & \bm{1}_{1} & \bm{2} & \bm{1} & \bm{1}_{1} \\
    \mu_L^+ \otimes \mu_C^+ \otimes \mu_R^+ & \bm{1}_{-1} & \bm{2} & \bm{1} & \bm{1}_{-1} \\
    \mu_L^+ \otimes \mu_C^+ \otimes \mu_R^+ & \bm{1}_{1} & \bm{1} & \bm{2} & \bm{1}_{-1} \\
    \mu_L^+ \otimes \mu_C^+ \otimes \mu_R^+ & \bm{1}_{-1} & \bm{1} & \bm{2} & \bm{1}_{1} 
  \end{array} \sfstop{\fstop}{65pt}
  \label{eq:momentmapoperators}
\end{equation}
Here, we have written the $\mathfrak{so}(4)$ attached to the central curve as
$\mathfrak{su}(2) \oplus \mathfrak{su}(2)$, and we have used subscripts
$L$, $R$, and $C$ to denote the operators coming from the Higgs branch
generators of the left, right, and central $(-1)$-curves, respectively. That
is, we observe eight extra moment-map operators coming from combinations of the
spinor representations, and it is easy to see from the charges, up to a linear redefinition of the $\mathfrak{u}(1)$s, that these come
from the standard regular maximal embedding of $\mathfrak{su}(2) \oplus
\mathfrak{u}(1)$ inside $\mathfrak{su}(3)$. Therefore, we have observed,
from this slightly unusual perspective with negative-rank gauge algebras on the
tensor branch, the reproduction of the expected global symmetry of the
associated SCFT: $\mathfrak{su}(3) \oplus \mathfrak{su}(3)$.

Now we can discuss what happens to these operators under the $\ZZ_2$ discrete gauging. For the non-spinorial operators, the story is the same as for the central $(-1)$-curve in the case where the flavor symmetry is classical: the $\mathfrak{so}(2) \oplus \mathfrak{so}(4) \oplus \mathfrak{so}(2)$ becomes $\mathfrak{so}(2) \oplus \mathfrak{u}(2)$. It is less obvious from first principles what happens to the operators coming from the gauge-invariant combinations of $\mu_L^+ \otimes \mu_C^+ \otimes \mu_R^+$; indeed, we shall return to this point when we discuss the wreathed 3d magnetic quiver for this discretely-gauged theory. We now have a proposal for the expected flavor symmetry of the $\ZZ_2$
discretely-gauged versions of 6d $(1,0)$ (Higgsed) conformal matter theories of
type $(A, A)$ and $(D, D)$. We test this proposal by considering the magnetic
quivers for the Higgs branch of certain conformal matter theories and wreathing
by the $\ZZ_2$ quiver automorphism, which we believe to be the dual of
the $\ZZ_2$ discrete symmetry. From the wreathed magnetic quiver, we can
determine the Coulomb branch Hilbert series, and thus extract the number of
moment map operators of the discretely-gauged 6d SCFT. For computational tractability, we are interested in cases where the total rank
of the gauge algebra in the magnetic quiver is not too large. Furthermore,
since we will use the wreathed analog of the monopole formula to determine the
Coulomb branch Hilbert series, we require that all the gauge nodes of the
(unwreathed) magnetic quiver be good in the sense of \cite{Gaiotto:2008ak}.
Due to these constraints, we consider a small set of explicit examples of 6d
SCFTs, which we list in Table \ref{tab:flavorsymm6d}, together with their
predicted flavor symmetries before and after discrete gauging.

\begin{landscape}
 \pagestyle{empty}
   \begin{longtable}{c|c|c|c|c}
      \caption{We list the curve configuration at the generic point of the tensor branch of each SCFT as well as the conjectural flavor symmetry before ($\mathfrak{f}$) and after  ($\mathfrak{f}_{\ZZ_2}$) discrete gauging. We write a ${\color{red}?}$ for $\mathfrak{f}_{\ZZ_2}$ in two rows where there are exceptional moment maps which obscure the expected flavor symmetry; these are predicted later from the wreathed magnetic quivers.}
    \label{tab:flavorsymm6d}\\    
              \# & Conformal Matter & Tensor Branch & $\mathfrak{f}$ & $\mathfrak{f}_{\ZZ_2}$ \\ \hhline{=|=|=|=|=}
  \endfirsthead
    \# & Conformal Matter & Tensor Branch & $\mathfrak{f}$ & $\mathfrak{f}_{\ZZ_2}$ \\ \hhline{=|=|=|=|=}
  \endhead
        1  & $A_{3}^{\so(6)}([3, 1^3], [3, 1^3])$ & $\overset{\mathfrak{su}(2)}{2}\,\,\overset{\mathfrak{su}(4)}{2}\,\,\overset{\mathfrak{su}(2)}{2}$  & $\mathfrak{su}(4)$ & $\mathfrak{usp}(4)$ \\ \hline
        2  & $A_{1}^{\so(6)}([1^6], [1^6])$ & $\overset{\mathfrak{su}(4)}{2}$  & $\mathfrak{su}(8)$ & $\mathfrak{usp}(8)$ \\\hline
        3  & $A_{5}^{\so(6)}([3^2], [3^2])$ & $\overset{\mathfrak{su}(2)}{2}\,\,\overset{\mathfrak{su}(3)}{2}\,\,\overset{\mathfrak{su}(4)}{2}\,\,\overset{\mathfrak{su}(3)}{2}\,\,\overset{\mathfrak{su}(2)}{2}$  & $\mathfrak{su}(2) \oplus \mathfrak{u}(1)^{\oplus 2}$ & $\mathfrak{su}(2) \oplus \mathfrak{u}(1)$  \\\hline
        4  & $A_{2}^{\so(8)}([3^2, 1^2], [3^2, 1^2])$ & $\overset{\mathfrak{su}(3)}{3}\,\,1\,\,\overset{\mathfrak{su}(3)}{3}$  & $\mathfrak{su}(3)^{\oplus 2}$& ${\color{red} ?}$  \\\hline
        5  & $A_{4}^{\so(8)}([5, 1^3], [5, 1^3])$ & $\overset{\mathfrak{su}(2)}{2}\,\,\overset{\mathfrak{so}(7)}{3}\,\,1\,\,\overset{\mathfrak{so}(7)}{3}\,\,\overset{\mathfrak{su}(2)}{2}$  & $\mathfrak{su}(2)^{\oplus 2} \oplus \mathfrak{u}(1)$ & $\mathfrak{su}(2) \oplus \mathfrak{u}(1)$ \\\hline
        6  & $A_{4}^{\so(8)}([5, 3], [5, 3])$ & $\overset{\mathfrak{su}(2)}{2}\,\,\overset{\mathfrak{g}_2}{3}\,\,1\,\,\overset{\mathfrak{g}_2}{3}\,\,\overset{\mathfrak{su}(2)}{2}$  & $\mathfrak{su}(2)$ & ${\color{red} ?}$ \\\hline
        7 &$A_{3}^{\so(8)}([4, 2,1^2], [4, 2,1^2])$ &$\overset{\su(3)}{3}\,\,1\,\,\overset{\so(8)}{4}\,\,1\,\,\overset{\su(3)}{3}$ & $\so(2)^{\oplus 4}$ & $\so(2)^{\oplus 2}$\\\hline
        8  & $A_{2}^{\so(10)}([3^3, 1], [3^3, 1])$ & $\overset{\mathfrak{g}_2}{3}\,\,\overset{\mathfrak{su}(2)}{1}\,\,\overset{\mathfrak{g}_2}{3}$  & $\mathfrak{so}(6)$ & ${\mathfrak{u}(3)}$\\\hline
        9  & $A_{4}^{\so(10)}([5, 3, 1^2], [5, 3, 1^2])$ & $\overset{\su(3)}{3}\,\,1\,\,\overset{\so(9)}{4}\,\,\overset{\su(2)}{1}\,\,\overset{\so(9)}{4}\,\,1\,\,\overset{\su(3)}{3}$ & $\mathfrak{u}(1)^{\oplus 3}$ & $\mathfrak{u}(1)^{\oplus 2}$ \\\hline
        10  & $A_{4}^{\so(10)}([5^2], [5^2])$ & $\overset{\su(2)}{2}\,\,\overset{\so(7)}{3}\,\,\overset{\su(2)}{1}\,\,\overset{\so(7)}{3}\,\,\overset{\su(2)}{2}$ & $\mathfrak{so}(4)$ &  $\mathfrak{u}(2)$ \\\hline
        11 &$A_{3}^{\so(10)}([4, 3,2,1], [4,3, 2,1])$ &$\overset{\mathfrak{g}_2}{3}\,\,\overset{\su(2)}{1}\,\,\overset{\so(10)}{4}\,\,\overset{\su(2)}{1}\,\,\overset{\mathfrak{g}_2}{3}$ & $\su(2)^{\oplus 2}$ & $\su(2)$ \\\hline
        12  & $A_{5}^{\so(10)}([5^2], [5^2])$ & $\overset{\su(2)}{2}\,\,\overset{\so(7)}{3}\,\,\overset{\su(2)}{1}\,\,\overset{\so(10)}{4}\,\,\overset{\su(2)}{1}\,\,\overset{\so(7)}{3}\,\,\overset{\su(2)}{2}$ & $\mathfrak{so}(2)^{\oplus 2}$ &  $\mathfrak{so}(2)$ \\\hline
        13 & $A_{6}^{\so(10)}([7, 1^3], [7, 1^3])$ & $\overset{\su(2)}{2}\,\,\overset{\so(7)}{3}\,\,1\,\,\overset{\so(9)}{4}\,\,\overset{\su(2)}{1}\,\,\overset{\so(9)}{4}\,\,1\,\,\,\overset{\so(7)}{3}\,\,\overset{\su(2)}{2}$ & $\mathfrak{su}(2)^{\oplus 2} \oplus \mathfrak{u}(1)$ & $\mathfrak{su}(2) \oplus \mathfrak{u}(1)$ \\\hline
        14 & $A_{6}^{\so(10)}([7, 3], [7, 3])$ & $\overset{\su(2)}{2}\,\,\overset{\mathfrak{g}_2}{3}\,\,1\,\,\overset{\so(9)}{4}\,\,\overset{\su(2)}{1}\,\,\overset{\so(9)}{4}\,\,1\,\,\,\overset{\mathfrak{g}_2}{3}\,\,\overset{\su(2)}{2}$ & $\mathfrak{u}(1)$ & $\mathfrak{u}(1)$ \\\hline
        15 & $A_{1}^{\su(3)}([1^3], [1^3])$ & $\overset{\mathfrak{su}(3)}{2}$  & $\mathfrak{su}(6)$ & $\mathfrak{usp}(6)$ \\\hline
        16 & $A_{3}^{\su(3)}([2, 1], [2, 1])$ & $\overset{\mathfrak{su}(2)}{2}\,\,\overset{\mathfrak{su}(3)}{2}\,\,\overset{\mathfrak{su}(2)}{2}$  & $\mathfrak{su}(2) \oplus \mathfrak{u}(1)^{\oplus 2}$ & $\mathfrak{su}(2) \oplus \mathfrak{u}(1)$  \\\hline
        17 & $A_{3}^{\su(4)}([2, 1^2], [2, 1^2])$ & $\overset{\mathfrak{su}(3)}{2}\,\,\overset{\mathfrak{su}(4)}{2}\,\,\overset{\mathfrak{su}(3)}{2}$  & $\mathfrak{su}(2)^{\oplus 3} \oplus \mathfrak{u}(1)^{\oplus 2}$ & $\mathfrak{su}(2)^{\oplus 2} \oplus \mathfrak{u}(1)$  
\end{longtable}
\end{landscape}

\section{Coulomb Branch of Wreathed Quivers}
\label{sec:wreathedCB}

In this section, we review the wreathing procedure in the context of 3d $\mathcal{N}=4$ quiver gauge theories. Introduced in \cite{Bourget:2020bxh}, it has been shown in a series of works \cite{Argyres:2016yzz, Hanany:2018vph, Hanany:2018cgo, Hanany:2018dvd, Bourget:2020bxh, Hanany:2023uzn,Giacomelli:2024sex,Grimminger:2024mks} that quivers wreathed by a discrete group $\Gamma$ are magnetic quivers for the Higgs branches of certain 4d $\mathcal{N}=2$ SCFTs discretely gauged by $\Gamma$. 

Let us first review the definition of the Coulomb branch Hilbert series. Consider a 3d $\mathcal{N}=4$ quiver gauge theory $\mathcal{X}$ with (reductive) gauge group $G$ connected by a set of edges associated with bifundamental hypermultiplets. The Coulomb branch Hilbert series is given by \cite{Cremonesi:2013lqa}
\begin{equation}
    \HS[\text{CB of $\mathcal{X}$}](t)=\frac{1}{|W|}\sum_{\mathbf{m}}\sum_{\gamma\in W(\mathbf{m})}\frac{t^{2\Delta(\mathbf{m})}}{\det(\ID-t^2\gamma)}\coma
\end{equation}
where $W$ is the Weyl group of $G$, and $\Delta(\mathbf{m})$ is the dimension of the monopole operator with magnetic flux $\mathbf{m}$, generically given by \cite{Gaiotto:2008ak,Cremonesi:2013lqa}
\begin{equation}
    \Delta(\mathbf{m}) = -\sum_{\alpha \in \Delta_+}|\alpha(\mathbf{m})|+\frac{1}{2}\sum_{i=1}^{N_f}\sum_{\rho_i\in\mathcal{R}_i}|\rho_i(\mathbf{m})|\coma
\end{equation}
where $\alpha\in\Delta_+$ are the positive roots of the gauge group $G$ and $\rho_i$ are the weights of the irreducible matter field representation $\mathcal{R}_i$ under the gauge group. Respectively, they represent the vector multiplet and hypermultiplet contributions to the dimension of the monopole operators. The summation of the magnetic fluxes depends on the gauge group $G$, e.g., for a quiver with only unitary gauge nodes, the summation is over $\mathbf{m}\in \ZZ^r$, with $r$ being the rank of the gauge group. In this work, we will consider only certain representations for unitary and orthosymplectic groups, which we list in Table \ref{tab:conformaldimensions}, together with the corresponding contributions to the conformal dimension of the monopole operators \cite{Cremonesi:2013lqa,Bourget:2020xdz}. 

\begin{table}[!htp]
    \centering
    \begin{subtable}[t]{\textwidth}
    \centering
    \renewcommand*{\arraystretch}{2}
        \begin{tabular}{c|c}
       Group  &  $\displaystyle -\sum_{\alpha \in \Delta_+}|\alpha(\mathbf{m})|$ \\\hhline{=|=}
       $\U(N)_\mathbf{m}$ & $\displaystyle -\sum_{i<j}^N|m_i-m_j|$\\\hline
       $\SO(2N)_\mathbf{m}$ & $\displaystyle -\sum_{i<j}^N\left(|m_i+m_j|+|m_i-m_j|\right)$\\\hline
       $\USp(2N)_\mathbf{m}$ & $\displaystyle -\sum_{i<j}^N\left(|m_i+m_j|+|m_i-m_j|\right)-2\sum_{i=1}^N|m_i|$
    \end{tabular}
    \caption{Vector multiplet contribution.}
    \end{subtable}\\[4em]
    \begin{subtable}[t]{\textwidth}
    \centering
    \renewcommand*{\arraystretch}{2}
    \begin{tabular}{c|c}
       Representation & $\displaystyle \frac{1}{2}\sum_{i=1}^{N_f}\sum_{\rho_i\in\mathcal{R}_i}|\rho_i(\mathbf{m})|$\\\hhline{=|=}
       Bifundamental of $\U(N)_\mathbf{m}\times \U(M)_\mathbf{n}$ & $\displaystyle \frac{1}{2}\sum_{i=1}^N\sum_{j=1}^M|n_i-m_j|$\\\hline
       Bifundamental of $\SO(2N)_\mathbf{m}\times \USp(2M)_\mathbf{n}$ & $\displaystyle \frac{1}{2}\sum_{i=1}^N\sum_{j=1}^M\left(|n_i+m_j|+|n_i-m_j|\right)$\\\hline
       Adjoint of $\U(N)_\mathbf{m}$ & $\displaystyle \sum_{i<j}^N|m_i-m_j|$\\\hline
       Antisymmetric $\Lambda^2$ of $\USp(2N)_\mathbf{m}$ & $\displaystyle \sum_{i<j}^N\left(|m_i+m_j|+|m_i-m_j|\right)$
    \end{tabular}
    \caption{Hypermultiplet contribution.}
    \end{subtable}
    \caption{Contributions to the conformal dimension $\Delta(\mathbf{m})$ in the monopole formula \cite{Cremonesi:2013lqa,Bourget:2020xdz}, for magnetic fluxes $\mathbf{m}=(m_1,\ldots,m_N)$ and $\mathbf{n}=(n_1,\ldots,n_M)$. The subscripts on the groups denote the magnetic fluxes in the lattice associated to that group.}
    \label{tab:conformaldimensions}
\end{table}

The prefactor
\begin{equation}
    P_G(t,\mathbf{m}) = \frac{1}{|W|}\sum_{\gamma\in W(\mathbf{m})}\frac{1}{\det(\ID-t^2\gamma)}\coma
\end{equation}
is a classical contribution that counts the gauge invariant operators of the residual gauge group unbroken by the magnetic flux $\mathbf{m}$. The computation of this contribution is reviewed in Appendix \ref{sec:prefactors}, while in Section \ref{sec:wreathedprefactors}, we will discuss how this is modified when a $\ZZ_2$ wreathing is considered.

The action of the wreathing can be implemented at the level of the Coulomb branch Hilbert series of a 3d $\mathcal{N}=4$ SCFT by acting with $\Gamma$ on the Weyl group $W$ of the gauge group of the quiver, and on the summations over the magnetic fluxes. Suppose that such a quiver possesses a diagram automorphism by the finite group $\Gamma$, i.e., $\Gamma$ leaves $\Delta(\mathbf{m})$ invariant. Then it is possible to consider the wreathing by $\Gamma$ of the quiver, obtaining the wreathed quiver, which we refer to as $\widetilde{\mathcal{X}}$. The general expression for the Hilbert series is 
\begin{equation}
    \HS\left[\text{CB of $\widetilde{\mathcal{X}}$}\right](t)=\frac{1}{|W_\Gamma|}\sum_{\mathbf{n}}\sum_{\gamma\in W_\Gamma(\mathbf{n})}\frac{t^{2\Delta(\mathbf{n})}}{\det(\ID-t^2\gamma)}\coma
\end{equation}
where we have denoted $W_\Gamma = W\wr \Gamma$. It is also important to note that the summation over $\mathbf{n}$ and the prefactors associated to the residual gauge symmetry may differ after wreathing. In fact, if the quiver has topological symmetries, the wreathing by $\Gamma$ acts on a subset of the gauge nodes of the quiver, breaking the symmetry to its diagonal subgroup. The topological symmetry is a subset of the original symmetry.

For our purposes, one computes the wreathing of the total residual Weyl group $W_\Gamma(\mathbf{m})$ for a given choice of fluxes $\mathbf{m}$. In practice, $W_\Gamma(\mathbf{m})$ is given by all the matrices in $W\wr \Gamma$ that leave $\mathbf{m}$ invariant. In the case of unitary quivers, $W$ is usually a product of $S_N$ factors, that being the Weyl group for a $\U(N)$ gauge group. This way of computing the Coulomb branch Hilbert series for wreathed unitary quivers has been explored in recent works \cite{Bourget:2020bxh,Arias-Tamargo:2021ppf,Giacomelli:2024sex,Grimminger:2024mks}, while in this paper we are interested in applying such techniques to orthosymplectic quivers. 

\subsection{Prefactors for Hilbert Series of \texorpdfstring{\boldmath{$\ZZ_2$}}{Z2}-wreathed Quivers}
\label{sec:wreathedprefactors}

In Appendix \ref{sec:prefactors}, we have summarized the method for computing the prefactor in the Hilbert series for a standard unwreathed quiver. In this section, we extend the discussion in the appendix to determine the prefactor contributions to the Hilbert series for a $\ZZ_2$-wreathed quiver. The procedure is easily generalized to more sophisticated wreathing; however we postpone a systematic discussion to future work. 

Let us consider how the classical prefactor is modified when we consider the wreathing of a gauge group $G$ by some discrete group $\Gamma$. As it has been explained in Section \ref{sec:wreathedCB}, $\Gamma$ must be a symmetry of the quiver, i.e., it leaves $\Delta(\mathbf{m})$ invariant. This means that the resulting theory must have a Hilbert series, in which we have to sum over only those operators that are invariant under the action of $\Gamma$. This must be also reflected at the level of the classical prefactor, which generally will be of the following form:
\begin{equation}
    P_{G\,\wr\, \Gamma}(t,\mathbf{n}) = \frac{1}{|W_\Gamma(\mathbf{n})|}\sum_{\gamma \in W_\Gamma(\mathbf{n})}\frac{1}{\det\left(\ID-t^2\gamma\right)}\coma
\end{equation}
where $W_\Gamma(\mathbf{n})$ is the wreathed Weyl group of $G$ left invariant by the choice of fluxes. 

One way to determine the prefactor is to generate the full resulting group $W_\Gamma$ obtained by wreathing the Weyl group of $G$, and select for each choice of fluxes $\mathbf{n}$ the elements that leave the fluxes invariant. However, this operation is computationally intensive, and one must be aware that leaving the lattice $\mathbf{n}$ as it is originally for a given quiver, generically results in overcounting. It is then more efficient trying to subdivide the summation in various contributions, referred to as chambers, depending on the choice of fluxes. Determining the minimal set of contributions that gives the correct result without overcounting is generally difficult, and it depends both on $\Gamma$ and the group $G$ involved in the wreathing. In the following, we limit ourselves to discuss the chambers for $\Gamma \simeq \ZZ_2$, explaining how the prefactor is obtained for $\U(N)$, $\USp(2N)$ and $\SO(2N)$ groups.\footnote{The prefactors for $\SU(N)$ and $\SO(2N+1)$ can be obtained trivially from $\U(N)$ and $\USp(2N)$ respectively.}

Since we consider $\ZZ_2$ wreathing, we start from a quiver that contains two gauge nodes $G$, respectively, with magnetic fluxes $\mathbf{m}=(m_1,\ldots, m_N)$ and $\mathbf{n}=(n_1,\ldots,n_N)$, which are identified under the action of the $\ZZ_2$. The $\ZZ_2$ wreathing acts on the magnetic fluxes by exchanging 
\begin{equation}
    m_i \leftrightarrow n_i\fstop
\end{equation}
As we discuss in Appendix \ref{sec:prefactors}, for any group $G$ being $\U(N)$, $\USp(2N)$ and $\SO(2N)$, one can restrict the computation to a Weyl chamber where the magnetic fluxes are ordered, e.g., for $\U(N)$ as $m_1\geq \ldots \geq m_N$. The only difference between the groups is the domain of the fluxes. To avoid overcounting, for the $\ZZ_2$ wreathing of these groups, one can divide the summations by imposing pairwise ordering on the magnetic fluxes of the two groups. 

Let us first explain via an example of taking a $\ZZ_2$ wreathing of two $\U(2)$ groups. We define $\mathbf{m}=(m_1,m_2)$ and $\mathbf{n}=(n_1,n_2)$ as the fugacities associated to the two $\U(2)$s subjected to the conditions that $m_1\geq m_2$ and $n_1\geq n_2$, as explained in Appendix \ref{sec:prefactors}. As explained in \cite[Page 29]{Bourget:2020bxh}, we need to define under which conditions $(m_1,m_2)\geq (n_1,n_2)$. This is obtained by a lexicographic order such that
\begin{equation}
    (n_1,n_2)\leq (m_1,m_2) \Longleftrightarrow n_2<m_2 \text{ or }  \left(n_2=m_2 \text{ and } n_1\leq m_1\right)\fstop
\end{equation}
The full Hilbert series can then be obtained by restricting the fluxes to the following four chambers 
\begin{equation}
\renewcommand*{\arraystretch}{1.2}
\begin{array}{c}
\text{Restriction}\\
\hhline{=}
    n_2< m_2\coma m_2\leq m_1\coma n_2\leq n_1 \\
    n_2= m_2\coma n_1<m_1\coma m_2\leq m_1\coma n_2\leq n_1 \\
    m_2= n_2\coma m_1=n_1\coma n_2<n_1 \\
    m_2= n_2\coma m_1=n_1\coma n_2=n_1
\end{array}\sfstop{\fstop}{38pt}
\end{equation}
One can see that, in this way, the whole lattice invariant under the $\ZZ_2$ wreathing is covered, but isolating the first two chambers guarantees that $W_\Gamma(\mathbf{m},\mathbf{n})$ is the same as $W(\mathbf{m},\mathbf{n})$, without any further constraints coming from the wreathing. On the other hand, the last two contributions generate further cyclic groups among the fugacities that modify the prefactor. In particular, the prefactor reduces to the one generated by the product of the Weyl group $S_2$ of a single $\U(2)$ with two distinct $S_2$ actions that exchange the fluxes. We can call $(12)$ the points of the $S_2$ symmetric group associated to the fugacities $(m_1,m_2)$ and $(34)$ the points for $(n_1,n_2)$, so, for the example at hand, the prefactors for the various contributions are generated by the group elements in 
\begin{equation}
\renewcommand*{\arraystretch}{1.2}
\begin{array}{c|c}
\text{Restriction} & \text{Cyclic Groups}\\
\hhline{=|=}
    n_2< m_2\coma m_2\leq m_1\coma n_2\leq n_1 & \varprod_i S_{\lambda_i(\mathbf{m})}^{12}\times \varprod_j S_{\lambda_j(\mathbf{n})}^{34}\\
    n_2= m_2\coma n_1<m_1\coma m_2\leq m_1\coma n_2\leq n_1 & \varprod_i S_{\lambda_i(\mathbf{m})}^{12}\times \varprod_j S_{\lambda_j(\mathbf{n})}^{34}\\
    m_2= n_2\coma m_1=n_1\coma n_2<n_1 & S_2^{13}\times S_2^{24}\times S_1^{3} \times S_1^{4}\\
    m_2= n_2\coma m_1=n_1\coma n_2=n_1& S_2^{13}\times S_2^{24}\times S_2^{34}
\end{array}\sfstop{\coma}{43pt}
\label{eq:U2restrfactor}
\end{equation}
where, as in Appendix \ref{sec:prefactors}, we have introduced $\lambda(\mathbf{m})$ as the partition that encodes how many fluxes $m_i$ are equal, and $\lambda_i(\mathbf{m})$ are the components of such partition.  The generalization to $\U(N)$ groups is straightforward by listing all possible pairwise identifications among the fluxes, while keeping the ordering of the fluxes for the single $\U(N)$ groups. The prefactor can be read directly from the restrictions on the fluxes by generating the group obtained by the cyclic permutation of the identified fluxes and the Weyl group of a single $\U(N)$ group.

The discussion is not much different for the $\USp(2N)$ and $\SO(2N)$ groups, because the restriction on the fluxes is the same as for the $\U(N)$ case, and the prefactors are again those obtained by generating the group obtained by the cyclic permutation of the identified fluxes, and the symmetric groups associated to the residual Weyl symmetry of a single $\USp(2N)$ or $\SO(2N)$ group. As in Appendix \ref{sec:prefactors}, we can introduce a matrix $\mathbf{T}^2$ to correctly count the Casimirs for the $\USp(2N)$ or $\SO(2N)$ groups, with entries equal to $t^4$ or $t^{2\lambda_0(\mathbf{m})}$ whenever necessary. For these reasons, it is convenient to divide the summation, distinguishing between fluxes being zero or not. Consider, for instance, the $\ZZ_2$ wreathing of two $\USp(4)$ groups with fluxes $\mathbf{m}$ and $\mathbf{n}$. The restriction is 
\begin{equation}\label{eq:USp4divisioninchambers}
\renewcommand*{\arraystretch}{1.2}
    \begin{array}{c}
    \text{Restriction}\\ 
\hhline{=}
n_2< m_2\coma m_2\leq m_1 \coma n_2\leq n_1\\ 
n_2= m_2\coma n_1<m_1 \coma m_2\leq m_1\coma n_2\leq n_1\\
m_2= n_2\coma m_1=n_1 \coma 0\neq n_2<n_1\\
m_2= n_2\coma m_1=n_1 \coma 0 = n_2<n_1\\
m_2= n_2\coma m_1=n_1 \coma 0\neq n_2=n_1\\
m_2= n_2\coma m_1=n_1 \coma 0 = n_2=n_1\\
    \end{array} \sfstop{\fstop}{60pt}
\end{equation}

In this way, even if not strictly necessary, one can easily read the group generating the prefactor from each restricted choice of fluxes.

\section{Utilizing the \texorpdfstring{\boldmath{$A_3 \cong D_3$}}{A3 = D3} Isomorphism}
\label{sec:A3D3}

Now that we have explained how to determine the prefactor for the $\ZZ_2$ wreathing of orthosymplectic magnetic quivers, we want to test it in 3d $\mathcal{N}=4$ theories for which we have unitary and orthosymplectic quiver descriptions. There are a variety of known constructions that lead to quivers with unitary and orthosymplectic descriptions (see, for example, \cite{Hanany:2023tvn,Bennett:2024llh,Akhond:2020vhc}). In this paper, we engineer such pairs by considering the magnetic quivers for the Higgs branches of 6d $(1,0)$ conformal matter theories of type $(\mathfrak{su}(4), \mathfrak{su}(4))$ and $(\mathfrak{so}(6), \mathfrak{so}(6))$. In particular, the magnetic quivers take the forms in equations \eqref{eqn:AtypeHB} and \eqref{eqn:DtypeHB}, and the Lie algebra isomorphism $\su(4) \cong \so(6)$ implies an isomorphism between the nilpotent orbits, which is summarized in Table \ref{tbl:A3vsD3}. Moreover, we note that we are computing the Hilbert series of the Coulomb branch using the dimension formula for the monopole operators introduced in \cite{Cremonesi:2013lqa}. Thus, we must restrict ourselves to quivers where each node is individually good in the sense of \cite{Gaiotto:2008ak}; this means that we avoid $T_O[\so(6)]$ theories where any symplectic node is bad. We will consider the same restriction when we move to higher-rank orthosymplectic quivers in Section \ref{sec:moreexamples}.

\begin{table}[t]
    \centering
    \renewcommand{\arraystretch}{1.2}
        \begin{tabular}{c|c|c|c}
            $O_{A_3}$ & $T_{O_{A_3}}[\su(4)]$ & $O_{D_3}$ & $T_{O_{D_3}}[\so(6)]$ \\\hhline{=|=|=|=}
            $[1^4]$ & $\begin{gathered}
    \begin{tikzpicture}[baseline=0,font=\footnotesize]
        \node[node, label=below:{$1$}] (A3) {};
      \node[node, label=below:{$2$}] (A4) [right=8mm of A3] {};
        \node[node, label=below:{$3$}] (A6) [right=8mm of A4] {};
        \node[flavor, label=below:{$4$}] (A7) [right=8mm of A6] {};
    
        \draw (A3.east) -- (A4.west);
      \draw (A4.east) -- (A6.west);
      \draw (A6.east) -- (A7.west);
    \end{tikzpicture}\end{gathered}$ 
 & $[1^6]$ & $\begin{gathered}
      \begin{tikzpicture}[baseline=0,font=\footnotesize]
        \node[node, label=below:{$2$},fill=red] (A3) {};
      \node[node, label=below:{$2$},fill=blue] (A4) [right=8mm of A3] {};
        \node[node, label=below:{$4$},fill=red] (A5) [right=8mm of A4] {};
        \node[node, label=below:{$4$},fill=blue] (A6) [right=8mm of A5] {};
        \node[flavor, label=below:{$6$},fill=red] (A7) [right=8mm of A6] {};
    
        \draw (A3.east) -- (A4.west);
      \draw (A4.east) -- (A5.west);
      \draw (A5.east) -- (A6.west);
      \draw (A6.east) -- (A7.west);
    \end{tikzpicture} 
  \end{gathered}$ 
  \\\hline
            $[2, 1^2]$ & $\begin{gathered}
    \begin{tikzpicture}[baseline=0,font=\footnotesize]
        \node[node, label=below:{$1$}] (A3) {};
      \node[node, label=below:{$2$}] (A4) [right=8mm of A3] {};
        \node[flavor, label=below:{$4$}] (A7) [right=8mm of A4] {};
    
        \draw (A3.east) -- (A4.west);
      \draw (A4.east) -- (A7.west);
    \end{tikzpicture} 
  \end{gathered}$ 
  & $[2^2, 1^2]$ & $\begin{gathered}
      \begin{tikzpicture}[baseline=0,font=\footnotesize]
        \node[node, label=below:{$2$},fill=red] (A5) {};
        \node[node, label=below:{$4$},fill=blue] (A6) [right=8mm of A5] {};
        \node[flavor, label=below:{$6$},fill=red] (A7) [right=8mm of A6] {};
    
      \draw (A5.east) -- (A6.west);
      \draw (A6.east) -- (A7.west);
    \end{tikzpicture} 
  \end{gathered}$ 
 \\\hline
            $[2^2]$ & $\begin{gathered}
    \begin{tikzpicture}[baseline=0,font=\footnotesize]
        \node[node, label=below:{$2$}] (A3) {};
        \node[flavor, label=below:{$4$}] (A7) [right=8mm of A3] {};
    
        \draw (A3.east)  -- (A7.west);
    \end{tikzpicture} 
  \end{gathered}$ 
  &  $[3, 1^3]$ & $\begin{gathered}
      \begin{tikzpicture}[baseline=0,font=\footnotesize]
        \node[node, label=below:{$2$},fill=red] (A5) {};
        \node[node, label=below:{$2$},fill=blue] (A6) [right=8mm of A5] {};
        \node[flavor, label=below:{$6$},fill=red] (A7) [right=8mm of A6] {};
    
      \draw (A5.east) -- (A6.west);
      \draw (A6.east) -- (A7.west);
    \end{tikzpicture} 
  \end{gathered}$ 
  \\\hline
            $[3, 1]$ & $\begin{gathered}
    \begin{tikzpicture}[baseline=0,font=\footnotesize]
        \node[node, label=below:{$1$}] (A3) {};
        \node[flavor, label=below:{$4$}] (A7) [right=8mm of A3] {};
    
        \draw (A3.east)  -- (A7.west);
    \end{tikzpicture} 
  \end{gathered}$ 
  & $[3^2]$ & $\begin{gathered}
      \begin{tikzpicture}[baseline=0,font=\footnotesize]
        \node[node, label=below:{$2$},fill=blue] (A6) {};
        \node[flavor, label=below:{$6$},fill=red] (A7) [right=8mm of A6] {};
    
      \draw (A6.east) -- (A7.west);
    \end{tikzpicture} 
  \end{gathered}$ 
  \\\hline
            $[4]$ & $\begin{gathered}
    \begin{tikzpicture}[baseline=0,font=\footnotesize]
        \node[flavor, label=below:{$4$}] (A7)  {};
    \end{tikzpicture} 
  \end{gathered}$ 
 & $[5, 1]$ & $\begin{gathered}
      \begin{tikzpicture}[baseline=0,font=\footnotesize]
        \node[flavor, label=below:{$6$},fill=red] (A7) {};
    \end{tikzpicture} 
 \end{gathered}$ 
        \end{tabular}
    \caption{How the nilpotent orbits of $\mathfrak{su}(4)$ and $\so(6)$ are related across the Lie algebra isomorphism. We have also written the unitary and orthosymplectic Lagrangian quivers associated with the $T_O[\mathfrak{g}]$ theories for each nilpotent orbit.}
    \label{tbl:A3vsD3}
\end{table}

In this section, hence, we consider the magnetic quivers for the Higgs branches and the $\ZZ_2$ wreathing of the Higgsed conformal matter theories
\begin{equation}
\begin{split}
    A_3^{\mathfrak{su}(4)}([2^2],[2^2])&\simeq A_3^{\mathfrak{so}(6)}([3,1^3],[3,1^3]) \coma \\
    A_1^{\mathfrak{su}(4)}([1^4],[1^4])&\simeq A_1^{\mathfrak{so}(6)}([1^6],[1^6])  \coma \\
    A_5^{\mathfrak{su}(4)}([3,1],[3,1])&\simeq A_5^{\mathfrak{so}(6)}([3^2],[3^2])\fstop
\end{split}
\end{equation}
These tests are important for understanding the restrictions on the fluxes and the prefactors in the case of orthosymplectic quivers and, following these cross-checks, we discuss the $\ZZ_2$ wreathing of orthosymplectic quivers that do not admit a unitary description in Section \ref{sec:moreexamples}.

\subsubsection*{Theory \hyperlink{so6-th1}{1}: \texorpdfstring{$A_{3}^{\su(4)}([2^2], [2^2])$}{A3SU4} and \texorpdfstring{$A_{3}^{\so(6)}([3,1^3], [3,1^3])$}{A3SO6}}
\label{sec:A3su42222}

From the discussion in Section \ref{sec:flavor} and the flavor symmetries reported in Table \ref{tab:flavorsymm6d}, we know that $A_{3}^{\su(4)}([2^2], [2^2])$ has an $\su(4)$ flavor symmetry. This can be confirmed by considering the magnetic quiver for the Higgs branch and computing the refined Hilbert series. The magnetic quiver is
\begin{equation}
    \begin{tikzpicture}[baseline=0,font=\footnotesize]
      \node[node, label=below:{$2$}] (A4) {};
      \node[node, label=below:{$4$}] (N3) [right=6mm of A4] {};
      \node[node, label=right:{$4$}] (Nu) [above=4mm of N3] {};
        \node[node, label=below:{$2$}] (A5) [right=6mm of N3] {};
    
      \draw (A4.east) -- (N3.west);
      \draw (N3.east) -- (A5.west);    
      \draw (N3.north) -- (Nu.south);
      \draw (Nu) to[out=130, in=410, looseness=12] (Nu);
      \node (U1) [right=6mm of A5] {$/\U(1)$};
    \end{tikzpicture} \,,
    \label{eq:A3SU4}
\end{equation}
where, by $/\U(1)$, we mean that there is an overall $\U(1)$ under which none of the matter spectrum is charged, and which must be decoupled. Let us call $z_1$, $z_2$ the topological fugacities for the two $\U(2)$ nodes respectively, while $z_3$ is the topological fugacity of the central $\U(4)$ node, and $z_4$ the one for the $\U(4)$ node with the adjoint field. These fugacities must satisfy
\begin{equation}
    z_1 z_2 z_3^2z_4^2=1\coma
    \label{eq:fugcondA3-decoupling}
\end{equation}
due to the decoupling of the overall $\U(1)$. We can, then, write the Hilbert series for this example as:
\begin{equation}
    \begin{split}
       \HS(t,z_i) = &\, \left(1-t^2\right) \sum_{a_1\geq a_2 \geq -\infty} \sum_{b_1\geq b_2 \geq b_3\geq b_4\geq -\infty} \sum_{c_1\geq c_2 \geq -\infty} \sum_{d_1\geq d_2 \geq d_3 \geq d_4 -\infty} P_{\U(2)}(t,\mathbf{a})\\& 
       P_{\U(4)}(t,\mathbf{b})P_{\U(2)}(t,\mathbf{c})P_{\U(4)}(t,\mathbf{d})t^{2\Delta}z_1^{a_1+a_2}z_3^{b_1+b_2+b_3+b_4}z_2^{c_1+c_2}z_4^{d_1+d_2+d_3+d_4}\coma
    \end{split}
\end{equation}
where the conformal dimension of the monopole operators are given by
\begin{equation}
    \begin{split}
        \Delta = & -| a_1-a_2|-\sum_{i=1}^4|b_i-b_{i+1}|-| c_1-c_2|+\frac{1}{2}\sum_{i=1}^2\sum_{j=1}^4\left(|a_i-b_j|+|b_j-c_i|\right)+\\
                 & +\frac{1}{2}\sum_{i=1}^4\sum_{j=1}^4|b_i-d_j|\coma
    \end{split}
\end{equation}
and while the prefactors are computed as explained in Appendix \ref{sec:UNPrefactor}. At the first relevant order, the Hilbert series reads\footnote{The Plethystic Exponential (PE) of a multivariate function $f(t_1,\ldots t_k)$ (vanishing at the origin) is defined as \[\PE\left[f(t_1,\ldots t_k)\right] = \exp\left(\sum_{n=1}^\infty \frac{1}{n}f(t_1^n,\ldots, t_k^n)\right)\fstop\] }

\begin{equation}
\scalebox{0.97}{$\displaystyle
\begin{aligned}
    \HS(t,z_i) = \PE&\left[t^2 \left(\frac{z_1^{1/2}}{z_2^{1/2} z_4}+z_1^{1/2} z_2^{1/2} z_4+\frac{z_1^{1/2} z_2^{1/2}}{z_4}+\frac{z_2^{1/2}}{z_1^{1/2} z_4}+\frac{z_2^{1/2} z_4}{z_1^{1/2}}+\right.\right.\\
    & \left.\left.+\frac{1}{z_1^{1/2} z_2^{1/2} z_4}+\frac{z_1^{1/2} z_4}{z_2^{1/2}}+\frac{z_4}{z_1^{1/2} z_2^{1/2}}+z_1+\frac{1}{z_1}+z_2+\frac{1}{z_2}+3\right)+\mathcal{O}(t^4)\right]\coma
\end{aligned}$
}
\end{equation}
where we have defined
\begin{equation}
    z_3 = \frac{1}{z_1^{1/2}z_2^{1/2}z_4} \,,
\end{equation} 
in order to satisfy equation \eqref{eq:fugcondA3-decoupling}.

We now rewrite the three remaining fugacities in terms of three new fugacities, $x$, $y$, and $z$, defined as follows:
\begin{equation}\label{eqn:dora}
    z_1 = x^2\coma z_2 = y^{-2} \coma z_4 =z^2\fstop
\end{equation}
Then, we can rewrite the Hilbert series as
\begin{equation}
  \begin{aligned}
    &\HS(t,x,y,z) = \\&\qquad \PE\left[t^2 \left(1+\chi_{[2]}^{\su(2)}(x)+\chi_{[2]}^{\su(2)}(y)+\left(z^2+\frac{1}{z^2}\right) \chi_{[1]}^{\su(2)}(x)\chi_{[1]}^{\su(2)}(y)\right)+\mathcal{O}(t^4)\right]\coma
  \end{aligned}
  \label{eq:A4SU4-HS}
\end{equation}
where $\chi^{\mathfrak{g}}_{[\cdots]}(\cdot)$ denotes the character of the irreducible representation of $\su(2)$ with highest weight $[\cdots]$ written in terms of the fugacities given in the parentheses. In fact, the coefficient of $t^2$ is nothing other than the character of the adjoint representation of $\su(4)$, using the branching
\begin{equation}
  \begin{aligned}
    \su(4) &\rightarrow \su(2)_x \oplus \su(2)_y \oplus \mathfrak{u}(1)_z \,, \\
    \mathbf{15} &\rightarrow (\mathbf{2}, \mathbf{2})_2 \oplus (\mathbf{3}, \mathbf{1})_0 \oplus (\mathbf{1}, \mathbf{3})_0 \oplus (\mathbf{1}, \mathbf{1})_0 \oplus (\mathbf{2}, \mathbf{2})_{-2} \,,
  \end{aligned}
\end{equation}
where we have written the associated fugacity as a subscript on the algebras. As expected, this signals the enhanced global symmetry.\footnote{For this example, we could have defined $z_2=y^2$, and the result would have been the same. This is because $y$ is the fugacity associated to an $\su(2)$ flavor symmetry, and there is no difference between the fundamental and anti-fundamental representations for this algebra. However, as we will see in the next unitary example, and we comment below, the choice of fugacity we use here is the one that generalizes when we consider wreathing of theories with higher-rank flavor symmetries.}

We can also confirm the prediction for the $\ZZ_2$-discretely-gauged flavor symmetry in Table \ref{tab:flavorsymm6d}, by considering a $\ZZ_2$ wreathing of equation \eqref{eq:A3SU4}, by following the procedure described in \cite{Arias-Tamargo:2022nlf}. In particular, the range of summations over the magnetic fluxes needs to be restricted and the prefactors are modified as explained in the previous section, according to which a cyclic group is generated. We make the identification $z_1=z_2$. Explicitly, the range of summations is restricted as follows
\begin{equation}
\renewcommand*{\arraystretch}{1.2}
    \begin{array}{c}
    \text{Restriction} \\
    \hhline{=}
    a_2<b_2\\
    a_2=b_2 \coma a_1<b_1\\
    a_2 = b_2 \coma a_1 = b_1 \coma a_2<a_1\\
    a_2 = b_2 \coma a_1 = b_1 \coma a_2 = a_1
    \end{array}
 \sfstop{\coma}{42pt}
\end{equation}
with the cyclic groups generated as in equation \eqref{eq:U2restrfactor}. The prefactors for the two $\U(4)$ groups are the same as in the unwreathed case. As discussed in Section \ref{sec:wreathedprefactors}, the division of the summation is done to be able to easily read off the prefactors, since it makes more explicit the group generated by the choice of fluxes, when we use the expression of the prefactor as in \cite{Bourget:2020bxh}.  

Once again, the decoupling of the overall $\U(1)$ allows us to define
\begin{equation}
    z_3 = \frac{1}{z_1 z_4}\coma
\end{equation}
and we obtain the Hilbert series for the $\ZZ_2$-wreathed quiver:
\begin{equation}
    \HS_{\wr\,\ZZ_2}(t,x,z) = \PE\left[t^2 \left(1+\chi_{[2]}^{\su(2)}(x)+\left(z^2+\frac{1}{z^2}\right)\chi_{[2]}^{\su(2)}(x)\right)+\mathcal{O}(t^4)\right] \,.
\end{equation}
Here we have replaced the fugacities $z_1$ and $z_4$ as in equation \eqref{eqn:dora}, and collected the $t^2$ coefficient into characters. We can see that the $t^2$ coefficient is nothing other than the character of the adjoint representation of $\usp(4)$ under the decomposition
\begin{equation}
    \begin{aligned}
      \usp(4) &\rightarrow \su(2) \oplus \mathfrak{u}(1) \,, \\
      \bm{10} &\rightarrow \bm{3}_2 \oplus \bm{3}_{-2} \oplus \bm{3}_0 \oplus \bm{1}_0 \,.
    \end{aligned}
\end{equation}
Thus, the Coulomb branch Hilbert series indicates that the Coulomb symmetry after wreathing is $\mathfrak{usp}(4)$, which is exactly what we would expect for the flavor symmetry of the 6d SCFT after discrete gauging, as discussed in Section \ref{sec:discretegaugingCM}.

One perspective on this $\ZZ_2$ wreathing is that we start with the $\su(2)_x \oplus \su(2)_y \oplus \mathfrak{u}(1)_z$ in the unwreathed case, and the wreathing identifies the fundamental representation of $\su(2)_x$ with the anti-fundamental representation of $\su(2)_y$. In this case, the distinction between fundamental and anti-fundamental is immaterial, but when considering higher-rank unitary quivers, the identification between the fundamental and anti-fundamental has consequences in the matter content of the theory after wreathing or discrete gauging.

We can now repeat this analysis for the orthosymplectic quiver realization of the same Coulomb branch, and extract the same information about the Coulomb symmetries before and after wreathing. Consider the orthosymplectic realization of the magnetic quiver of the Higgs branch of $A_{3}^{\so(6)}([3,1^3], [3,1^3])$:
\begin{equation}\label{eq:A3so6quiver}
    \begin{tikzpicture}[baseline=0,font=\footnotesize]
        \node[node, label=below:{$2$},fill=red] (A3) {};
      \node[node, label=below:{$2$},fill=blue] (A4) [right=6mm of A3] {};
      \node[node, label=below:{$6$},fill=red] (N3) [right=6mm of A4] {};
      \node[node, label=right:{$8$},fill=blue] (Nu) [above=4mm of N3] {};
        \node[node, label=below:{$2$},fill=blue] (A5) [right=6mm of N3] {};
        \node[node, label=below:{$2$},fill=red] (A6) [right=6mm of A5] {};
    
        \draw (A3.east) -- (A4.west);
      \draw (A4.east) -- (N3.west);
      \draw (N3.east) -- (A5.west);    
      \draw (N3.north) -- (Nu.south);
      \draw (Nu) to[out=130, in=410, looseness=12] (Nu);
      \draw (A5.east) -- (A6.west);
      \node (Z2) [right=6mm of A6] {$(/\ZZ_2)$};
    \end{tikzpicture} \,.
\end{equation}

This quiver has an overall $\ZZ_2$ that acts trivially on the matter fields; this can be considered as the center of the $\SO(6)$ group that is not screened by the bifundamental fields. This gives rise to a $\ZZ_2^{[1]}$ 1-form symmetry. This symmetry can be gauged or not, as discussed in Section \ref{sec:higherformsym}. The gauging of such a symmetry gives rise to a $\ZZ_2^{[0]}$ 0-form symmetry in the 3d quiver, and vice versa. We call $\omega$ the fugacity associated to the $\ZZ_2^{[0]}$ 0-form symmetry, obtained by gauging the $\ZZ_1^{[1]}$ 1-form symmetry in 3d. As explained in Section \ref{sec:higherformsym}, we will generally keep the fugacity $\omega$ explicit in the Hilbert series that we write.

The Hilbert series of the unwreathed quiver again takes the form of a sum over each of the gauge nodes:
\begin{equation}
    \begin{split}
       \HS(t,\omega) = &\, \sum_{\epsilon\in\lbrace 0,1\rbrace} \sum_{a\in\ZZ+\frac{1}{2}\epsilon} \sum_{b\geq\frac{1}{2}\epsilon} \sum_{c_1\geq c_2 \geq |c_3|,\, c_3\in\ZZ+\frac{1}{2}\epsilon}  \sum_{d\geq\frac{1}{2}\epsilon} \sum_{e\in\ZZ+\frac{1}{2}\epsilon} \sum_{f_1\geq f_2 \geq f_3\geq f_4\geq \frac{1}{2}\epsilon} P_{\SO(2)}(t,a)\\
       &P_{\USp(2)}(t,b)P_{\SO(6)}(t,\mathbf{c})P_{\USp(2)}(t,d)P_{\SO(2)}(t,e)P_{\USp(8)}(t,\mathbf{f})t^{2\Delta}\omega^\epsilon\coma
    \end{split}
\end{equation}
where
    \begin{align}
        \Delta = & -2| b|-\sum_{i=1}^3\sum_{j=i+1}^3\left(|c_i+c_j|+|c_i-c_j|\right)-2|d|-2\sum_{i=1}^4|f_i|+|a+b|+|a-b|\nonumber\\
        &+\sum_{i=1}^3\left(|b+c_i|+|b-c_i|\right)+\sum_{i=1}^3\left(|d+c_i|+|d-c_i|\right)+|e+d|+|e-d|\\
        &+\sum_{i=1}^3\sum_{j=1}^4\left(|c_i-f_j|+|c_i+f_j|\right) \,.\nonumber
    \end{align}
The prefactors are computed as explained in Section \ref{sec:wreathedprefactors}. In this case, we are computing the unrefined Hilbert series, where we have not included the fugacities for the Coulomb symmetries. Up to the first relevant order, the Hilbert series reads
\begin{equation}
    \HS(t,\omega)  = \PE\left[t^2(7+8\omega)+t^4(13+16\omega)+\mathcal{O}(t^6)\right]\fstop
    \label{eq:A3SO6-HS}
\end{equation}
We can see that the Hilbert series coincide with the unrefined Hilbert series in equation \eqref{eq:A4SU4-HS} provided the identification of the fugacity $\omega = z^2 = z^{-2}$. In fact, one can see the $\ZZ_2^{[0]}$ 0-form symmetry in the unitary quiver as a $\ZZ_2$ subgroup of the $\mathfrak{u}(1)$ symmetry. 

For completeness, we also consider the theory obtained from the gauging of the $\ZZ_2^{[0]}$ 0-form symmetry by summing over the elements of $\ZZ_2$:\footnote{A one-dimensional irreducible representation of the group $\ZZ_n$ is given by the map $\rho:\ZZ_n \rightarrow \CC^\times$ with $\rho(m) = e^{\frac{2\pi i m}{n}}$. The fugacity $\omega$ signal states charged under the $\ZZ_2$ group, and gauging a finite group is equivalent to sum over all elements of the group (divided by the dimension of the group), that, in our case, are $\left\{1, e^{\pi i}\right\}$.}
\begin{equation}
    \frac{1}{2}\sum_{\omega \in\{1,-1\}} \HS(t,\omega) = \PE\left[7t^2+49t^4+\mathcal{O}(t^6)\right]\coma
\end{equation}
which corresponds to the theory with a global $\ZZ_2^{[1]}$ 1-form symmetry. This theory is not associated to the 3d mirror of $A_3^{\so(6)}([3,1^3],[3,1^3])$, but it still exists as an 3d $\mathcal{N}=4$ quiver.

We can now consider the $\ZZ_2$ wreathing of the orthosymplectic quiver in equation \eqref{eq:A3so6quiver}. The wreathing proceeds with the identification of the various contributions to the prefactor as explained above. We can call $a$ and $e$ the magnetic fluxes associated to the two $\so(2)$s while $b$ and $d$ are the magnetic fluxes associated to the $\usp(2)$ algebras involved in the wreathing. To be able to compute the prefactors from the value of the fluxes, we restrict the computation of the Hilbert series as follows: 
\begin{equation}
\renewcommand*{\arraystretch}{1.2}
    \begin{array}{c}
    \text{Restriction}\\ 
   \hhline{=}
a<e \\
a=e\coma b<d \\
a=e\coma b=d  \\
    \end{array}\sfstop{\fstop}{31pt}
\end{equation}
Once again, the division in the various contributions is only done to be able to read the prefactors explicitly from the choice of fluxes. The wreathing of such a quiver gives
\begin{equation}
    \HS_{\wr\,\ZZ_2}(t)  = \PE\left[t^2(4+6\omega)+t^4(19+18\omega)+\mathcal{O}(t^6)\right]\coma
\end{equation}
which is compatible with the results obtained from the unitary version of the quiver, signaling the identification of the $\su(2)$ flavor symmetries inside the $\su(4)$ unwreathed flavor, leading to a $\usp(4)$ enhanced global symmetry.

\subsubsection*{Theory \hyperlink{so6-th2}{2}: \texorpdfstring{$A_{1}^{\su(4)}([1^4], [1^4])$}{A1SU4} and \texorpdfstring{$A_{1}^{\so(6)}([1^6], [1^6])$}{A1SO6}}
\label{sec:A1su41414}

Another theory that admits both a unitary and orthosymplectic realization is the magnetic quiver for the Higgs branch of $A_{1}^{\su(4)}([1^4], [1^4])$, associated to Theory 2 in Table \ref{tab:flavorsymm6d}. This time, the 6d theory predicts a flavor symmetry of $\su(8)$, which can be confirmed by computing the Hilbert series of the following magnetic quiver:
\begin{equation}
    \begin{tikzpicture}[baseline=0,font=\footnotesize]
        \node[node, label=below:{$1$}] (A1) {};
      \node[node, label=below:{$2$}] (A2) [right=6mm of A1] {};
      \node[node, label=below:{$3$}] (A3) [right=6mm of A2] {};
      \node[node, label=below:{$4$}] (N4) [right=6mm of A3] {};
      \node[node, label=right:{$2$}] (Nu) [above=4mm of N4] {};
      \node[node, label=below:{$3$}] (A4) [right=6mm of N4] {};
      \node[node, label=below:{$2$}] (A5) [right=6mm of A4] {};
      \node[node, label=below:{$1$}] (A6) [right=6mm of A5]{};
    
        \draw (A1.east) -- (A2.west);
        \draw (A2.east) -- (A3.west);
        \draw (A3.east) -- (N4.west);
        \draw (N4.east) -- (A4.west);
        \draw (A4.east) -- (A5.west);
        \draw (A5.east) -- (A6.west);  
        \draw (N4.north) -- (Nu.south);
      \draw (Nu) to[out=130, in=410, looseness=12] (Nu);
      \node (U1) [right=6mm of A6] {$/\U(1)$};
    \end{tikzpicture} \,.
\end{equation}
The Hilbert series, with the appropriate fugacities obtained similarly as in the previous section, reads
\begin{equation}
    \begin{split}
        \HS(t,\mathbf{x},\mathbf{y},z) = &\,\PE\left[t^2\left(1+\chi^{\su(4)}_{[1,0,1]}(\mathbf{x})+\chi^{\su(4)}_{[1,0,1]}(\mathbf{y})+\right.\right.\\
        &\,\,\,\,\,\,\,\,\,\left.\left.+z^{-2}\chi^{\su(4)}_{[1,0,0]}(\mathbf{x})\chi^{\su(4)}_{[0,0,1]}(\mathbf{y})+z^{2}\chi^{\su(4)}_{[0,0,1]}(\mathbf{x})\chi^{\su(4)}_{[1,0,0]}(\mathbf{y}) \right)+\mathcal{O}(t^4)\right]\coma
    \end{split}
\end{equation}
compatible with the branching rule of $\su(8)$ into $\su(4)_\mathbf{x}\oplus\su(4)_\mathbf{y}\oplus\mathfrak{u}(1)_z$ as
\begin{equation}
    \mathbf{63} = (\mathbf{1},\mathbf{1})_0 +(\mathbf{15},\mathbf{1})_0 +(\mathbf{1},\mathbf{15})_0 +(\mathbf{4},\overline{\mathbf{4}})_{-2}+(\overline{\mathbf{4}},\mathbf{4})_2 \,.
\end{equation}
Therefore, we see the expected $\su(8)$ global symmetry predicted by the F-theory construction.

We can perform a $\ZZ_2$ wreathing of the two horizontal tails, computing the prefactors as explained above. The resulting Hilbert series is
\begin{equation}
    \begin{split}
        \HS_{\wr\,\ZZ_2}(t,\mathbf{x},z) = &\PE\left[t^2\left(1+\chi^{\su(4)}_{[1,0,1]}(\mathbf{x})+z^{-2}\chi^{\su(4)}_{[2,0,0]}(\mathbf{x})+z^{2}\chi^{\su(4)}_{[0,0,2]}(\mathbf{x}) \right)+\mathcal{O}(t^4)\right]\coma
    \end{split}
\end{equation}
which is compatible with identifying the fundamental representation of $\su(4)_\mathbf{x}$ with the anti-fundamental representation of $\su(4)_\mathbf{y}$. This is precisely the expected identification from the discrete gauging perspective on the 6d $(1,0)$ SCFT. This result shows that the wreathing projects $\su(8)$ down to $\usp(8)$, decomposed as $\su(4)_\mathbf{x}\oplus \mathfrak{u}(1)_z$, corresponding to the branching 
\begin{equation}
    \mathbf{36} = \mathbf{1}_0+\mathbf{15}_0 +\mathbf{10}_{-2}+\mathbf{10}_2 \coma
\end{equation}
confirming the expectation in Table \ref{tab:flavorsymm6d}.

The same theory, in its guise as $A_1^{\so(6)}([1^6],[1^6])$ conformal matter, admits also an orthosymplectic quiver, which is
\begin{equation}
    \begin{tikzpicture}[baseline=0,font=\footnotesize]
        \node[node, label=below:{$2$},fill=red] (A1) {};
      \node[node, label=below:{$2$},fill=blue] (A2) [right=6mm of A1] {};
      \node[node, label=below:{$4$},fill=red] (A3) [right=6mm of A2] {};
      \node[node, label=below:{$4$},fill=blue] (A4) [right=6mm of A3] {};
      \node[node, label=below:{$6$},fill=red] (N3) [right=6mm of A4] {};
      \node[node, label=right:{$4$},fill=blue] (Nu) [above=4mm of N3] {};
        \node[node, label=below:{$4$},fill=blue] (A5) [right=6mm of N3] {};
      \node[node, label=below:{$4$},fill=red] (A6) [right=6mm of A5] {};
      \node[node, label=below:{$2$},fill=blue] (A7) [right=6mm of A6] {};
      \node[node, label=below:{$2$},fill=red] (A8) [right=6mm of A7] {};
    
        \draw (A1.east) -- (A2.west);
        \draw (A2.east) -- (A3.west);
        \draw (A3.east) -- (A4.west);
      \draw (A4.east) -- (N3.west);
      \draw (N3.east) -- (A5.west);   
      \draw (A5.east) -- (A6.west);
      \draw (A6.east) -- (A7.west);
      \draw (A7.east) -- (A8.west);
      \draw (N3.north) -- (Nu.south);
      \draw (Nu) to[out=130, in=410, looseness=12] (Nu);
      \node (Z2) [right=6mm of A8] {$(/\ZZ_2)$};
    \end{tikzpicture} \,.
\end{equation}
The computation of the Hilbert series leads to
\begin{equation}
    \HS(t,\omega) = \PE\left[t^2(31+32\omega)+\mathcal{O}(t^4)\right]\coma
\end{equation}
which is compatible with the unrefined Hilbert series of the unitary quiver by choosing to refine only for a $\ZZ_2^{\omega}\subset \U(1)_z$. The $\ZZ_2$ wreathing proceeds by dividing the computations in contributions as described in the section above, computing for each restricted choice of fluxes the corresponding prefactor. The division in chambers for the $\so(2)$ and $\usp(2)$ factors is similar to the previous section, while the chambers for $\usp(4)$ are given in equation \eqref{eq:USp4divisioninchambers}, from which one can also extract the chambers for the $\so(4)$ nodes (keeping in mind the modification of the $\mathbf{T}^2$ matrix as explained in Appendix \ref{sec:SO2NPrefactor}). The result is 
\begin{equation}
    \HS_{\wr\,\ZZ_2}(t,\omega) = \PE\left[t^2(16+20\omega)+\mathcal{O}(t^4)\right]\coma
\end{equation}
which is precisely the expected result: there are $36$ operators generating the Coulomb symmetry after wreathing, compatible with the anticipated $\mathfrak{usp}(8)$ flavor symmetry.

\subsubsection*{Theory \hyperlink{so6-th3}{3}: \texorpdfstring{$A_5^{\su(4)}([3,1],[3,1])$}{A4SU4} and \texorpdfstring{$A_{5}^{\so(6)}([3^2], [3^2])$}{A5SO6}}
\label{sec:A4su43131}

We can briefly discuss Theory \hyperlink{so6-th3}{3}, since it is not significantly different from the two previous theories. In Section \ref{sec:flavor}, we reviewed the way to read off the flavor symmetry at generic point in the tensor branch for a 6d SCFT. For this specific configuration, as listed in Table \ref{tab:flavorsymm6d}, the theory has an $\su(2)\oplus \mathfrak{u}(1)\oplus\mathfrak{u}(1)$ flavor symmetry before the wreathing. This can also be confirmed by computing the Hilbert series of the unitary realization of such a theory corresponding to $A_5^{\su(4)}([3,1],[3,1])$ conformal matter whose magnetic quiver is 
    \begin{equation}
        \begin{tikzpicture}[baseline=0,font=\footnotesize]
      \node[node, label=below:{$1$}] (A4) {};
      \node[node, label=below:{$4$}] (N3) [right=6mm of A4] {};
      \node[node, label=right:{$6$}] (Nu) [above=4mm of N3] {};
        \node[node, label=below:{$1$}] (A5) [right=6mm of N3] {};
    
      \draw (A4.east) -- (N3.west);
      \draw (N3.east) -- (A5.west);    
      \draw (N3.north) -- (Nu.south);
      \draw (Nu) to[out=130, in=410, looseness=12] (Nu);
      \node (U1) [right=6mm of A5] {$/\U(1)$};
    \end{tikzpicture} \,.
    \label{eq:A4su43131}
    \end{equation}
    The Hilbert series at the first leading orders is 
    \begin{equation}
        \text{HS}(t) = \PE \left[5t^2+16t^4+\mathcal{O}(t^5)\right]\coma
    \end{equation}
    confirming the flavor symmetry. By performing the wreathing of such a quiver, the two $\mathfrak{u}(1)$ are identified, while the $\su(2)\simeq \usp(2)$ remains untouched. At the level of the Hilbert series computation for the wreathed quiver above, we can precisely see this effect obtaining
    \begin{equation}
        \text{HS}_{\wr\,\ZZ_2}(t) = \PE \left[4t^2+13t^4+\mathcal{O}(t^5)\right]\fstop
    \end{equation}
    The computation of the Hilbert series for Theory \hyperlink{so6-th3}{3} in its orthosymplectic formulation is used to further confirm the proposed procedure to compute wreathing of orthosymplectic quivers. We obtain 
    \begin{equation}
        \begin{split}
            \HS(t, \omega) = &\,\PE\left[t^2 (3+2 \omega)+t^4 (8+8 \omega)+\mathcal{O}(t^5)\right]\coma\\
      \HS_{\wr\,\ZZ_2}(t, \omega) =&\,\PE\left[t^2 (2+2 \omega)+t^4 (7+6 \omega)+\mathcal{O}(t^5)\right]\coma
        \end{split}
    \end{equation}
    where, after unrefining the Hilbert series, we find perfect agreement with the unitary counterpart.

\section{Hilbert Series for \texorpdfstring{\boldmath{$\ZZ_2$}}{Z2}-wreathed Quivers}
\label{sec:moreexamples}

In this section, we are going to provide further evidence of how the flavor symmetry is changed by the $\ZZ_2$ wreathing, confirming the expectations described in Section \ref{sec:discretegaugingCM}. We first discuss all the theories in Table \ref{tbl:exampleswreathing}, which are those that have good orthosymplectic quiver realizations. Theories from \hyperlink{so6-th1}{1} to \hyperlink{so6-th3}{3} in Table \ref{tbl:exampleswreathing} have been extensively discussed in \cref{sec:A3D3}. Their flavor symmetries match the expectation in Table \ref{tab:flavorsymm6d}, both from the unitary and the orthosymplectic quiver perspective and they have been used as a benchmark for our proposal of wreathing of orthosymplectic quivers. Some additional unitary theories are shown in Table \ref{tbl:workingsu3su4}. The computation of the Hilbert series of the unitary quiver confirms the expectation on the flavor symmetry of the theory, matching the last three entries of Table \ref{tab:flavorsymm6d}.

\subsubsection*{Theory \hyperlink{so8-th4}{4}: \texorpdfstring{$A_{2}^{\so(8)}([3^2,1^2], [3^2,1^2])$}{A2SO8}}

Theory \hyperlink{so8-th4}{4} is the first theory that we discuss that does not admit a unitary quiver. However, as we discussed in Section \ref{sec:discretegaugingCM}, it also represents one of the handful of theories where the flavor symmetry after the discrete gauging is unclear. The reason is that the central $(-1)$-curve is undecorated, and there exists E-string flavor attached to it. When performing the $\ZZ_2$ discrete gauging, we expect an enhanced flavor symmetry. We will use the Hilbert series to confirm such an expectation, completing the discussion we started in Section \ref{sec:discretegaugingCM}.

Let us first consider the unwreathed theory, for which we know from Section \ref{sec:flavor} and Table \ref{tab:flavorsymm6d} that it has an $\su(3)\oplus\su(3)$ flavor symmetry, obtained by the relations in equation \eqref{eqn:commutants}. We can confirm such an expectation from the computation of the Hilbert series, i.e.,
    \begin{equation}
        \HS(t, \omega) = \PE\left[t^2 (8 \omega +8)+t^4 (32 \omega +32) +\mathcal{O}(t^5)\right]\coma
    \end{equation}
    after unrefining it. 
     After wreathing,
     the Hilbert series computation gives
    \begin{equation}
        \HS_{\wr\,\ZZ_2}(t, \omega) =\PE\left[t^2 (6 \omega +5)+t^4 \left(26 \omega +29\right)+\mathcal{O}\left(t^5\right)\right]\fstop
    \end{equation}
    Unrefining the Hilbert series, we see that the dimension of the flavor symmetry algebra is $11$. From Section \ref{sec:discretegaugingCM}, this theory is one of those where one needs to understand the action of the $\ZZ_2$ GS automorphism on the $\frac{1}{2}$-BPS operator spectrum. In particular, it is not clear what happens to the operators coming from the gauge-invariant combinations of $\mu_L^+\otimes \mu_C^+\otimes \mu_R^+$ in equation \eqref{eq:momentmapoperators}. We would expect that the flavor symmetry is either $\su(2)\oplus \su(3)$ or $\usp(4)\oplus \mathfrak{u}(1)$, but we leave the precise identification of the flavor symmetry algebra, which would require the full refinement of the Hilbert series, to future work.

\subsubsection*{Theory \hyperlink{so8-th5}{5}: \texorpdfstring{$A_{4}^{\so(8)}([5,1^3], [5,1^3])$}{A4SO8}}

More straightforward is the analysis of Theory \hyperlink{so8-th5}{5}. The flavor symmetry for this theory, before the wreathing, is $\su(2)\oplus \su(2)\oplus \mathfrak{u}(1)$, which is confirmed by the computation of the Hilbert series of the magnetic quiver, i.e.,
    \begin{equation}
        \HS(t, \omega) =\, \PE\left[7 t^2+t^4 (8 \omega +5)+\mathcal{O}(t^5)\right]\fstop
    \end{equation}
    Even though we expect an E-string flavor, this time we do not have an enhancement of the symmetry, so the expectation is that the wreathing identifies the two $\su(2)$ as we have already seen in Section \ref{sec:A3su42222}, while leaving the $\mathfrak{u}(1)$ untouched via $\mathfrak{so}(2) \rightarrow \mathfrak{u}(1)$. The computation of the wreathed Hilbert series leads to
    \begin{equation}
        \HS_{\wr\,\ZZ_2}(t, \omega) =\PE\left[4 t^2+t^4 (6 \omega +11)+\mathcal{O}(t^5)\right]\coma
    \end{equation}
    which confirms the expectation of a four-dimensional Coulomb symmetry.

\subsubsection*{Theory \hyperlink{so8-th6}{6}: \texorpdfstring{$A_{4}^{\so(8)}([5,3], [5,3])$}{A4SO8}}

Theory \hyperlink{so8-th6}{6}, instead, should be discussed similarly to Theory \hyperlink{so8-th4}{4}. The 6d theory has a flavor symmetry of $\su(2)$, 
and, as expected, the Hilbert series reads
    \begin{equation}
        \HS(t, \omega) = \PE\left[t^2 (2 \omega +1)+t^4 (8 \omega +11)+\mathcal{O}\left(t^{5}\right)\right]\fstop
    \end{equation}
    From the 6d perspective, this $\mathfrak{su}(2)$ arises from the fact that some of the gauge-invariant combinations involving spinorial generators of the rank one $(D, D)$ conformal matter building blocks contribute moment maps after gauging. This is similar to Theory \hyperlink{so8-th4}{4}; as such, we do not have an a priori proposal for the global symmetry after discrete gauging.
    The computation of the wreathed Hilbert series leads to
    \begin{equation}
        \HS_{\wr\,\ZZ_2}(t, \omega) =\PE\left[t^2 (2 \omega +1)+t^4 (6 \omega +8)+\mathcal{O}\left(t^{5}\right)\right]\coma
    \end{equation}
    which indicates that the wreathing did not affect the flavor symmetry of the theory. Nevertheless, we can see that the higher-order operator spectrum has been modified.

\subsubsection*{Theory \hyperlink{so8-th7}{7}: \texorpdfstring{$A_{3}^{\so(8)}([4, 2,1^2], [4, 2,1^2])$}{A3SO8}}

From now on, all the flavor symmetries of the wreathed theories follow from the discussion in Section \ref{sec:discretegaugingCM}. For instance, Theory \hyperlink{so8-th7}{7} has four $\so(2)$ factors as the flavor symmetry, which is reflected in the computation of the Hilbert series
    \begin{equation}
        \HS(t, \omega) = \PE\left[4t^2+t^4(19+8\omega)+\mathcal{O}(t^5)\right]\fstop
    \end{equation}
    After wreathing, as explained in Section \ref{sec:discretegaugingCM}, the four $\so(2)$s are identified in pairs, leading to effectively an $\so(2)\oplus \so(2)$ flavor symmetry, confirmed by the computation of the Hilbert series of the wreathed quiver, as follows:
    \begin{equation}
         \HS_{\wr\,\ZZ_2}(t, \omega) = \PE\left[2t^2+t^4(15+6\omega)+\mathcal{O}(t^5)\right]\fstop
    \end{equation}

\begin{landscape}
 \pagestyle{empty}
   \renewcommand{\arraystretch}{1.1}
   \begin{longtable}{c|c|c|c}
       \caption{Orthosymplectic magnetic quivers of the theories in Table \ref{tab:flavorsymm6d}.}\label{tbl:exampleswreathing}\\    
            \# & Conformal Matter & Tensor Branch & Magnetic Quiver \\
            \hhline{=|=|=|=}
  \endfirsthead
  \# & Conformal Matter & Tensor Branch & Magnetic Quiver \\\hhline{=|=|=|=} 
  \endhead
  \hypertarget{so6-th1}{} $1$ &  $A_{3}^{\so(6)}([3, 1^3], [3, 1^3])$ & $\begin{gathered}
               \overset{\su(2)}{2}\,\,\overset{\su(4)}{2}\,\,\overset{\su(2)}{2}
           \end{gathered}$ & 
    \begin{tikzpicture}[baseline=0,font=\footnotesize]
        \node[node, label=below:{$2$},fill=red] (A3) {};
      \node[node, label=below:{$2$},fill=blue] (A4) [right=6mm of A3] {};
      \node[node, label=below:{$6$},fill=red] (N3) [right=6mm of A4] {};
      \node[node, label=right:{$8$},fill=blue] (Nu) [above=4mm of N3] {};
        \node[node, label=below:{$2$},fill=blue] (A5) [right=6mm of N3] {};
        \node[node, label=below:{$2$},fill=red] (A6) [right=6mm of A5] {};
    
        \draw (A3.east) -- (A4.west);
      \draw (A4.east) -- (N3.west);
      \draw (N3.east) -- (A5.west);    
      \draw (N3.north) -- (Nu.south);
      \draw (Nu) to[out=130, in=410, looseness=12] (Nu);
      \draw (A5.east) -- (A6.west);
    \end{tikzpicture} 
 \\\hline
  \hypertarget{so6-th2}{} $2$ &  $A_{1}^{\so(6)}([1^6], [1^6])$ & $\begin{gathered}
               \overset{\su(4)}{2}
           \end{gathered}$ & 
    \begin{tikzpicture}[baseline=0,font=\footnotesize]
        \node[node, label=below:{$2$},fill=red] (A1) {};
      \node[node, label=below:{$2$},fill=blue] (A2) [right=6mm of A1] {};
      \node[node, label=below:{$4$},fill=red] (A3) [right=6mm of A2] {};
      \node[node, label=below:{$4$},fill=blue] (A4) [right=6mm of A3] {};
      \node[node, label=below:{$6$},fill=red] (N3) [right=6mm of A4] {};
      \node[node, label=right:{$4$},fill=blue] (Nu) [above=4mm of N3] {};
        \node[node, label=below:{$4$},fill=blue] (A5) [right=6mm of N3] {};
      \node[node, label=below:{$4$},fill=red] (A6) [right=6mm of A5] {};
      \node[node, label=below:{$2$},fill=blue] (A7) [right=6mm of A6] {};
      \node[node, label=below:{$2$},fill=red] (A8) [right=6mm of A7] {};
    
        \draw (A1.east) -- (A2.west);
        \draw (A2.east) -- (A3.west);
        \draw (A3.east) -- (A4.west);
      \draw (A4.east) -- (N3.west);
      \draw (N3.east) -- (A5.west);   
      \draw (A5.east) -- (A6.west);
      \draw (A6.east) -- (A7.west);
      \draw (A7.east) -- (A8.west);
      \draw (N3.north) -- (Nu.south);
      \draw (Nu) to[out=130, in=410, looseness=12] (Nu);
    \end{tikzpicture} 
  \\\hline
  \hypertarget{so6-th3}{} $3$ & $A_{5}^{\so(6)}([3^2], [3^2])$ & $\begin{gathered}
               \overset{\su(2)}{2}\,\,\overset{\su(3)}{2}\,\,\overset{\su(4)}{2}\,\,\overset{\su(3)}{2}\,\,\overset{\su(2)}{2}
           \end{gathered}$ & 
    \begin{tikzpicture}[baseline=0,font=\footnotesize]
      \node[node, label=below:{$2$},fill=blue] (A4) {};
      \node[node, label=below:{$6$},fill=red] (N3) [right=6mm of A4] {};
      \node[node, label=right:{$12$},fill=blue] (Nu) [above=4mm of N3] {};
        \node[node, label=below:{$2$},fill=blue] (A5) [right=6mm of N3] {};
    
      \draw (A4.east) -- (N3.west);
      \draw (N3.east) -- (A5.west);    
      \draw (N3.north) -- (Nu.south);
      \draw (Nu) to[out=130, in=410, looseness=12] (Nu);
    \end{tikzpicture} 
   \\\hline
   \hypertarget{so8-th4}{} $4$ & $A_{2}^{\so(8)}([3^2, 1^2], [3^2, 1^2])$ & $
               \overset{\su(3)}{3}\,\,1\,\,\overset{\su(3)}{3} $ & 
    \begin{tikzpicture}[baseline=0,font=\footnotesize]
        \node[node, label=below:{$2$},fill=red] (A3) {};
      \node[node, label=below:{$4$},fill=blue] (A4) [right=6mm of A3] {};
      \node[node, label=below:{$8$},fill=red] (N3) [right=6mm of A4] {};
      \node[node, label=right:{$6$},fill=blue] (Nu) [above=4mm of N3] {};
        \node[node, label=below:{$4$},fill=blue] (A5) [right=6mm of N3] {};
        \node[node, label=below:{$2$},fill=red] (A6) [right=6mm of A5] {};
        \draw (A3.east) -- (A4.west);
      \draw (A4.east) -- (N3.west);
      \draw (N3.east) -- (A5.west);    
      \draw (N3.north) -- (Nu.south);
      \draw (Nu) to[out=130, in=410, looseness=12] (Nu);
      \draw (A5.east) -- (A6.west);
    \end{tikzpicture} 
  \\\hline
  \hypertarget{so8-th5}{} $5$ & $A_{4}^{\so(8)}([5, 1^3], [5, 1^3])$ & $\begin{gathered}
               \overset{\su(2)}{2}\,\,\overset{\so(7)}{3}\,\,1\,\,\overset{\so(7)}{3}\,\,\overset{\su(2)}{2}
           \end{gathered}$ &
    \begin{tikzpicture}[baseline=0,font=\footnotesize]
        \node[node, label=below:{$2$},fill=red] (A3) {};
      \node[node, label=below:{$2$},fill=blue] (A4) [right=6mm of A3] {};
      \node[node, label=below:{$8$},fill=red] (N3) [right=6mm of A4] {};
      \node[node, label=right:{$10$},fill=blue] (Nu) [above=4mm of N3] {};
        \node[node, label=below:{$2$},fill=blue] (A5) [right=6mm of N3] {};
        \node[node, label=below:{$2$},fill=red] (A6) [right=6mm of A5] {};
        \draw (A3.east) -- (A4.west);
      \draw (A4.east) -- (N3.west);
      \draw (N3.east) -- (A5.west);    
      \draw (N3.north) -- (Nu.south);
      \draw (Nu) to[out=130, in=410, looseness=12] (Nu);
      \draw (A5.east) -- (A6.west);
    \end{tikzpicture} 
  \\\hline
  \hypertarget{so8-th6}{} $6$ & $A_{4}^{\so(8)}([5, 3], [5, 3])$ & $\begin{gathered}
               \overset{\su(2)}{2}\,\,\overset{\mathfrak{g}_2}{3}\,\,1\,\,\overset{\mathfrak{g}_2}{3}\,\,\overset{\su(2)}{2}
           \end{gathered}$ & 
    \begin{tikzpicture}[baseline=0,font=\footnotesize]
      \node[node, label=below:{$2$},fill=blue] (A4) {};
      \node[node, label=below:{$8$},fill=red] (N3) [right=6mm of A4] {};
      \node[node, label=right:{$10$},fill=blue] (Nu) [above=4mm of N3] {};
        \node[node, label=below:{$2$},fill=blue] (A5) [right=6mm of N3] {};
      \draw (A4.east) -- (N3.west);
      \draw (N3.east) -- (A5.west);    
      \draw (N3.north) -- (Nu.south);
      \draw (Nu) to[out=130, in=410, looseness=12] (Nu);
    \end{tikzpicture} 
  \\\hline
    \hypertarget{so8-th7}{} 7 & $A_{3}^{\so(8)}([4, 2,1^2], [4, 2,1^2])$  &$\overset{\su(3)}{3}\,1\,\overset{\so(8)}{4}\,1\,\overset{\so(3)}{3}$ & 
    \begin{tikzpicture}[baseline=0,font=\footnotesize]
        \node[node, label=below:{$2$},fill=red] (A3) {};
      \node[node, label=below:{$4$},fill=blue] (A4) [right=6mm of A3] {};
      \node[node, label=below:{$8$},fill=red] (N3) [right=6mm of A4] {};
      \node[node, label=right:{$8$},fill=blue] (Nu) [above=4mm of N3] {};
        \node[node, label=below:{$4$},fill=blue] (A5) [right=6mm of N3] {};
        \node[node, label=below:{$2$},fill=red] (A6) [right=6mm of A5] {};
    
        \draw (A3.east) -- (A4.west);
      \draw (A4.east) -- (N3.west);
      \draw (N3.east) -- (A5.west);    
      \draw (N3.north) -- (Nu.south);
      \draw (Nu) to[out=130, in=410, looseness=12] (Nu);
      \draw (A5.east) -- (A6.west);
    \end{tikzpicture} 
    \\\hline
    \hypertarget{so10-th8}{} $8$  &    $A_{2}^{\so(10)}([3^3, 1], [3^3, 1])$ & $\begin{gathered}
               \overset{\mathfrak{g}_2}{3}\,\,\overset{\su(2)}{1}\,\,\overset{\mathfrak{g}_2}{3}
           \end{gathered}$ & 
    \begin{tikzpicture}[baseline=0,font=\footnotesize]
        \node[node, label=below:{$4$},fill=red] (A3) {};
      \node[node, label=below:{$6$},fill=blue] (A4) [right=6mm of A3] {};
      \node[node, label=below:{$10$},fill=red] (N3) [right=6mm of A4] {};
      \node[node, label=right:{$6$},fill=blue] (Nu) [above=4mm of N3] {};
        \node[node, label=below:{$6$},fill=blue] (A5) [right=6mm of N3] {};
        \node[node, label=below:{$4$},fill=red] (A6) [right=6mm of A5] {};
    
        \draw (A3.east) -- (A4.west);
      \draw (A4.east) -- (N3.west);
      \draw (N3.east) -- (A5.west);    
      \draw (N3.north) -- (Nu.south);
      \draw (Nu) to[out=130, in=410, looseness=12] (Nu);
      \draw (A5.east) -- (A6.west);
    \end{tikzpicture} 
  \\\hline
  \hypertarget{so10-th9}{} $9$  & $A_{4}^{\so(10)}([5, 3, 1^2], [5, 3, 1^2])$ & $\begin{gathered}
               \overset{\su(3)}{3}\,\,1\,\,\overset{\so(9)}{4}\,\,
               \overset{\su(2)}{1}\,\,
               \overset{\so(9)}{4}\,\,
               1\,\,\overset{\su(3)}{3}
           \end{gathered}$ & 
    \begin{tikzpicture}[baseline=0,font=\footnotesize]
        \node[node, label=below:{$2$},fill=red] (A3) {};
      \node[node, label=below:{$4$},fill=blue] (A4) [right=6mm of A3] {};
      \node[node, label=below:{$10$},fill=red] (N3) [right=6mm of A4] {};
      \node[node, label=right:{$10$},fill=blue] (Nu) [above=4mm of N3] {};
        \node[node, label=below:{$4$},fill=blue] (A5) [right=6mm of N3] {};
        \node[node, label=below:{$2$},fill=red] (A6) [right=6mm of A5] {};
    
        \draw (A3.east) -- (A4.west);
      \draw (A4.east) -- (N3.west);
      \draw (N3.east) -- (A5.west);    
      \draw (N3.north) -- (Nu.south);
      \draw (Nu) to[out=130, in=410, looseness=12] (Nu);
      \draw (A5.east) -- (A6.west);
    \end{tikzpicture} 
  \\\hline
  \hypertarget{so10-th10}{} $10$  & $A_{4}^{\so(10)}([5^2], [5^2])$ & $\begin{gathered}
               \overset{\su(2)}{2}\,\,\overset{\so(7)}{3}\,\,
               \overset{\su(2)}{1}\,\,
               \overset{\so(7)}{3}\,\,
               \overset{\su(2)}{2}
           \end{gathered}$ & 
    \begin{tikzpicture}[baseline=0,font=\footnotesize]
      \node[node, label=below:{$4$},fill=blue] (A4) {};
      \node[node, label=below:{$10$},fill=red] (N3) [right=6mm of A4] {};
      \node[node, label=right:{$10$},fill=blue] (Nu) [above=4mm of N3] {};
        \node[node, label=below:{$4$},fill=blue] (A5) [right=6mm of N3] {};
    
      \draw (A4.east) -- (N3.west);
      \draw (N3.east) -- (A5.west);    
      \draw (N3.north) -- (Nu.south);
      \draw (Nu) to[out=130, in=410, looseness=12] (Nu);
    \end{tikzpicture} 
  \\\hline
     \hypertarget{so10-th11}{}   11 & $A_{3}^{\so(10)}([4, 3,2,1], [4, 3,2,1])$ &$\overset{\mathfrak{g}_2}{3}\,\,\overset{\su(2)}{1}\,\,\overset{\so(10)}{4}\,\,\overset{\su(2)}{1}\,\,\overset{\mathfrak{g}_2}{3}$ &
     \begin{tikzpicture}[baseline=0,font=\footnotesize]
        \node[node, label=below:{$4$},fill=red] (A3) {};
      \node[node, label=below:{$6$},fill=blue] (A4) [right=6mm of A3] {};
      \node[node, label=below:{$10$},fill=red] (N3) [right=6mm of A4] {};
      \node[node, label=right:{$8$},fill=blue] (Nu) [above=4mm of N3] {};
        \node[node, label=below:{$6$},fill=blue] (A5) [right=6mm of N3] {};
        \node[node, label=below:{$4$},fill=red] (A6) [right=6mm of A5] {};
    
        \draw (A3.east) -- (A4.west);
      \draw (A4.east) -- (N3.west);
      \draw (N3.east) -- (A5.west);    
      \draw (N3.north) -- (Nu.south);
      \draw (Nu) to[out=130, in=410, looseness=12] (Nu);
      \draw (A5.east) -- (A6.west);
    \end{tikzpicture} 
  \\\hline
  \hypertarget{so10-th12}{} $12$  & $A_{5}^{\so(10)}([5^2], [5^2])$ & $\begin{gathered}
               \overset{\su(2)}{2}\,\,\overset{\so(7)}{3}\,\,
               \overset{\su(2)}{1}\,\,
               \overset{\so(10)}{4}\,\,
               \overset{\su(2)}{1}\,\,
               \overset{\so(7)}{3}\,\,
               \overset{\su(2)}{2}
           \end{gathered}$ & 
    \begin{tikzpicture}[baseline=0,font=\footnotesize]
      \node[node, label=below:{$4$},fill=blue] (A4) {};
      \node[node, label=below:{$10$},fill=red] (N3) [right=6mm of A4] {};
      \node[node, label=right:{$12$},fill=blue] (Nu) [above=4mm of N3] {};
        \node[node, label=below:{$4$},fill=blue] (A5) [right=6mm of N3] {};
    
      \draw (A4.east) -- (N3.west);
      \draw (N3.east) -- (A5.west);    
      \draw (N3.north) -- (Nu.south);
      \draw (Nu) to[out=130, in=410, looseness=12] (Nu);
    \end{tikzpicture} 
  \\\hline
  \hypertarget{so10-th13}{} $13$ & $A_{6}^{\so(10)}([7, 1^3], [7, 1^3])$ & $\begin{gathered}
               \overset{\su(2)}{2}\,\,\overset{\so(7)}{3}\,\,1\,\,\overset{\so(9)}{4}\,\,
               \overset{\su(2)}{1}\,\,
               \overset{\so(9)}{4}\,\,
               1\,\,\overset{\so(7)}{3}\,\,
               \overset{\su(2)}{2}
           \end{gathered}$ & 
    \begin{tikzpicture}[baseline=0,font=\footnotesize]
        \node[node, label=below:{$2$},fill=red] (A3) {};
      \node[node, label=below:{$2$},fill=blue] (A4) [right=6mm of A3] {};
      \node[node, label=below:{$10$},fill=red] (N3) [right=6mm of A4] {};
      \node[node, label=right:{$14$},fill=blue] (Nu) [above=4mm of N3] {};
        \node[node, label=below:{$2$},fill=blue] (A5) [right=6mm of N3] {};
        \node[node, label=below:{$2$},fill=red] (A6) [right=6mm of A5] {};
    
        \draw (A3.east) -- (A4.west);
      \draw (A4.east) -- (N3.west);
      \draw (N3.east) -- (A5.west);    
      \draw (N3.north) -- (Nu.south);
      \draw (Nu) to[out=130, in=410, looseness=12] (Nu);
      \draw (A5.east) -- (A6.west);
    \end{tikzpicture} 
  \\\hline
  \hypertarget{so10-th14}{} $14$  & $A_{6}^{\so(10)}([7, 3], [7, 3])$ & $\begin{gathered}
               \overset{\su(2)}{2}\,\,\overset{\mathfrak{g}_2}{3}\,\,1\,\,\overset{\so(9)}{4}\,\,
               \overset{\su(2)}{1}\,\,
               \overset{\so(9)}{4}\,\,
               1\,\,\overset{\mathfrak{g}_2}{3}\,\,
               \overset{\su(2)}{2}
           \end{gathered}$ &
    \begin{tikzpicture}[baseline=0,font=\footnotesize]
      \node[node, label=below:{$2$},fill=blue] (A4) {};
      \node[node, label=below:{$10$},fill=red] (N3) [right=6mm of A4] {};
      \node[node, label=right:{$14$},fill=blue] (Nu) [above=4mm of N3] {};
        \node[node, label=below:{$2$},fill=blue] (A5) [right=6mm of N3] {};
    
      \draw (A4.east) -- (N3.west);
      \draw (N3.east) -- (A5.west);    
      \draw (N3.north) -- (Nu.south);
      \draw (Nu) to[out=130, in=410, looseness=12] (Nu);
    \end{tikzpicture} 
\end{longtable}
\end{landscape} 
    
\subsubsection*{Theory \hyperlink{so10-th8}{8}: \texorpdfstring{$A_{2}^{\so(10)}([3^3,1], [3^3,1])$}{A2SO10}}

In Table \ref{tab:flavorsymm6d}, we have that Theory \hyperlink{so10-th8}{8} has an $\so(6)$ flavor symmetry. This is confirmed also by the computation of the Hilbert series at leading order, as follows: 
    \begin{equation}
         \HS(t, \omega) = \PE\left[15t^2+8\omega t^3+33t^4+\mathcal{O}(t^5)\right]\fstop
    \end{equation}
    The wreathed Hilbert series is 
    \begin{equation}
        \HS_{\wr\,\ZZ_2}(t, \omega) = \PE\left[9t^2+6\omega t^3+40t^4+\mathcal{O}(t^5)\right]\fstop
    \end{equation}
    The flavor symmetry $\so(6)$ is coming only from the central $(-1)$-curve. As we explained in Section \ref{sec:discretegaugingCM}, under discrete gauging, such symmetry is projected to $\mathfrak{u}(3)$, which is also confirmed by the computation of the Hilbert series above.

\subsubsection*{Theory \hyperlink{so10-th9}{9}: \texorpdfstring{$A_{4}^{\so(10)}([5,3,1^2], [5,3,1^2])$}{A4SO10}}

Theory \hyperlink{so10-th9}{9} has a flavor symmetry of $\mathfrak{u}(1)\oplus\mathfrak{u}(1)\oplus\mathfrak{u}(1)$, confirmed by the computation of its Hilbert series:
    \begin{equation}
        \HS(t, \omega) = \PE\left[3 t^2+13t^4+\mathcal{O}(t^5)\right]\fstop
    \end{equation}
    The expectation is that the flavor symmetry after the $\ZZ_2$ wreathing becomes $\mathfrak{u}(1)\oplus\mathfrak{u}(1)$, with the identification of two of the $\mathfrak{u}(1)$s. This is confirmed also by the computation of the corresponding wreathed Hilbert series, i.e.,
    \begin{equation}
         \HS_{\wr\,\ZZ_2}(t, \omega) =\PE\left[2 t^2+9t^4+\mathcal{O}(t^5)\right]\fstop
    \end{equation}

\subsubsection*{Theory \hyperlink{so10-th10}{10}: \texorpdfstring{$A_{4}^{\so(10)}([5^2], [5^2])$}{A4SO10}}

The flavor symmetry for Theory \hyperlink{so10-th10}{10} is $\so(4)$, once again, attached to the $(-1)$-curve of the central node. The flavor symmetry is confirmed by the computation of the Hilbert series of the 3d magnetic quiver theory, i.e.,
    \begin{equation}
         \HS(t, \omega) = \PE\left[6t^2+2\omega t^3+10 t^4+\mathcal{O}(t^5)\right]\fstop
    \end{equation}
    According to Section \ref{sec:discretegaugingCM}, this flavor symmetry is projected to $\mathfrak{u}(2)$ after discrete gauging, and this is reflected also at the level of Hilbert series of the wreathed quiver:
    \begin{equation}
        \HS_{\wr\,\ZZ_2}(t, \omega) =\PE\left[4t^2+2\omega t^3+10 t^4+\mathcal{O}(t^5)\right]\fstop
    \end{equation}

\subsubsection*{Theory \hyperlink{so10-th11}{11}: \texorpdfstring{$A_{3}^{\so(10)}([4, 3,2,1], [4, 3,2,1])$}{A3SO10}}

A similar conclusion can be drawn for Theory \hyperlink{so10-th11}{11}, that has $\su(2)\oplus\su(2)$ flavor symmetry, as confirmed at the level of Hilbert series
    \begin{equation}
        \HS(t, \omega) = \PE\left[6t^2+27t^4+\mathcal{O}(t^5)\right]\coma
    \end{equation}
    that after wreathing gets identified into a single $\su(2)$, leading to the Hilbert series of the wreathed quiver starting from
    \begin{equation}
         \HS_{\wr\,\ZZ_2}(t, \omega) = \PE\left[3t^2+22t^4+\mathcal{O}(t^5)\right]\fstop
    \end{equation}

\subsubsection*{Theory \hyperlink{so10-th12}{12}: \texorpdfstring{$A_{5}^{\so(10)}([5^2], [5^2])$}{A4SO10}}

Theory \hyperlink{so10-th12}{12}  has an $\so(2)\oplus\so(2)$ flavor symmetry, as confirmed at the level of Hilbert series
    \begin{equation}
        \HS(t, \omega) = \PE\left[2 t^2+10 t^4+\mathcal{O}(t^5)\right]\coma
    \end{equation}
    and after discrete gauging has been identified into a single $\so(2)$, consistent with the Hilbert series of the wreathed quiver which starts from
    \begin{equation}
         \HS_{\wr\,\ZZ_2}(t, \omega) = \PE\left[t^2+8t^4+\mathcal{O}(t^5)\right]\fstop
    \end{equation}
    
\subsubsection*{Theory \hyperlink{so10-th13}{13}: \texorpdfstring{$A_{6}^{\so(10)}([7,1^3], [7,1^3])$}{A6SO10}}

Theory \hyperlink{so10-th13}{13} instead has a flavor symmetry of $\su(2)\oplus\su(2)\oplus\mathfrak{u}(1)$, and its Hilbert series also reflect such a flavor symmetry, giving
    \begin{equation}
         \HS(t, \omega) = \PE\left[7 t^2+3t^4+\mathcal{O}(t^5)\right]\fstop
    \end{equation}
    We can confirm that the flavor symmetry for the wreathed theory is consistent with $\su(2)\oplus\mathfrak{u}(1)$ by computing the corresponding wreathed Hilbert series:
    \begin{equation}
        \HS_{\wr\,\ZZ_2}(t, \omega) =\PE\left[4 t^2+9t^4+\mathcal{O}(t^5)\right]\fstop
    \end{equation}

\subsubsection*{Theory \hyperlink{so10-th14}{14}: \texorpdfstring{$A_{6}^{\so(10)}([7,3], [7,3])$}{A6SO10}}

Finally, Theory \hyperlink{so10-th14}{14} has a single $\mathfrak{u}(1)$ as flavor symmetry. The Hilbert series for this theory is simply
    \begin{equation}
        \HS(t, \omega) = \PE\left[t^2+5t^4+\mathcal{O}(t^5)\right]\coma
    \end{equation}
    and the $\mathfrak{u}(1)$ is preserved after the wreathing. We obtain a Hilbert series which is given by
    \begin{equation}
         \HS_{\wr\,\ZZ_2}(t, \omega) = \PE\left[t^2+4t^4+\mathcal{O}(t^5)\right]\,.
    \end{equation}
    Once again, we see that, even though the flavor symmetry remains unchanged, the higher-order operator content reorganizes.

 \begin{table}[t]
   \centering
   \begin{tabular}{c|c|c|c}
  \# & Conformal Matter & Tensor Branch & Magnetic Quiver  \\
  \hhline{=|=|=|=}
    \hypertarget{su3-th15}{} $15$  &    $A_{1}^{\su(3)}([1^3], [1^3])$ & $
               \overset{\su(3)}{2}
           $ & 
    \begin{tikzpicture}[baseline=0,font=\footnotesize]
        \node[node, label=below:{$1$}] (A3) {};
      \node[node, label=below:{$2$}] (A4) [right=6mm of A3] {};
      \node[node, label=below:{$3$}] (N3) [right=6mm of A4] {};
      \node[node, label=right:{$2$}] (Nu) [above=4mm of N3] {};
        \node[node, label=below:{$2$}] (A5) [right=6mm of N3] {};
        \node[node, label=below:{$1$}] (A6) [right=6mm of A5] {};
    
        \draw (A3.east) -- (A4.west);
      \draw (A4.east) -- (N3.west);
      \draw (N3.east) -- (A5.west);    
      \draw (N3.north) -- (Nu.south);
      \draw (Nu) to[out=130, in=410, looseness=12] (Nu);
      \draw (A5.east) -- (A6.west);
    \end{tikzpicture} 
  \\\hline
  \hypertarget{su3-th16}{} $16$  & $A_{3}^{\su(3)}([2, 1], [2, 1])$ & $
               \overset{\su(2)}{2}\,\,\overset{\su(3)}{2}\,\,\overset{\su(2)}{2}
           $ & 
    \begin{tikzpicture}[baseline=0,font=\footnotesize]
      \node[node, label=below:{$1$}] (A4) {};
      \node[node, label=below:{$3$}] (N3) [right=6mm of A4] {};
      \node[node, label=right:{$4$}] (Nu) [above=4mm of N3] {};
        \node[node, label=below:{$1$}] (A5) [right=6mm of N3] {};
    
      \draw (A4.east) -- (N3.west);
      \draw (N3.east) -- (A5.west);    
      \draw (N3.north) -- (Nu.south);
      \draw (Nu) to[out=130, in=410, looseness=12] (Nu);
    \end{tikzpicture} 
  \\\hline
  \hypertarget{su4-th17}{} $17$ & $A_{3}^{\su(4)}([2, 1^2], [2, 1^2])$ & $
               \overset{\su(3)}{2}\,\,\overset{\su(4)}{2}\,\,\overset{\su(3)}{2}
          $ & 
    \begin{tikzpicture}[baseline=0,font=\footnotesize]
        \node[node, label=below:{$1$}] (A3) {};
      \node[node, label=below:{$2$}] (A4) [right=6mm of A3] {};
      \node[node, label=below:{$4$}] (N3) [right=6mm of A4] {};
      \node[node, label=right:{$4$}] (Nu) [above=4mm of N3] {};
        \node[node, label=below:{$2$}] (A5) [right=6mm of N3] {};
        \node[node, label=below:{$1$}] (A6) [right=6mm of A5] {};
    
        \draw (A3.east) -- (A4.west);
      \draw (A4.east) -- (N3.west);
      \draw (N3.east) -- (A5.west);    
      \draw (N3.north) -- (Nu.south);
      \draw (Nu) to[out=130, in=410, looseness=12] (Nu);
      \draw (A5.east) -- (A6.west);
    \end{tikzpicture} 
    \end{tabular}
 \caption{The unitary magnetic quivers for the theories in Table \ref{tab:flavorsymm6d} which do not admit (good) orthosymplectic magnetic quivers.}\label{tbl:workingsu3su4}    
\end{table}

\subsubsection*{Theories \hyperlink{su3-th15}{15}, \hyperlink{su3-th16}{16} and \hyperlink{su4-th17}{17}}

While the principal focus in this section has been on orthosymplectic quivers and their wreathing, we now briefly include some examples for which we do not have (good) orthosymplectic descriptions, but only unitary descriptions. These examples again demonstrate that the Coulomb symmetry as determined by the Hilbert series of the wreathed quiver matches the expectations from the 6d tensor branch analysis in Section \ref{sec:discretegaugingCM}. We have written the 6d SCFTs and the magnetic quivers for their Higgs branches in Table \ref{tbl:workingsu3su4}.

Theory \hyperlink{su3-th15}{15} is expected to have an $\mathfrak{su}(6)$ global symmetry before $\ZZ_2$ discrete gauging and a $\mathfrak{usp}(6)$ global symmetry after the gauging. This is replicated in the Hilbert series of the Coulomb branch of the respective magnetic quivers:
\begin{equation}
 \begin{aligned}
      \HS(t) = &\,\PE\left[35 t^2+20 t^3-35 t^4+\mathcal{O}(t^5)\right]\coma\\
      \HS_{\wr\,\ZZ_2}(t) =&\,\PE\left[21 t^2+20 t^3+105 t^4+\mathcal{O}(t^5)\right]\fstop
  \end{aligned}
\end{equation}

Theory \hyperlink{su3-th16}{16} has a non-Abelian $\mathfrak{su}(2)$ global symmetry attached to the central $(-2)$-curve in the tensor branch configuration, as well as Abelian symmetries localized on the left and the right of the configuration. We expect that the $\mathfrak{su}(2) \rightarrow \mathfrak{usp}(2)$ and the two $\mathfrak{u}(1)$s are identified under the discrete gauging; we see this also from the Coulomb branch Hilbert series:
\begin{equation}
    \begin{aligned}
      \HS(t) = &\,\PE\left[5 t^2+12 t^3+12 t^4+\mathcal{O}(t^5)\right]\coma\\
      \HS_{\wr\,\ZZ_2}(t) =&\,\PE\left[4 t^2+8 t^3+11 t^4+\mathcal{O}(t^5)\right]\fstop
  \end{aligned}
\end{equation}

Finally, we come to Theory \hyperlink{su4-th17}{17}. This theory has a realization as Higgsed $(\mathfrak{su}(4), \mathfrak{su}(4))$ conformal matter, and thus it does have a orthosymplectic description for the Higgs branch; however, this quiver contains symplectic gauge nodes which are not good, and thus the application of the monopole formula fails. Regardless, we can determine the Coulomb branch Hilbert series of the unitary magnetic quiver, before and after wreathing: 
\begin{equation}
    \begin{aligned}
      \HS(t) = &\,\PE\left[11 t^2+16t^3+\mathcal{O}(t^4)\right]\coma\\
      \HS_{\wr\,\ZZ_2}(t) =&\,\PE\left[7 t^2+6t^3+\mathcal{O}(t^4)\right]\fstop
  \end{aligned}
\end{equation}
We find, as expected, that the change in the number of R-charge $2$ operators is consistent with the modification of the global symmetry as follows:
\begin{equation}
    \mathfrak{su}(2)^{\oplus 3} \oplus \mathfrak{u}(1)^{\oplus 2} \quad \rightarrow \quad  \mathfrak{su}(2)^{\oplus 2} \oplus \mathfrak{u}(1) \,.
\end{equation}

\section{Discussion}\label{sec:disc}

In this work, we focused on 6d $(1,0)$ SCFTs that have a $\ZZ_2$ Green--Schwarz automorphism, and we studied the Higgs branch of such theories after the gauging of this discrete symmetry. This study has been developed first at the level of the tensor branch in 6d of the theory, and confirmed by computing the Coulomb branch Hilbert series of the wreathed magnetic quivers for their Higgs branches. While the matching between the $\mathbb{Z}_2$ discrete gauging and the $\mathbb{Z}_2$ wreathing is remarkable, we have only just started to scratch the surface of the study of discrete symmetries and their gauging in the landscape of 6d $(1,0)$ SCFTs.

\paragraph{\uline{Refinement and the 6d flavor symmetry}} The 6d proposal for the continuous flavor symmetry of the discretely-gauged theory is almost completely clear from the discussion in Section \ref{sec:discretegaugingCM}. However, whenever there is a central undecorated $(-1)$-curve (meaning that an E-string flavor symmetry is attached to it), the resulting flavor symmetry is not always easily determined. The computation of the Hilbert series gives the expected dimension of the flavor symmetry algebra, however, this is coarse information and the precise identification of the algebra is lacking. One way out is to compute the refined Hilbert series, by using a different prescription for the computation of the Hilbert series, such as proposed in \cite{Cremonesi:2014kwa}. In \cite{Harding:2025aaa}, this approach is being explored, with an expanded investigation of the wreathing procedure, extending the prescription proposed in \cite{Grimminger:2024mks}. This analysis will lay the groundwork for a broader generalization, enabling the wreathing of unitary and orthosymplectic quivers beyond $\ZZ_2$.

\paragraph{\uline{Geometric realization}} As we have discussed, the geometric construction of 6d $(1,0)$ SCFTs \cite{Heckman:2013pva,Heckman:2015bfa} is based on the compactification of F-theory on singular elliptically-fibered Calabi--Yau threefolds that satisfy specific properties. The strength of this approach lies in its ability to systematically construct a sufficiently non-singular elliptically-fibered Calabi--Yau manifold, which encodes the tensor branch effective field theory; this manifold can then be contracted to a singular Calabi--Yau that engineers the corresponding superconformal theory. Moreover, this framework provides an algorithm to classify all possible tensor branch Calabi--Yau manifolds. Consequently, it has been proposed that this geometric construction is not just a method for engineering 6d $(1,0)$ SCFTs but also serves as a classification --- suggesting that every such SCFT arises from an F-theory compactification on a Calabi--Yau threefold. The discretely-gauged 6d $(1,0)$ SCFTs that we have proposed in this paper constitute a challenge to this proposal; we must ask what are the elliptically-fibered Calabi--Yau threefolds that engineer these novel SCFTs in F-theory. We expect that it is, in fact, the same Calabi--Yau as for the non-discretely-gauged SCFT, paired with the action of a geometric automorphism that implements the gauging of the (geometric) $\mathbb{Z}_2$ Green--Schwarz automorphism. We expect to tackle this geometric question in future work.

\paragraph{\uline{Hasse diagrams for wreathed quivers}}
The Higgs branch of an eight-supercharge SCFT is expected to have the structure of a symplectic singularity \cite{Beauville_2000,Kaledin_2006}. A symplectic singularity, $X$, admits a finite stratification into symplectic leaves, with partial ordering via inclusion $X_0 \subset X_1 \subset \cdots \subset X$, and the transverse slice between each leaf and the total space $X$ corresponds to the Higgs branch of an SCFT which can be obtained via Higgsing of the original SCFT \cite{Bourget:2019aer}. The stratification can be captured concisely in a Hasse diagram. To understand the pattern of Higgsing in a given theory, it is necessary to understand the stratification of the Higgs branch as a symplectic singularity. In particular, we can ask when there exists a Higgs branch renormalization group flow from the discretely-gauged theory $\widetilde{\mathcal{T}}$ to another 6d $(1,0)$ SCFT $\mathcal{T}'$, which may or may not be discretely-gauged. For example, it seems reasonable to expect that there exist flows between 
\begin{equation}\label{eqn:sparx}
    \widetilde{A}_{N-1}^{\mathfrak{g}}(O, O) \quad \rightarrow \quad \widetilde{A}_{N-1}^{\mathfrak{g}}(O', O') \,,
\end{equation}
whenever $O' < O$ in the partial ordering on nilpotent orbits of $\mathfrak{g}$. Here, as usual, the tilde indicates the discrete gauging of the $\mathbb{Z}_2$ Green--Schwarz automorphism. Algorithms have been developed \cite{Cabrera:2018ann,Bourget:2023dkj,Bourget:2024mgn,Lawrie:2024wan,Cabrera:2019dob,Lawrie:2024zon} to determine the stratification structure of 3d $\mathcal{N}=4$ Lagrangian quivers satisfying certain restrictive properties; the extension of these algorithms to wreathed quivers would allow us to extract the stratification of the Higgs branch of the discretely-gauged conformal matter theories, and thus determine which theories are related by Higgs branch renormalization group flows. Alternatively, the geometric expectations from the six-dimensional perspective, such as in equation \eqref{eqn:sparx}, may guide in the development of such algorithms, similarly to the approach in \cite{Lawrie:2024wan}.

\paragraph{\uline{Stiefel--Whitney twists and 4d $\mathcal{N}=2$ SCFTs}} Here, we have considered the Higgs branches of discretely-gauged 6d $(1,0)$ SCFTs. For theories in four, five, or six dimensions with eight-supercharges, when they are compactified on a circle the Higgs branch is not modified \cite{Argyres:1996eh}; therefore, we have equally studied the Higgs branch of circle compactifications of the 6d $(1,0)$ SCFTs in question. Let us focus on 4d $\mathcal{N}=2$ SCFTs obtained via a $T^2$ compactification of 6d $(1,0)$; it is sometimes possible to turn on a discrete $\mathbb{Z}_\ell$, for $\ell = 2, \ldots, 6$, Stiefel--Whitney twist along the torus. The resulting theories have been discussed in \cite{Apruzzi:2020pmv,Ohmori:2018ona,Giacomelli:2024dbd,Giacomelli:2020jel,Heckman:2020svr,Giacomelli:2020gee,Bourget:2020mez,Heckman:2022suy}, and generally they give rise to novel classes of 4d $\mathcal{N}=2$ SCFTs.\footnote{Recent results in \cite{Giacomelli:2024ycb} have demonstrated that all class $\mathcal{S}$ theories of A-type with regular untwisted punctures arise through torus-compactification, without Stiefel--Whitney twist, and further mass deformation of certain 6d $(1,0)$ SCFTs. Incorporating the Stiefel--Whitney twist into such an analysis would appear to thus be a powerful approach for the study of the landscape of 4d $\mathcal{N}=2$ SCFTs.} Magnetic quivers for the Higgs branches of such Stiefel--Whitney twisted SCFTs have been generally proposed in \cite{Giacomelli:2024dbd}; these quivers, perhaps after mass deformations (i.e., Fayet--Iliopoulos deformations in the magnetic quiver), often have discrete diagram automorphisms, and thus it is natural to consider the wreathing of the magnetic quivers, and to speculate that these will be magnetic quivers for certain discretely-gauged versions of the Stiefel--Whitney twisted 4d $\mathcal{N}=2$ SCFTs. We leave the thorough exploration of this proposal for the future.

\paragraph{\uline{Non-invertible symmetries}} It is well-known that the gauging of discrete 0-form symmetries, such as the $\mathbb{Z}_2$ Green--Schwarz automorphisms that we discuss in this article, can lead to non-invertible symmetries, a form of generalized global symmetry \cite{Gaiotto:2014kfa}, in the discretely-gauged theory \cite{Tachikawa:2017gyf}. This phenomenon has been explored in 4d gauge theories, where the gauging of the outer-automorphism of the gauge algebra leads to a 1-form symmetry becoming non-invertible \cite{Arias-Tamargo:2022nlf,Bhardwaj:2022yxj}; this is straightforward to see using the technology that was developed in \cite{Heidenreich:2021xpr,Rudelius:2020orz}, and which is intimately related to the anomalies of the outer-automorphisms \cite{Henning:2021ctv}. 
To give a few more examples: mixed anomalies between 0-form and 1-form symmetries for Argyres--Douglas theories have been explored in \cite{Carta:2023bqn} and for a wide class of $\mathcal{N}=1$ Lagrangian theories in \cite{Kang:2024elv}; the discrete gauging of the 0-form symmetry is then again expected to give rise to a non-invertible 1-form symmetry in the discretely-gauged theory.

In the 6d $(1,0)$ SCFTs in which we are interested, a 2-form global symmetry, if it exists and has a mixed anomaly with the $\mathbb{Z}_2^{[0]}$ 0-form symmetry, would be expected to become non-invertible under the discrete gauging. Before it makes sense to discuss the 2-form symmetry of a 6d SCFT, it is necessary to select a polarization of the intermediate defect group, which can be determined from the tensor branch \cite{DelZotto:2015isa}, of the theory; this is a necessary data to have a well-defined quantum field theory (as opposed to a relative quantum field theory \cite{Freed:2012bs}). Unlike four dimensions, the Dirac pairing in six dimensions is symmetric which means that the intermediate defect group does not always admit a polarization; the process of determining polarizations for 6d SCFTs has been studied in detail in \cite{Gukov:2020btk,Lawrie:2023tdz}.\footnote{The discussion here should not be confused with that of \cite{Lawrie:2023tdz}; in that reference, the authors construct non-invertible 0-form symmetries out of the Green--Schwarz automorphisms of certain 6d $(1,0)$ SCFTs by gauging instead the 2-form symmetry, following the duality defect construction of \cite{Choi:2021kmx,Kaidi:2022uux,Kaidi:2021xfk}.} To give a simple example, the Higgsed conformal matter theory which we have called Theory \hyperlink{so6-th1}{1}, i.e.,
\begin{equation}
    A_3^{\mathfrak{su}(4)}([2^2], [2^2]) \,,
\end{equation}
has a $\mathbb{Z}_2$ Green--Schwarz automorphism and admits a polarization of the intermediate defect group that leads to a $\mathbb{Z}_2$ 2-form symmetry. As such, we would expect that the discretely-gauged theory, whose magnetic quiver was given in equation \eqref{eq:A3SU4}, has a non-invertible 2-form symmetry. It would be interesting to explore the reflection of such generalized symmetries in the associated magnetic quiver.\footnote{Connections between generalized symmetries in higher-dimensional theories and the magnetic quivers for their Higgs branches have been explored, for example, in \cite{Mekareeya:2022spm,Bhardwaj:2023zix,Nawata:2023rdx}.}

\subsection*{Acknowledgements}

We thank Guillermo Arias-Tamargo, Antoine Bourget, Jacques Distler, Simone Giacomelli, William Harding, Monica Jinwoo Kang, Lorenzo Mansi, and  Noppadol Mekareeya for helpful discussions. We are grateful to Lorenzo Mansi for insightful comments on a previous draft of this article. C. L.~and A. M.~thank the Simons Summer Workshop 2024 for hospitality
during an important stage of this work; C. L.~further thanks the Department of Physics at UW--Madison for hospitality when the final phase of this work was completed. C. L.~acknowledges support from DESY (Hamburg, Germany), a member of the Helmholtz Association HGF; C. L.~also acknowledges the Deutsche Forschungsgemeinschaft under Germany's Excellence Strategy - EXC 2121 ``Quantum Universe'' - 390833306 and the Collaborative Research Center - SFB 1624 ``Higher Structures, Moduli Spaces, and Integrability'' - 506632645. T. L.~is supported by HORIZON-MSCA-2023-SE-01-101182937-HeI. A. M.~is supported in part by the DOE grant DE-SC0017647. Until September 2024, A. M.~was supported in part by Deutsche Forschungsgemeinschaft under Germany's Excellence Strategy EXC 2121 Quantum Universe 390833306 and by Deutsche Forschungsgemeinschaft through a German--Israeli Project Cooperation (DIP) grant ``Holography and the Swampland''.  

\appendix

\section{Prefactors for Coulomb Branch Hilbert Series}
\label{sec:prefactors}

In this section, we are going to review how to compute the contribution to the Hilbert series given by the residual gauge group, which is left unbroken by a given choice of magnetic fluxes. This will serve as a starting point to explain how the procedure is modified in case of $\ZZ_2$ wreathing that we have explained in Section \ref{sec:wreathedprefactors}.

\subsection{\texorpdfstring{$\U(N)$}{U(N)} Prefactor}
\label{sec:UNPrefactor}

Let us start from the unitary group $\U(N)$ and we review the discussion in \cite[Appendix A]{Cremonesi:2013lqa}. For a given choice of fluxes, $\mathbf{m}\in \ZZ^N$, with $m_1\geq \ldots \geq m_N$, it is possible to define a partition $\lambda(\mathbf{m})$ that encodes how many fluxes $m_i$ are equal. Hence, $\lambda(\mathbf{m})$ is a partition of $N$ associated to the magnetic flux $\mathbf{m}$, extended with zeros to be of length $N$. Let us call $\lambda_i(\mathbf{m})$ the components of $\lambda(\mathbf{m})$, satisfying $\sum_i \lambda_i(\mathbf{m})=N$. The commutant of the monopole flux that gives the residual gauge group is then given by $\prod_{i=1}^N\U(\lambda_i(\mathbf{m}))$. The classical Casimir contribution is 
\begin{equation}
    P_{\U(N)}(t,\mathbf{m}) =\prod_{j=1}^NZ_{\lambda_j(\mathbf{m})}^{\U} \coma \text{ with } Z_l^{\U} = \begin{dcases} \prod_{i=1}^l\frac{1}{1-t^{2i}} & l\geq 1\coma\\ 1 & l=0\fstop\end{dcases}
\end{equation}
It is possible to rewrite the classical Casimir contribution in a way that is better suitable for wreathing as follows: one can consider the Weyl group of $\U(N)$, i.e., $S_N$, and construct the $N\times N$ matrix representation of the group. The commutant of the monopole flux $\mathbf{m}$ is given by
\begin{equation}
    W(\mathbf{m})=\left\{\left.g\in  S_N\right| g\cdot \mathbf{m} = \mathbf{m}\right\}\coma
\end{equation}
so that for a given choice of fluxes, the prefactor is given by
\begin{equation}
    P_{\U(N)}(t,\mathbf{m}) = \frac{1}{|W(\mathbf{m})|}\sum_{\gamma \in W(\mathbf{m})}\frac{1}{\det\left(\ID-t^2\gamma\right)}\fstop
    \label{eq:secondwayprefactor}
\end{equation}
The advantage of the first description for the classical contribution is that it makes explicit that for a given $\U(N)$ gauge groups, the choice of fluxes breaks the gauge group into smaller $\U(M)$ factors, and the prefactor can then be computed recursively. The second description, instead, makes a clearer connection with the Weyl group and its elements, with $W(\mathbf{m})$ selecting those matrices of $S_N$ that admit an eigenvector $\mathbf{m}$ with unitary eigenvalue. However, the two prescriptions together allow us to see $W(\mathbf{m})$ as the group given by the product of $S_{\lambda_j(\mathbf{m})}$ symmetric groups.

In \cite[Appendix A]{Cremonesi:2013lqa}, the authors used the definition of the prefactor for $\U(N)$ also to construct the prefactors for $\USp$ and $\SO$ groups. Our aim is then to express the prefactors using equation \eqref{eq:secondwayprefactor}.

\subsection{\texorpdfstring{$\USp(2N)$}{USp(2N)} Prefactor}\label{sec:USp2NPrefactor}

For $\USp(2N)$ (and analogously $\SO(2N+1)$), one defines the flux vector $\mathbf{m}$, such that $m_1\geq \ldots \geq m_N\geq 0$. The relevant part is that the choice of flux will break the residual gauge group $\USp(2N)$ to products of $\U(M)$ and $\USp(2M)$ factors, depending on which fluxes are vanishing or not. As in \cite{Cremonesi:2013lqa}, let us call $\lambda_0(\mathbf{m})$ the number of vanishing magnetic fluxes, and let us collect all the other fluxes into a partition $\lambda(\mathbf{m})$ counting how many fluxes are equal, but non-vanishing.\footnote{Once again, the partition is extended with zeros so that it has length $N$.} By construction, we have $\lambda_0(\mathbf{m})+\sum_i \lambda_i(\mathbf{m}) = N$. The residual gauge group is then $\prod_{i=1}^N\U(\lambda_i(\mathbf{m}))\times \USp(2\lambda_0(\mathbf{m}))$. The classical prefactor is 
\begin{equation}
    P_{\USp(2N)}(t,\mathbf{m}) =Z_{\lambda_0(\mathbf{m})}^{\USp}\prod_{j=1}^NZ_{\lambda_j(\mathbf{m})}^{\U} \coma \text{ with } Z_l^{\USp} = \begin{dcases} \prod_{i=1}^l\frac{1}{1-t^{4i}} & l\geq 1\coma\\ 1 & l=0\fstop\end{dcases}
\end{equation}
In rewriting this contribution using equation \eqref{eq:secondwayprefactor}, one can still consider the full Weyl group of $\USp(2N)$, i.e., $S_N\ltimes \ZZ_2^N$, and consider the matrix elements that are left invariant by a given choice of matrices. However, the observation in \cite{Cremonesi:2013lqa} is that we can generate that group by considering a product of $S_{\lambda_j(\mathbf{m})}$ and $S_{\lambda_0(\mathbf{m})}$ groups but with the Casimirs in $S_{\lambda_0(\mathbf{m})}$ with double the charge. This means that one can consider the $N\times N$ matrix realization of the group elements in $W(\mathbf{m})=\varprod_j S_{\lambda_j(\mathbf{m})}\times S_{\lambda_0(\mathbf{m})}$ and replace equation \eqref{eq:secondwayprefactor} with
\begin{equation}
    P_{\USp(2N)}(t,\mathbf{m}) = \frac{1}{|W(\mathbf{m})|}\sum_{\gamma \in W(\mathbf{m})}\frac{1}{\det\left(\ID-\mathbf{T}^2\cdot\gamma\right)}\coma
\end{equation}
where
\begin{equation}
    \mathbf{T}^2 = \diag\left(\vphantom{t^2}\right.\underbrace{t^2,\ldots,t^2}_{\sum_j \lambda_j(\mathbf{m})},\underbrace{t^4,\ldots,t^4}_{\lambda_0(\mathbf{m})}\left.\vphantom{t^2}\right)\fstop
\end{equation}

As already mentioned, the prefactor for $\SO(2N+1)$ is the same as for $\USp(2N)$, so we can now discuss $\SO(2N)$. 

\subsection{\texorpdfstring{$\SO(2N)$}{SO(2N)} Prefactor}\label{sec:SO2NPrefactor}

The magnetic fluxes $\mathbf{m}$ for $\SO(2N)$ groups are now restricted to be $m_1\geq \ldots \geq m_{N-1}\geq |m_N|$, so one defines a vector $\mathbf{n} = (m_1,\ldots,m_{N-1},|m_{N}|)$, which depends on the magnetic fluxes $\mathbf{m}$, and we can proceed as in Appendix \ref{sec:USp2NPrefactor}. We call $\lambda_0(\mathbf{n})$ the number of vanishing $\mathbf{n}$ fluxes, and $\lambda(\mathbf{n})$ the partition counting how many fluxes are equal, but non-vanishing. The residual gauge group is $\prod_{i=1}^N\U(\lambda_i(\mathbf{n}))\times \SO(2\lambda_0(\mathbf{n}))$, giving the prefactor
\begin{equation}
    P_{\SO(2N)}(t,\mathbf{n}) =Z_{\lambda_0(\mathbf{n})}^{\SO}\prod_{j=1}^NZ_{\lambda_j(\mathbf{n})}^{\U} \coma \text{ with } Z_l^{\SO} = \begin{dcases} \frac{1}{1-t^{2l}}\prod_{i=1}^{l-1}\frac{1}{1-t^{4i}} & l\geq 1\coma\\ 1 & l=0\fstop\end{dcases}
\end{equation}
Once again, one can construct the Weyl group of $\SO(2N)$, i.e., $S_N\ltimes \ZZ_2^{N-1}$, or by analogy to the $\USp$ case above, we can consider $W(\mathbf{n})=\varprod_j S_{\lambda_j(\mathbf{n})}\times S_{\lambda_0(\mathbf{n})-1}\times S_1$, leading to
\begin{equation}
    P_{\SO(2N)}(t,\mathbf{n}) = \frac{1}{|W(\mathbf{n})|}\sum_{\gamma \in W(\mathbf{n})}\frac{1}{\det\left(\ID-\mathbf{T}^2\cdot\gamma\right)}\coma
\end{equation}
where
\begin{equation}
    \mathbf{T}^2 = \diag\left(\vphantom{t^{2\lambda_0(\mathbf{n})}}\right.\underbrace{t^2,\ldots,t^2}_{\sum_j \lambda_j(\mathbf{n})},\underbrace{t^4,\ldots,t^4}_{\lambda_0(\mathbf{n})-1},t^{2\lambda_0(\mathbf{n})}\left.\vphantom{t^{2\lambda_0(\mathbf{n})}}\right)\fstop
\end{equation}

\bibliographystyle{sortedbutpretty}
\bibliography{references}
	
\end{document}